\documentclass[12pt]{article}

\usepackage{amsmath,amssymb,epsfig,subfigure}

\usepackage{graphicx}
\usepackage{cite}
\makeatletter

\def\unit{{1\kern-.65ex {\rm l}}}
\def\1{{1\kern-.65ex {\rm l}}}

\def\Im{\mathop{\mathrm{Im}}\nolimits}

\def\Re{\mathop{\mathrm{Re}}\nolimits}
\def\Res{\mathop{\mathrm{Res}}\limits}

\def\tr{\mathop{\mathrm{tr}}\nolimits}

\def\Bracket#1{{\left\langle{#1}\right\rangle}}
\let\ev\bracket

\def\CA{{\cal A}}
\def\CB{{\cal B}}

\def\CF{{\cal F}}

\def\CM{{\cal M}}
\def\CN{{\cal N}}
\def\CO{{\cal O}}

\def\CW{{\cal W}}

\def\bbC{{\mathbb{C}}}

\def\bbP{{\mathbb{P}}}

\def\bbR{{\mathbb{R}}}

\newcount\hour \newcount\minute
\hour=\time \divide \hour by 60
\minute=\time
\def\now{%
\ifnum \hour<13
  \ifnum \hour=0 \advance \hour by 12 \number\hour:\else \number\hour:\fi%
     \ifnum \minute<10 0\fi%
     \number\minute%
\ A.M.%
\else \advance \hour by -12 \number\hour:%
  \ifnum \minute<10 0\fi%
  \number\minute%
  \ P.M.%
\fi%
}
\makeatother
\begin{document}
\baselineskip=18pt  
\numberwithin{equation}{section}  
\allowdisplaybreaks  
\thispagestyle{empty}
\vspace*{-2cm} 
\begin{flushright}
{\tt arXiv:0804.4006}\\
CALT-68-2682\\
ITFA-2008-14
\end{flushright}
\vspace*{0.8cm} 
\begin{center}
{\LARGE Nonsupersymmetric Flux Vacua and\\[1ex] Perturbed $\CN=2$ Systems\\}
 \vspace*{1.5cm}
 Lotte Hollands$^1$, Joseph Marsano$^2$, Kyriakos Papadodimas$^1$, and Masaki
Shigemori$^1$\\
 \vspace*{1.0cm} 
 $^1$ 
 Institute for Theoretical Physics, University of Amsterdam\\
 Valckenierstraat 65, 1018 XE Amsterdam, The Netherlands\\[1ex]
 $^2$ California Institute of Technology 452-48, Pasadena, CA 91125, USA\\
 \vspace*{0.6cm} 
 \verb|lholland_at_science.uva.nl|,
 \verb|marsano_at_theory.caltech.edu|,
 \verb|kpapado_at_science.uva.nl|,
 \verb|mshigemo_at_science.uva.nl|
\end{center}
\vspace*{.5cm}
\noindent

We geometrically engineer $\CN=2$ theories perturbed by a superpotential
by adding 3-form flux with support at infinity to local Calabi-Yau
geometries in type IIB\@.  This allows us to apply the formalism of
Ooguri, Ookouchi, and Park [arXiv:0704.3613] to demonstrate that, by
tuning the flux at infinity, we can stabilize the dynamical complex
structure moduli in a metastable, supersymmetry-breaking configuration.
Moreover, we argue that this setup can arise naturally as a limit of a
larger Calabi-Yau which separates into two weakly interacting regions;
the flux in one region leaks into the other, where it appears to be
supported at infinity and induces the desired superpotential.  In our
endeavor to confirm this picture in cases with many 3-cycles, we also
compute the CIV-DV prepotential for arbitrary number of cuts up to fifth
order in the glueball fields.
 
%
%

\newpage
\setcounter{page}{1} 
\tableofcontents

\section{Introduction}

Over the last years much progress has been made in studying flux
compactifications in string theory; see \cite{Douglas:2006es,
  Denef:2008wq} for recent reviews. By now there is strong evidence
that there is a huge number of supersymmetric vacua with negative
cosmological constant in which all scalar moduli are stabilized, the
so called landscape of string theory.  Typical constructions start
with a warped Calabi-Yau compactification of type IIB string theory to
four dimensions.  Some of the scalar moduli are stabilized by the
addition of fluxes through the compact cycles of the internal manifold
and others by various quantum effects.  Since supersymmetry is broken
in the real world, to make contact with phenomenology it is necessary
to extend the previous constructions to non-supersymmetric
(meta)stable vacua with small positive cosmological constant. For this
we need to understand the mechanism of supersymmetry breaking in
string theory. By now several methods of supersymmetry breaking for
string vacua have been proposed, such as the introduction of
anti-branes \cite{Kachru:2002gs, Kachru:2003aw}, or simply the
existence of metastable points of the flux-induced potential
\cite{Saltman:2004sn, Denef:2004cf}. The main drawback of these
constructions is that, in most cases, they are not under complete
quantitative control.

While the question of supersymmetry breaking should be ultimately
understood in an honest compactification, that is in a theory including
gravity in four dimensions, it is technically easier to study simpler
systems where the gravitational dynamics has been decoupled from the
gauge theory degrees of freedom. This typically happens in the limit
where a local singularity develops in the Calabi-Yau manifold. In such a
situation all the interesting dynamics related to the degrees of freedom
of the singularity takes place at energy scales much lower than the four
dimensional Planck scale. Assuming that supersymmetry breaking is
related to these light degrees of freedom, it is then possible to zoom
in towards the singularity and forget about the rest of the
Calabi-Yau. This leads us to the study of supersymmetry breaking and
string phenomenology in the context of local Calabi-Yau geometries
possibly with the addition of probe D3-branes\cite{Verlinde:2005jr,
Buican:2007is, Aganagic:2006ex, Argurio:2006ny, Argurio:2007qk}.

Meanwhile a new important aspect of supersymmetry breaking in gauge
theories was developed after the discovery of Intriligator, Seiberg and
Shih \cite{Intriligator:2006dd} that even simple supersymmetric gauge
theories can exhibit dynamical supersymmetry breaking in metastable
vacua. From a phenomenological point of view this possibility is quite
attractive, and a lot of activity has been concentrated around
extensions of the ISS model and various related string theory
constructions \cite{Franco:2006es, Ooguri:2006pj, Ooguri:2006bg,
Franco:2006ht, Bena:2006rg, Dine:2006gm, Giveon:2007fk,
Intriligator:2007py, Giveon:2007ef, Giveon:2007ew} (see also
\cite{Intriligator:2007cp}).  A certain class of gauge theories where
supersymmetry breaking in metastable vacua can be studied with good
control is that of ${\cal N}=2$ gauge theories perturbed by a small
superpotential, initiated by \cite{Ooguri:2007iu}. In such theories the
exact K\"ahler metric on the moduli space is known, which allows one to
compute the scalar potential produced by the perturbation of the theory
by a small superpotential exactly to first order in the perturbation. It
was shown that generically there are metastable supersymmetry breaking
vacua generated by appropriate superpotentials. We will refer to this as
the OOP mechanism for supersymmetry breaking in ${\cal N}=2$ theories.

String theory in a local Calabi-Yau singularity realizes geometric
aspects of supersymmetric gauge theories. In particular the question
of supersymmetry breaking in these two systems should be related. The first
goal of our paper is to make this connection more precise by giving a
geometric realization of the OOP supersymmetry breaking mechanism in
IIB on a local Calabi-Yau singularity. To realize OOP one first has to
engineer the (IR of the) ${\cal N}=2$ gauge theory and then to find a
way of introducing the appropriate superpotential.  The first step is
achieved by the standard geometric engineering of ${\cal N}=2$ gauge
theories by IIB on noncompact Calabi-Yau manifolds
\cite{Klemm:1996bj, Katz:1996fh}. It is well known that the
moduli space of the Calabi-Yau compactification encodes the geometry
of the Coulomb branch of the gauge theory and that the Seiberg-Witten
solution can be rederived by the complex geometry of the Calabi-Yau.

The introduction of superpotential to this system is less
straightforward and, to our knowledge, has not been studied in the
literature before, in this context.  Our main proposal is that the
superpotential can be introduced by turning on 3-form flux in the local
Calabi-Yau, which is not piercing its compact 3-cycles, but which is
growing in the noncompact direction of the Calabi-Yau. In other words,
it is flux which has support at infinity. While this flux is not
directly piercing the compact cycles we show that, once appropriately
regularized, it does introduce an effective superpotential for the
complex structure moduli, which is generalization of the usual
Gukov-Vafa-Witten superpotential \cite{Michelson:1996pn, Gukov:1999ya,
Taylor:1999ii} to 3-form flux with noncompact support. This is a way to
introduce a general superpotential in a geometrically engineered ${\cal
N}=2$ gauge theory. In particular, we explain that in certain cases it
is possible to engineer the OOP-type superpotential, which guarantees
the existence of metastable, supersymmetry breaking vacua for the
complex structure moduli.

\begin{figure}[h!]
\begin{quote}
 \begin{center}
 \epsfig{file=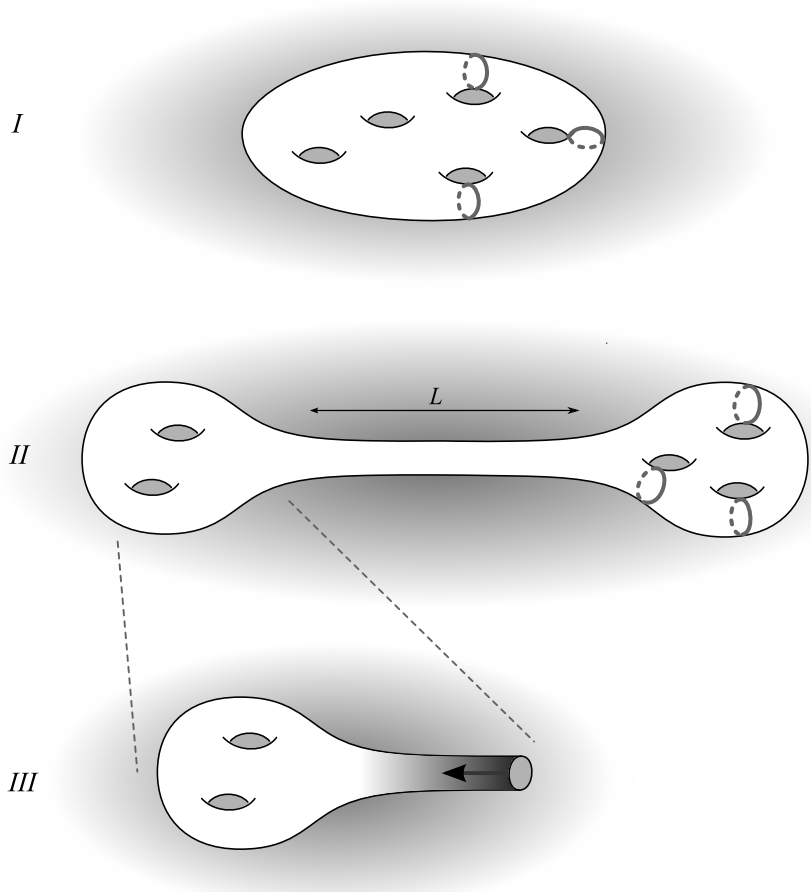,width=.70\textwidth}
 \caption{Idea of the paper: In I we start with a generic
  Calabi-Yau with flux piercing through some of its 3-cycles, while making the
  distance between the cycles with and without flux very large in 
  II. This is seen as flux from infinity in the left sector without compact
  flux in III, and generates an OOP-like potential in that sector. 
  }
 \label{fig:overview}
 \end{center}
\end{quote}
\end{figure}

The second goal of our paper is to find a ``natural'' way to generate
the supersymmetry breaking flux configurations described above, starting
from a more standard setup. In this process, we also clarify the meaning
of flux which has noncompact support and the various subtleties related
to it\footnote{For example the fact that naively the total energy
density in four dimensions diverges.}. The natural interpretation of the
flux described in the previous paragraph emerges once we embed the
previous supersymmetry-breaking local singularity into a bigger IIB
compactification with standard flux of compact support.  As shown in
figure~\ref{fig:overview}, the physical idea is to start with a
Calabi-Yau manifold with a set of three-cycles which are isolated from
the other three-cycles by a large distance. We turn on 3-form flux on
all cycles except for the isolated set. While the flux that we have
turned on is not piercing the isolated cycles, it does leak into their
region\footnote{This means that the 3-form field strength is nonzero in
the region around the isolated set of 3-cycles, but once integrated over
one of these 3-cycles the integral is zero.}  and it produces a
potential for their complex structure moduli. In the limit where the
distance between the two sets of cycles of the Calabi-Yau becomes very
large, which we will refer to as the \emph{factorization limit}, the
flux leaking towards the isolated set will start to look like the flux
coming from infinity, mentioned in the previous paragraph. In this way
we manage to embed the scenario of the previous paragraph into a well
defined system. While this factorization idea should work even in the
case where the total Calabi-Yau is a compact manifold divided into two
parts, \footnote{Because of no-go theorems \cite{Maldacena:2000mw,
deWit:1986xg, Wesley:2008de}, in such compact setups one will need extra
ingredients such as O-planes, which we do not consider in the present
paper.} in this paper we will only analyze it in the local case, as it
is technically easier.

As a check of this,  we consider the example of a local Calabi-Yau
based on a hyperelliptic Riemann surface. In this case the factorization
can be studied more explicitly.  Matrix model techniques can be
used to compute the prepotential in the factorization limit. Our results
verify the general intuition of the last paragraph.

The plan of this paper is as follows.  In section
\ref{sec:generalcase}, we review some general aspects of IIB flux
compactifications and the potentials that can be generated by noncompact
fluxes in the local limit.  We also discuss the general mechanism by
which such fluxes may be used to stabilize complex structure moduli in
metastable supersymmetry-breaking configurations.
After this, we turn in section \ref{sec:OOPlocalCY} to the study of
local Calabi-Yau geometries based on Riemann surfaces, providing a more
detailed description of the generation of metastable vacua in this
context and providing an explicit example.  Section
\ref{sec:factorization} then addresses the second point of this paper,
namely the ability to obtain our local geometries with noncompact fluxes
from larger Calabi-Yau with compact ones by taking a factorization
limit.  Section \ref{sec:ex2} supplements this general discussion by
providing an explicit demonstration, using matrix model techniques, of
the factorization limit in the class of local geometries studied in
section \ref{sec:OOPlocalCY}.  Finally, we finish in section
\ref{sec:conclusions} with some concluding remarks concerning the
generalization of our story to other $\CN=2$ contexts, such as $M$ and
$F$-theory compactifications.  Some supplementary material and technical
details are contained in four appendices.
In appendix \ref{app:prepot}, in particular, we study prepotential for
the Cachazo-Intriligator-Vafa/Dijkgraaf-Vafa geometry using matrix model
as a step towards confirming the above scenario of realizing metastable
vacua by factorization.  We compute the prepotential for an arbitrary
number of cuts up to fifth order in glueball field.

\bigskip
As this paper was being prepared for publication, a paper
appeared \cite{Aganagic:2008qa} where a similar system was studied as an
example from a different perspective.

\section{IIB Compactifications with Flux at Infinity}
\label{sec:generalcase}

\subsection{Compact Calabi-Yau}
\label{subsec:iibcompact}

Compactification of IIB on a Calabi-Yau threefold $\CM$ leads to an
${\cal N}=2$ supergravity theory in 4d. The number of vector
multiplets is $h^{2,1}$, their scalar components correspond to the
complex structure moduli of $\CM$. We also have $h^{1,1}+1$
hypermultiplets, whose scalars correspond to the K\"ahler moduli of
$\CM$ and the axion-dilaton. The two sets of multiplets are decoupled,
and in the rest of the paper we will concentrate on the dynamics of
the vector multiplets.

A Calabi-Yau threefold has a nowhere vanishing holomorphic $(3,0)$ form
$\Omega$ which is unique up to scale. Consider a symplectic basis of
3-cycles $\{{\cal A}^I,{\cal B}_J\}$ with
$I,J=0,1,\dots,h^{2,1}$. We define the periods of $\Omega$ as
\begin{equation}
  X^I = \int_{\CA^I}\Omega,\qquad F_I = \int_{\CB_I} \Omega.
\label{xfperiods}
\end{equation}
The ${\cal A}$-periods $X^I$ are projective coordinates on the complex
structure moduli space of $\CM$, and the $F_I$ are functions of $X^I$.  The
metric on the complex structure moduli space is \emph{special K\"ahler}
and the K\"ahler potential is given by
\begin{equation}
  K = -\log\left( i \int \Omega \wedge \overline{\Omega}\right).
\label{Kcomplex}
\end{equation}
This is an exact result which does not receive any $\alpha'$ or $g_s$
corrections.

The easiest way to lift the phenomenologically unrealistic moduli space
of such compactifications is to turn on fluxes through the compact
cycles of the Calabi-Yau. In the case of IIB we can turn on RR and NS-NS
3-form flux $F_3$ and $H_3$ through the 3-cycles of the threefold. This
generates a superpotential for the complex structure moduli
\cite{Michelson:1996pn, Gukov:1999ya, Taylor:1999ii} given by
\begin{equation}
  W = \int G_3 \wedge \Omega,
\label{gvw}
\end{equation}
where $G_3 = F_3 -\tau H_3$ and $\tau = C_0 + i/g_s$. The scalar potential is computed by the standard ${\cal N}=1$
supergravity expression\footnote{In this expression the indices $a,b$
  run over complex structure moduli, K\"ahler moduli and the
  axion-dilaton. We denote by $\widetilde{K}$ the total K\"ahler
  potential for all moduli and by $K$, as in \eqref{Kcomplex}, the one for the
  complex structure moduli alone.}
\begin{equation}
  V= e^{\widetilde{K}} \left( G^{a\overline{b}}D_a W \overline{D_bW}
-3|W|^2\right),
\label{potsugra}
\end{equation}
where $G_{a\overline{b}}$ is the metric on the moduli space derived
from the K\"ahler potential $\widetilde{K}$, and where we have introduced the
K\"ahler covariant derivative $D_a W = \partial_a W +
( \partial_a\widetilde{K} ) W$.

The $F_3$ and $H_3$ fluxes generate charge for the $F_5$ form via a
Chern-Simons coupling in the 10d IIB supergravity action. The $F_5$
flux has nowhere to end, so we are lead to the tadpole cancellation
condition for IIB compactifications
\begin{equation}
  {1\over l_s^4}\int F_3 \wedge H_3 + Q_{D3} = 0,
\label{tadcanc}
\end{equation}
where $Q_{D3}$ receives positive contribution from probe D3 branes and
negative contribution from induced charge on D7 and orientifold
planes.

\subsection{Local Limit}
\label{subsec:locallimit}

It is well known that there are points on the complex structure moduli
space of Calabi-Yau manifolds where the manifold develops a singularity
\cite{Candelas:1990rm}.  The simplest example is the conifold
singularity, where we have a 3-cycle whose size goes to zero. More
generally, a more complicated set of cycles may become very small in
some region of the moduli space.  As we approach this region, the local
dynamics of the singularity decouples from the rest of the fields. What
this means is that in 4d the typical energy scale for the dynamics of
the fields corresponding to the singularity becomes much smaller than
any other scale, in particular much smaller than the Planck mass $M_p$
in 4 dimensions. In this sense, the dynamics of the singularity is
decoupled from gravity.  Moreover to study the relevant dynamics, we can
zoom in close to the singularity and forget about the rest of the
Calabi-Yau. In this limit the Calabi-Yau looks noncompact, and it
becomes technically easier to study the low energy dynamics.

A typical example of such a local Calabi-Yau is a complex manifold of
the form of a hypersurface in ${\mathbb C}^4$
\begin{equation}
\CM:\quad  uv -F(x,y) =0,
\label{noname}
\end{equation}
where $F(x,y)$ is a polynomial. In this case the holomorphic 3-form is
\begin{equation}
  \Omega = {du \over u} \wedge dx \wedge dy.
\label{omegafornoname}
\end{equation}
By taking the local limit to go from a compact Calabi-Yau to a
noncompact one, the structure of special geometry described above
reduces to what is called \emph{rigid} special geometry
\cite{Seiberg:1994rs, Billo:1998yr}, which is relevant for the low
energy dynamics of ${\cal N}=2$ gauge theories. In this case the
K\"ahler potential reduces to
\begin{equation}
  K = i\int \Omega \wedge \overline{\Omega},
\label{krigid}
\end{equation}
and the K\"ahler covariant derivative $D_i$ reduces to the ordinary
derivative $\partial_i$.

An important point is the distinction between normalizable and
non-normalizable complex structure moduli in the case of noncompact
Calabi-Yau manifolds. To be more precise let us consider the example
\eqref{noname}. The coefficients $\{t^i\}$ of the polynomial $P(x,y)$
characterize the complex structure of $\CM$, so $\{t^i\}$ are the
complex structure moduli of $\CM$. However not all of them are
dynamical.  Some of them control the complex structure of 3-cycles
which are localized in the ``interior'' of the singularity and are
dynamical, while others describe how the singularity is embedded in
the bigger Calabi-Yau and become frozen when we take the decoupling
limit.
 
To determine if a specific complex structure modulus $t$
is dynamical or not, one has to compute the corresponding K\"ahler metric
\begin{equation}
  g_{t\overline{t}} = \partial_t \overline{\partial_t}K = i 
\partial_t \overline{\partial_t}\int \Omega \wedge \overline{ \Omega}.
\label{metriccomplex}
\end{equation}
If this expression is finite, then the modulus $t$ is dynamical,
otherwise it is decoupled and should be treated as a parameter of the
theory.  We will refer to the first set of moduli as
\emph{normalizable} and to the second as \emph{non-normalizable}.  

\subsection{Adding Flux}
\label{subsec:addflux}

As in the compact case, the addition of fluxes to the local Calabi-Yau
introduces a superpotential for the moduli. The dynamics of the
K\"ahler moduli and the dilaton decouple, and we can concentrate on
the normalizable complex structure moduli. The superpotential is still
given by \eqref{gvw}, but now the scalar potential is computed by the
rigid ${\cal N}=2$ expression
\begin{equation}
  V = G^{i\overline{\jmath}}\partial_i W \overline{\partial_j W}=\int G_3
  \wedge * \overline{G_3}.
\label{potrigid}
\end{equation}
Since we are in a noncompact Calabi-Yau it is not necessary to impose
the tadpole cancellation condition. Instead, the quantity
\begin{equation}
  \int F_3 \wedge H_3
\label{tadrigid}
\end{equation}
represents the $F_5$ flux going off to infinity and remains constant
as we vary the moduli. We will use this to simplify the potential in the
next section.

In most treatments of fluxes in noncompact Calabi-Yau manifolds the
assumption is made that the flux is threading the compact cycles of the
singularity and is going to zero at infinity.  As we explained in the
introduction the goal of our paper is to study the dynamics in the case
where the flux is actually coming in from infinity and is not supported
on the compact three-cycles. Of course, in a local singularity inside a
bigger compact Calabi-Yau, what is meant by infinity is the rest of the
Calabi-Yau and we should think of flux coming from infinity as flux
leaking towards the singularity from the other compact cycles.

More precisely, in a noncompact Calabi-Yau we consider the vector space
$H^3(\CM)$ of harmonic 3-forms which do not necessarily have compact
support, so they can grow at infinity. The harmonic 3-forms of compact
support form a linear subspace $H^3_{\text{cpct}}(\CM)\subset H^3(\CM)$. There is a
natural way to define the complement subspace $H^3_{\infty}(\CM)\subset
H^3(\CM)$ as the harmonic forms with vanishing integrals on the compact
3-cycles\footnote{We should clarify that we are not interested in the
most general harmonic 3-form with noncompact support, but only in a
restricted subset characterized by 3-forms which grow in a
``controlled'' way at infinity. This means that we want to consider
forms which have at most a ``pole'' of finite order at infinity, and not
essential singularities.  This statement has a nice interpretation in
the example where we have a local Calabi-Yau based on a Riemann surface
that we will study later. Another way to state this restriction is that
we will consider harmonic 3-forms on a local Calabi-Yau which do have a
lift to the original Calabi-Yau that we started with before we took the
local limit near its singularity.\label{footnote:bura}}.  Then we have
the decomposition
\begin{equation}
  H^3(\CM) = H^3_{\infty}(\CM) \oplus H^3_{\text{cpct}}(\CM).
\label{formdec}
\end{equation}
We will also refer to the forms in $H^3_{\text{cpct}}(\CM)$ as harmonic 3-forms with
compact support and to those in $H^3_{\infty}(\CM)$ as 3-forms with support
at infinity.

Now we want to consider the case where the 3-form field strength that
we have turned on has support at infinity
\begin{equation}
  G_3 \in H_{\infty}^3(\CM),
\label{g3infinity}
\end{equation}
which means that $G_3$ has zero flux through the compact cycles
\begin{equation}
  \int_{\CA^i}G_3 = \int_{\CB_i}G_3 =0.
\label{g3zerocomp}
\end{equation}
The intuitive picture that one should keep in mind, is that this flux at
infinity represents usual flux piercing other 3-cycles which are very 
far away from the singularity in the big Calabi-Yau. As we will see in more
detail in the next section, in this case and if one zooms into the
local singularity it is a good approximation to treat the flux from the
distant 3-cycles as flux which ``diverges'' at infinity. In other words
both $H^3_\infty (\CM)$ and $H^3_{\text{cpct}}(\CM)$ correspond to the usual
$H^3_{\text{cpct}}(\widetilde{\CM})$ of the bigger Calabi-Yau $\widetilde{\CM}$ in
which the singularity $\CM$ develops.

What is maybe more surprising is that the 3-form flux $G_3$ with support at infinity generates a potential for the complex structure moduli of the singularity $\CM$,
even though it is not directly piercing the compact cycles of $\CM$, as can
be seen from \eqref{g3zerocomp}. 
Our starting point for the computation of this potential is the
energy stored in the 3-form field
\begin{equation}
  \widetilde{V} = \int G_3 \wedge * \overline{G_3}.
\label{vinf}
\end{equation}
Since $G_3$ has noncompact support, this is a divergent integral
meaning that the energy of the flux is infinite. This was to be
expected and is not really a problem, since we are interested in the
\emph{changes} of this energy as we vary the sizes of the 3-cycles in
the neighborhood of the singularity. We would like to throw away the
divergent, moduli independent piece of this quantity and keep the
finite, moduli dependent one. A nice way to achieve this is to use the
fact that the net $F_5$ form flux leaking off at infinity, being a
topological quantity, has to be kept constant as we vary the
moduli. It is easy to show that we can write
\begin{equation}
  \int G_3 \wedge \overline{G_3}= (\tau - \overline{\tau}) \int F_3 \wedge H_3,
\label{fhg}
\end{equation}
and the left hand side must be constant for the reason we explained. Since it is a constant we can subtract it from the potential and define
\begin{equation}
  V \equiv \int G_3 \wedge * \overline{G_3} - \int G_3 \wedge  \overline{G_3} .
\label{vregul}
\end{equation}
It is easy to show that this is equal to
\begin{equation}
  V = \int G_3^- \wedge * \overline{G_3^-},
\label{potf}
\end{equation}
where $G^-_3$ is the imaginary anti-self dual part of the $G_3$ flux
\begin{equation}
  * G^-_3 = -i G^-_3.
  \label{g3asd}
\end{equation}
The expression \eqref{potf} is the finite and moduli dependent piece
of the potential \eqref{vinf}.

\subsection{Simplifying the Potential}
\label{subsec:potsimpl}
In this section we simplify the expression \eqref{potf} for the
potential. In general we have the following relation between
the Hodge decomposition and the $*$ operator on a threefold 
\begin{equation}\begin{split}
  & * H^{3,0} = -i H^{3,0}, \, \qquad * H^{1,2} = -i H^{1,2},\\
  & * H^{2,1} = i H^{2,1}, \quad \qquad * H^{0,3}= i H^{0,3}.
\end{split}
\label{hodgestar}
\end{equation}
Before we proceed we would like to analyze the relation between
the decomposition \eqref{formdec} and the Hodge decomposition. In general
we have the following decomposition\footnote{Again, we are only considering
a certain subset of all harmonic 3-forms with noncompact support, as 
explained in footnote \ref{footnote:bura}.}
\begin{equation}
  H^3(\CM) =  H^{3,0}_{\infty}\oplus H^{3,0}_{\text{cpct}}\oplus
H^{2,1}_{\infty}\oplus
H^{2,1}_{\text{cpct}}\oplus \{ c.c.\}.
\label{hodgedec}
\end{equation}
Harmonic forms in $H^{p,q}_{\text{cpct}}$ have compact support, while those in
$H^{p,q}_{\infty}$ do not, and are chosen to have vanishing
$\mathcal{A}$-periods on the compact cycles\footnote{A harmonic
$(p,q)$-form cannot have vanishing periods on all compact cycles unless
it is identically zero.}.  Since we do not want to break supersymmetry
explicitly by the boundary conditions of the system, we want our
configuration to be supersymmetric at infinity, which means that the
flux at infinity has to be imaginary self dual so
\begin{equation}
  G_3 \in H^{2,1}_\infty \oplus H^{2,1}_{\text{cpct}} \oplus H^{1,2}_{\text{cpct}}.
\label{g3atinf}
\end{equation}
where the subscript $\infty$ means that we have to consider the 
elements of the cohomology with noncompact support. We pick a basis
\begin{equation}
  \Xi_m \in H^{2,1}_\infty,\qquad \Omega_i \in H^{2,1}_{\text{cpct}}
\label{formbasis}
\end{equation}
with the following periods
\begin{equation}\begin{split}
 & \int_{\CA^i} \Xi_m = 0, \qquad\qquad \int_{\CA^i} \Omega_j =
    \delta^i_{j},\\ &\int_{\CB_i}\Xi_m = K_{im},\,\,\,\quad
    \,\,\,\,\,\int_{\CB_i} \Omega_j = \tau_{ij},
\end{split}\label{periodbasis}\end{equation}
where $\tau_{ij}$ is the period matrix of the Calabi-Yau, and $K_{im}$ are
holomorphic functions of the normalizable-complex structure moduli.

The flux has an expansion of the form
\begin{equation}
  G_3 = T^m \Xi_m + h^i \Omega_i + \overline{l^i}\, \overline{\Omega_i}.
\label{formflux}
\end{equation}
The parameters $T^m$ are fixed by the boundary conditions and have
to be kept constant as we vary the normalizable moduli. 
We have also assumed that
\begin{equation}
  \int_{\CA^i}G_3 = \int_{\CB_i} G_3 =0.
\label{zerocompflux}
\end{equation}
which means
\begin{equation}\begin{split}
 & T^m\int_{\CA^i} \Xi_m + h^j \int_{\CA^i} \Omega_j +
    \overline{l^j}\int_{\CA^i} \overline{\Omega_j} =0\\ &
    T^m\int_{\CB_i} \Xi_m + h^j \int_{\CB_i} \Omega_j +
    \overline{l^j}\int_{\CB_i} \overline{\Omega_j} =0.
\label{compactfluxzero}
\end{split}\end{equation}
The first equation of \eqref{compactfluxzero} implies that
\begin{equation}
  \overline{l^j} = - h^j.
\label{blabla1}
\end{equation}
and the second
\begin{equation}
  h^i = - {1\over 2 i}\left({1\over \Im\tau}\right)^{ij}\left(K_{jm}T^m \right).
\label{blabla2}
\end{equation}
As we explained before, only the imaginary anti-self dual part of the
flux $ G_3^- = \overline{l^i}\,\overline{ \Omega_i}$ contributes to the
regularized potential and we have
\begin{align}
  V &= \int G_3^-\wedge \overline{G_3^-} \nonumber \\
&= {1\over 4}~ \overline{\left(K_{im} T^m \right)}\left({1\over \Im \tau}\right)^{ij} \left(
  K_{jn}T^n\right).
\label{finalpot}
\end{align}
In this final expression the period matrix $\tau^{ij}$ and $K_{im}$ are
functions of the normalizable complex structure moduli, while $T^m$'s
have to be considered as constants which play the role of external
parameters.  This potential is in general very complicated and can have
local nonsupersymmetric minima for appropriate choices of the parameters
$T^m$ as we will explain later\footnote{Although we do not discuss this
in the present paper, from the viewpoint of flux compactification it is
a natural generalization to consider fluxes through the compact
3-cycles, relaxing the condition \eqref{zerocompflux}.  Such flux will
make additional contribution to the superpotential of the form $N^i
F_i-\alpha_i X^i$, $\alpha_i=\int_{\CB_i}\Omega$, which cannot be
controlled by external parameters and makes realization of OOP-like
vacua more difficult.}.

\subsection{Properties of the Potential}
\label{subsec:potprops}

The potential \eqref{finalpot} should look somewhat familiar as it
shares the same basic structure as the scalar potential that arises when
one adds a small superpotential to Seiberg-Witten theory.  This
connection can be made even more transparent by noting that $K_{im}$ can
in general be written as a total derivative with respect to the special
coordinates $X^i$\,\,{\footnote{One quick way to see this is to use the
identity $\int_\CM\,\Xi_m\wedge \partial_i\overline{\Omega}=0$ to derive
$K_{im}\sim \int_{\partial \CM} \Lambda_m\wedge
\partial_i\overline{\Omega}$ for a 2-form $\Lambda_m$ satisfying
$d\Lambda_m=\Xi_m$ on the boundary (at infinity) of $\CM$.  Because the
divergent contributions to $\Lambda_m$ at infinity can be chosen
independent of the dynamical moduli, we can pull the derivative outside
of everything.}}
\begin{equation}K_{im}=\frac{\partial}{\partial
    X^i}\kappa_m(X^j),\qquad\qquad X^i=\oint_{{\cal{A}}^i}\,\Omega. \end{equation}
With this notation, \eqref{finalpot} takes the standard form
\begin{equation}V=\frac{1}{4}\overline{\left(\frac{\partial W_{\text{eff}}(X^k)}{\partial X^i}\right)}
\left({1\over\Im\tau}\right)^{ij}\left(\frac{\partial W_{\text{eff}}(X^k)}{\partial X^j}\right),\label{finalpotOOP}\end{equation}
where
\begin{equation}W_{\text{eff}}(X^k)=T^m\kappa_m(X^k)\end{equation}
is in fact proportional to the Gukov-Vafa-Witten superpotential induced by the flux $G_3$.

\subsubsection{OOP Mechanism}
\label{subsec:OOPmech}

Equation \eqref{finalpotOOP} makes manifest the relation between our
flux-induced potential \eqref{finalpot} and that which arises in
deformed Seiberg-Witten theory and allows us to utilize the technology
developed by Ooguri, Ookouchi, and Park \cite{Ooguri:2007iu} in that
context{\footnote{See also related work by Pastras
\cite{Pastras:2007qr}.}} for engineering supersymmetry-breaking vacua.
In particular, if we want to realize a nonsupersymmetric minimum at some
point $X^{i\,(0)}$ in the moduli space, the OOP procedure tells us to
first construct K\"ahler normal coordinates \cite{AlvarezGaume:1981hn,
Hull:1985pq, Higashijima:2000wz}, around $X^{i\,(0)}$
\begin{equation}z^i = \Delta X^i+\tilde{g}^{i\bar{\jmath}}\sum_{n=2}^{\infty}\frac{1}{n!}\partial_{i_3}\ldots\partial_{i_n}\tilde{\Gamma}_{j i_1 i_2}\Delta X^{i_1}\Delta X^{i_2}\ldots \Delta X^{i_n},\label{KNCdef}\end{equation}
where $\Delta X^i= X^i-X^{i\,(0)}$ and $\tilde{\ }$ means evaluation at $X=X^{(0)}$.  We then build the potential $V$ in \eqref{finalpotOOP} from a superpotential $W_{\text{eff}}$ consisting of a linear combination of the $z^i$
\begin{equation}W_{\text{eff}}=k_i z^i,\qquad\text{$k_i$: constant.}\label{WeffKNC}\end{equation}
Stability can then be demonstrated by expanding $V$ near $p$
\begin{equation}V=k_i\bar{k}_{\bar{\jmath}}\tilde{g}^{i\bar{\jmath}}+k_i\bar{k}_{\bar{\jmath}}\tilde{R}^{i\bar{\jmath}}_{\,\,\,\,k\bar{l}}z^k\bar{z}^{\bar{l}}+{\cal{O}}(z^3).\end{equation}
The curvature of special K\"ahler manifolds, of which the complex
structure moduli space is an example,  is positive definite at generic
points.  As a result, any potential of the form \eqref{finalpotOOP}
that agrees with \eqref{WeffKNC} near $X^{i\,(0)}$ to cubic order will engineer
a nontrivial vacuum at $X^{i\,(0)}$.{\footnote{For non-generic $X^{(0)}$, the curvature
    may have a zero eigenvalue in which case higher order agreement
    with \eqref{WeffKNC} is required (that $X^{i\,(0)}$ is a stable for
    superpotential exactly equivalent to \eqref{WeffKNC} will follow
    from the discussion below).}}

One can obtain a nice physical picture for this mechanism by noting,
as in \cite{Marsano:2007mt}, that the series \eqref{KNCdef} can be
summed exactly and inserted into \eqref{WeffKNC} to yield 
\begin{equation}
W_{\text{eff}}\sim e_i X^i+m^i F_i,\label{Wefffluxsup}
\end{equation}
where $e_i$ and $m^j$ satisfy 
\begin{equation}e_i+m^j\overline{\tilde{\tau}}_{ji}=0.\label{fluxcond}\end{equation}
From this, we see that the superpotential \eqref{Wefffluxsup} built from
K\"ahler normal coordinates is of precisely the form that we would have
obtained had we instead simply turned on compactly-supported fluxes
$m^i$ and $e_i$ threading the cycles ${\cal{A}}^i$ and ${\cal{B}}_i$,
respectively.  The condition \eqref{fluxcond}, however, combined with
the requirement that $\Im\tau$ be positive definite, implies that $e_i$
and $m^i$ can never satisfy the condition $e_i+m^j\tilde{\tau}_{ji}=0$
that is required for preservation of the manifest ${\cal{N}}=1$
supersymmetry.

It is well-known that the flux-induced superpotential
\eqref{Wefffluxsup} only breaks the full ${\cal{N}}=2$ supersymmetry
spontaneously, though, so there is a second ${\cal{N}}=1$ in the game
that is not manifest in this formalism.  The relation \eqref{fluxcond}
is, in fact, nothing other than the condition that the vacuum at
$X^{i\,(0)}$ preserves precisely these non-manifest supersymmetries
\cite{Antoniadis:1995vb,Marsano:2007mt}.  As such, the vacuum at
$X^{i\,(0)}$ in the presence of the superpotential \eqref{WeffKNC} is
stable for a good reason---it is secretly supersymmetric!

In general, our noncompactly supported fluxes will not generate
potentials with $W_{\text{eff}}$ exactly equivalent to \eqref{WeffKNC}.
Rather, the $W_{\text{eff}}$'s that arise are globally well-defined
functions on the moduli space{\footnote{Contrast this with
\eqref{Wefffluxsup}, which manifestly suffer from monodromies for
constant (non-transforming) $k_i$.}} which we can then tune to agree
with \eqref{WeffKNC} to cubic order within a neighborhood of the point
$X^{i\,(0)}$.  The delicate manner by which the superpotential
\eqref{WeffKNC} managed to realize a non-manifest ${\cal{N}}=2$
supersymmetry is crucially dependent on the full infinite series
expansion about $X^{i\,(0)}$ so, by failing to exactly reproduce
\eqref{WeffKNC}, we are able to explicitly break, at the level of the
Lagrangian, the half of supersymmetry which would otherwise have been
preserved by the vacuum at $X^{i\,(0)}$.  Stability of the $X^{i\,(0)}$
vacuum, on the other hand, depends only on the local behavior of
$W_{\text{eff}}$ so our procedure will retain this property, leaving us
with a locally stable supersymmetry-breaking vacuum.

In the end, what we are doing to engineer a supersymmetry-breaking
vacuum at $X^{i\,(0)}$ is actually a quite intuitive procedure.  We
first turn on a collection of noncompactly supported fluxes which
explicitly break half of the ${\cal{N}}=2$ supersymmetry.  We then tune
these fluxes so that, near $X^{i\,(0)}$, their interactions with the
dynamical complex structure moduli mimic those of the compactly
supported fluxes that would generate a vacuum at $X^{i\,(0)}$ which
preserves the opposite half of supersymmetries.
 
\subsubsection{Supersymmetric Vacua}

In addition to possessing supersymmetry-breaking vacua when the $T^m$
are suitably tuned, the potential \eqref{finalpotOOP} also typically
contains a wide variety of supersymmetric vacua.  As discussed in
\cite{Ooguri:2007iu}, these vacua fall into two different classes.
First, because there is no flux directly threading the compact cycles,
the energy cost associated with shrinking them is necessarily finite.
Because the period matrix $\tau_{ij}$ diverges, the potential vanishes
at these singular points and we obtain stable vacua which are in fact
supersymmetric.

This argument is of course rather crude because we are neglecting the
new light degrees of freedom that enter as 3-cycles degenerate but, as
is well-known, this is easily fixed.  In particular, the light D3 branes
which wrap the degenerating cycles give rise to hypermultiplets
\cite{Strominger:1995cz, Greene:1995hu} comprised of pairs of
${\cal{N}}=1$ chiral superfields $Q_i$ and $\tilde{Q}_i$ with bilinear
superpotential couplings to special coordinates.  For the simple case of
degenerating ${\cal{A}}^i$ cycles, the superpotential takes the
schematic form
\begin{equation}W=W_{\text{eff}}(X^i)+(Q\tilde{Q})_i X^i,\end{equation}
and allows a supersymmetric vacuum at $X^i=0$ through condensation of
$Q\tilde{Q}$
\begin{equation}(Q\tilde{Q})_i=-\frac{\partial W_{\text{eff}}}{\partial X^i}\left(X^j=0\right).\end{equation}

The second class of supersymmetric vacua correspond to solutions of the
$F$-term equations
\begin{equation}\partial_i W_{\text{eff}}(X^j)=0.\end{equation}
In general, there may be many solutions to these equations, as we will
explore later in the example of section \ref{sec:OOPex}.

\subsubsection{Lifetime of Supersymmetry-Breaking Vacua}

Because we have managed to achieve supersymmetry-breaking vacua while
freezing all non-normalizable moduli, the energies $V_0$ will in general
be finite and independent of the cutoff scale $\Lambda_0$ that we use to
regulate the local geometry.  This means that our vacua are truly
metastable, even within this local model, and can decay to any of the
supersymmetric vacua that exist in these models.  Because the number the
supersymmetric vacua is potentially large and their properties quite
model-dependent, it is difficult to make general statements about the
lifetime of our OOP vacua.  Nevertheless, we recall here one observation
from \cite{Ooguri:2007iu}, namely that the decay rates will in general
scale like
\begin{equation}e^{-S},\qquad S\sim \frac{(\Delta z)^4}{V_+},\end{equation}
where $\Delta z$ is the distance in field space between the initial and
final vacuum state and $V_+$ is the difference in their energies.  By
simultaneously scaling all $T^m$ by a common factor,
$T^m\rightarrow\epsilon\, T^m$, we can retain our supersymmetry-breaking
vacua while decreasing $V_+$ by the same factor, $V_+\rightarrow\epsilon
V_+$.  In this manner, we see that, just as with OOP vacua in deformed
Seiberg-Witten theory, these OOP flux vacua can be made arbitrarily
long-lived{\footnote{Because we should really think of the local Calabi-Yau as
sitting inside some larger compact geometry, one important caveat to
this statement of longevity is that the noncompact fluxes $T^m$ in
reality derive from a suitable set of compact fluxes in the full
Calabi-Yau.  This means that there will be a series of quantization
conditions that must be imposed that may affect the degree to which they
may be tuned.}}.

\section{Metastable Flux Vacua in Local Calabi-Yau}
\label{sec:OOPlocalCY} \label{sec:localCYnoncptflux}

In the previous section, we saw that, starting from a compact Calabi-Yau and
taking a decoupling limit, one ends up with a local Calabi-Yau with noncompact
flux with support at infinity, which is nothing but the flux leaking
from the rest of the full Calabi-Yau that have been decoupled, towards ``our''
local Calabi-Yau\@.  Furthermore, this noncompact flux induces potential
\eqref{finalpot} for the complex structure moduli in the local Calabi-Yau\@.
Depending on the noncompact flux, this potential can be very complicated
and create nonsupersymmetric metastable vacua in the local Calabi-Yau; the OOP
mechanism \cite{Ooguri:2007iu} reviewed in \ref{subsec:OOPmech} tells us
exactly how this can be done.
In this section, we will take specific examples of local Calabi-Yau and
demonstrate that one can generate such OOP vacua as IIB flux
geometries.

In subsection \ref{subsec:localCY}, we review constructions of typical
local Calabi-Yau geometries, taking Seiberg-Witten and Dijkgraaf-Vafa geometries
as examples.  The focus will be on the form of the potential for the
moduli which is induced by flux at infinity.
We also make remarks on the gauge
theory interpretation of the physics of these geometries.
In subsection \ref{sec:OOPex}, we proceed to an explicit
construction of metastable flux vacua in a Dijkgraaf-Vafa geometry, by
tuning superpotential appropriately.
In subsection \ref{subsec:tunability}, we estimate how much control of
flux at infinity is required to create OOP vacua.

\subsection{Local Calabi-Yau Based on Riemann Surface}
\label{subsec:localCY}

A large group of examples of noncompact Calabi-Yau manifold in IIB is
defined by an equation of the form
\begin{align}
 uv-F(x,y)=0,
 \label{generalRSCY}
\end{align}
where $x,y$ can both be variables in $\bbC$ or $\bbC^*$.  Compactifying
on such a Calabi-Yau leaves $\CN=2$ supersymmetry unbroken in four dimensions.
Important roles in these Calabi-Yau's are played by the underlying
one-dimensional complex curve in the $x,y$-plane defined by $F(x,y)=0$
\cite{Klemm:1996bj, Aganagic:2003qj}. In most of our examples this curve
is smooth, and we will refer to it as the Riemann surface $\Sigma$. The
total Calabi-Yau space will be named $\CM_{\Sigma}$. The holomorphic
3-form of $\CM_\Sigma$ is given, {\it e.g.}\ for $x,y\in\bbC$, by
\begin{align}
 \Omega = {du\wedge dx\wedge dy\over \partial f/\partial v}
 =\frac{du}{u} \wedge dx \wedge dy.
 \label{OmegaRS}
\end{align}

Notice that the total threefold can be described as a local (or
decompactified) elliptic fibration over the $x,y$-plane. Over generic
points in the base $x,y$-space, its fiber is described by a hyperboloid
satisfying the equation $uv = \mu$ where $\mu$ is nonzero, which may be
viewed as a decompactified compact torus, making its B-cycle very
large. On the other hand, when $(x,y)\in \Sigma$, the noncompact fiber
degenerates into a cone $uv=0$, which one obtains by decompactifying a
pinched torus, corresponding to an $A_0$ geometry.

Many important properties of the noncompact Calabi-Yau threefold
$\CM_{\Sigma}$ have an interpretation in terms of the underlying Riemann
surface $\Sigma$. For example, the compact 3-cycles $\{\CA^i,\CB_j\}$ in
$\CM_\Sigma$ are lifts of compact 1-cycles on $\Sigma$, which we 
denote by $\{A^i,B_j\}$.  If $(x,y) \in \bbC^2$, these 3-cycles may be
constructed by filling in a disk $D$ in $\bbC^2$ whose boundary
$\partial D$ is the 1-cycle on $\Sigma$. Now consider an $S^1$-fibration
over $D$ where $S^1$ is the compact circle in the $uv$-fiber. Since this
circle shrinks over $\Sigma$, the total 3-cycle has the topology of an
$S^3$. If one of the variables $x$ or $y$ is $\bbC^*$-valued, the disk
$D$ will be punctured. In such a situation differences of 1-cycles have
to be considered. We will see an example of this shortly. Notice that
the one-to-one correspondence between 3- and 1-cycles shows an
equivalence between the complex structure moduli on  $\CM_\Sigma$ and
$\Sigma$.

A basis of (2,1)-forms with compact support on $\CM_\Sigma$ is given by
derivatives of $\Omega$ with respect to the normalizable complex
structure moduli: $\{\Omega_i=\partial_i \Omega\}$.  If $\CM_{\Sigma}$
were compact, these derivatives $\partial_i$ would be K\"ahler covariant
derivatives $D_i$ on the moduli space. Being noncompact instead, the
moduli space is described by rigid special geometry and, as we saw
before, the covariant derivatives simplify into partial derivatives.
Another reduction over the compact 3-cycles in the Calabi-Yau shows that
all these compactly supported $(2,1)$-forms $\Omega_i$ reduce to a basis
of holomorphic 1-forms $\omega_i$ on $\Sigma$.  Similarly, $(1,2)$-forms
$\overline{\partial_i\Omega}$ in $\CM_\Sigma$ reduce to antiholomorphic
forms $\overline{\omega}_i$ on $\Sigma$.  $\omega_i$ satisfy the
following relations, which are reductions of \eqref{periodbasis}:
\begin{equation}
 {1\over 2\pi i}\int_{A^i} \omega_j = \delta^i_{j},\qquad
 {1\over 2\pi i}\int_{B_i} \omega_j = \tau_{ij}.
\label{prdomegaRS}
\end{equation}

The relation between the 3-cycles/3-forms on $\CM_{\Sigma}$ and the
1-cycles/1-forms on $\Sigma$ through the trivial $uv$-fibration being
understood, we can rewrite the various relations in section
\ref{sec:generalcase} in terms of the Riemann surface $\Sigma$.  First
of all, the holomorphic 3-form $\Omega$ of $\CM_\Sigma$, which is given
{\it e.g.}\ for $x,y\in\bbC$ by \eqref{OmegaRS}, is easily seen to
reduce to a meromorphic 1-form $\lambda = y\,dx$ on the Riemann surface
in this case \cite{Klemm:1996bj, Aganagic:2003qj}. The special coordinates
\eqref{xfperiods} parametrizing complex structure moduli are
\begin{align}
 X^i={1\over 2\pi i}\int_{A^i}\lambda,\qquad
 F_i={1\over 2\pi i}\int_{B_i}\lambda,
 \label{spclcoordRS}
\end{align}
and the K\"ahler potential \eqref{krigid} is given by
\begin{align}
 K=i\int_\Sigma \lambda\wedge\overline{\lambda}.
 \label{KahlerpotRS}
\end{align}
Recall that, in the special coordinates $\{X^i\}$, the moduli space
metric takes a particularly simple form:
\begin{align}
 ds^2=\left({\partial^2 K\over \partial X^i\partial\overline{X^j}}\right)
 dX^i d\overline{X^j}
 =(\Im \tau)_{ij}\, dX^i d\overline{X^j},
 \label{spclmtrcRS}
\end{align}
as can be shown using ${\partial_i \lambda}=\omega_i$ and
the Riemann bilinear relation. 

Now we want to consider a very small deformation of the system breaking
supersymmetry to $\CN=1$, thus generating a potential $V$ for the
moduli. As we saw before, this can be accomplished by turning on 3-form
flux $G_3$ with support at infinity in the local Calabi-Yau\@.  This flux an be
thought of as leaking from the other part of the full compact Calabi-Yau, which
has been frozen in the decoupling limit.
We assume that the decoupling limit was taken consistently with the
elliptic fibration structure; namely, we assume that the noncompact flux
is supported at the asymptotic infinities of $\Sigma$, while being
compact in the direction of the $uv$-fibers.

The basis of (2,1)-forms with noncompact support, $\{\Xi_m\}$, in the
Calabi-Yau $\CM_\Sigma$ descend to meromorphic 1-forms $\{\xi_m\}$ on the
Riemann surface $\Sigma$, satisfying the relations 
\begin{align}
\int_{A^i}\xi_m=0,\qquad
\int_{B_i}\xi_m=K_{im},\label{KIM_lclCY}
\end{align}
which are reductions of \eqref{periodbasis}.  Therefore, the
3-form flux with noncompact support, $G_3$, on $\CM_\Sigma$ as given in
\eqref{formflux} descends to a harmonic 1-form flux
\begin{equation}
 \begin{split}
 g&=g_H+\overline{g_A},\\
 g_H&= T^m \xi_m + h^i \omega_i,\qquad
 \overline{g_A}=\overline{l^i}\overline{\omega_i},
\end{split}
\label{1formRS}
\end{equation}
which will have poles at the punctures (or asymptotic legs) of
$\Sigma$. 
%
%
A 3-form flux $G_3$ in $\CM_\Sigma$ induces 
superpotential
\eqref{gvw}, which reduces to an integral on $\Sigma$:
\begin{equation}
  W = \int_{\Sigma} g \wedge \lambda,\label{gvwRS}
\end{equation}
while the associated scalar potential \eqref{potf} reduces to an
integral on $\Sigma$:
\begin{equation}
 V = \int_{\Sigma} g_A \wedge \overline{g_A}.
\label{scalarpotRS}
\end{equation}

If we require the condition \eqref{zerocompflux} that the flux
\eqref{1formRS} is zero through compact 3-cycles of $\CM_{\text{SW}}$, which
translates into
\begin{align}
 \int_{A^i}g=\int_{B_i}g=0,\label{0cptflxRS}
\end{align}
then by the exactly same argument we did for general Calabi-Yau's in the
previous section now reduced to the Riemann surface $\Sigma$ (or simply
by borrowing the result \eqref{finalpot}), we can rewrite
\eqref{scalarpotRS} in terms of periods on $\Sigma$:
\begin{equation}
  V = {1\over 4} \overline{(K_{im}T^m)}
   \left({1\over \Im \tau}\right)^{ij} 
   K_{jn}T^n.\label{scalarpotRS2}  
\end{equation}
For convenience, the relation between 3- and 1-forms in $\CM_\Sigma$ and
$\Sigma$ is summarized in Table \ref{3and1forms}.
\begin{table}[htb]
 \begin{quote}
\begin{center}
\renewcommand{\arraystretch}{1.5}
 \begin{tabular}{|c|c|c|c|}
  \hline
  & special forms
  & \begin{minipage}{5cm}\begin{center}noncompact flux\\inducing superpotential $W$\end{center}\end{minipage}
  & \begin{minipage}{4cm}\begin{center}compact flux\\entering potential $V$\end{center}\end{minipage}   \\
  \hline
  $\CM_\Sigma$ 
  & $\Omega\in H^{3,0}(\CM_\Sigma)$
  & $T^m \Xi_m\in H^{2,1}_{\infty}(\CM_\Sigma)  $ 
  & $G_3^-=\overline{l^i}\overline{\Omega_i}\in H^{1,2}_{\text{cpct}}(\CM_\Sigma)$ \\
  \hline
 $\Sigma$   
  & $\lambda\in H^{1,0}(\Sigma)$
  & $T^m\xi_m\in H^{1,0}_{\infty}(\Sigma)$ 
  & $\overline{g_A}=\overline{l^i}\overline{\omega_i}\in H^{0,1}_{\text{cpct}}(\Sigma)$   \\
  \hline
 \end{tabular}
\end{center}
\caption{The summary of the relation between
3-forms\protect\footnotemark{} in Calabi-Yau $\CM_\Sigma$ and 1-forms on
Riemann surface $\Sigma$\label{3and1forms}}
 \end{quote}
\end{table}
\footnotetext{As explained in footnote \ref{footnote:bura}, we do not
mean here that $\Xi_m$'s span a complete basis of 3-forms in $\CM_\Sigma$
with support at infinity; we are only considering a certain subset of all
3-forms diverging at infinity, which are the lifts of meromorphic
1-forms $\xi_m$ on $\CM$.}

%
%
%

Now we will turn to more specific examples of local Calabi-Yau
geometries based on Riemann surfaces which have been studied in the
context of string theory.

 
\subsubsection{Seiberg-Witten Geometries}
\label{subsec:SWgeo}

An illustrative example of the general Calabi-Yau's \eqref{generalRSCY}
is given by $SU(N)$ Seiberg-Witten (SW) geometries.  In type IIB, these
correspond to compactifications on noncompact Calabi-Yau threefold
$\CM_{\text{SW}}$ defined by
\begin{equation}
\CM_{\text{SW}}:\qquad
uv - F_{\text{SW}}(x,y) =0, \qquad x \in \bbC,\quad y \in \bbC^*,\label{SWCY}
\end{equation}
where the underlying Riemann surface $\Sigma_{\text{SW}}$ is a hyperelliptic curve 
\begin{equation}
\Sigma_{\text{SW}}:\qquad
F_{\text{SW}}(x,y) \equiv \Lambda^N \left( y + \frac{1}{y} \right) - P_N(x)=0
\label{SWcurve}
\end{equation}
and $P_N(x)$ is a polynomial of degree $N$ with the coefficient of
$x^{N-1}$ being zero:
\begin{align}
P_N(x)=\prod_{i=1}^N (x- \alpha_i),
 \qquad
\sum_{i=1}^N \alpha_i = 0.\label{defPN_SW}
\end{align}
The coefficients of $P_N(x)$, or equivalently $\alpha_i$, are
normalizable moduli, while $\Lambda$ is a fixed parameter.  The
holomorphic 3-form on $\CM_{\text{SW}}$ is $\Omega_{\text{SW}} = \frac{du}{u} \wedge dx
\wedge \frac{dy}{y}$ and reduces to \cite{Klemm:1996bj}
\begin{align}
 \lambda_{\text{SW}} = x \frac{dy}{y}
\label{SW1form}
\end{align}
on the Riemann surface $\Sigma_{\text{SW}}$.

Type IIB string theory compactified on the Calabi-Yau \eqref{SWCY}
without flux geometrically engineers \cite{Kachru:1995fv, Klemm:1996bj}
an $\CN=2$ Seiberg-Witten theory \cite{Seiberg:1994rs, Seiberg:1994aj}.
In particular, the $SU(N)$ Seiberg-Witten curve of gauge theory
\cite{Klemm:1994qs, Argyres:1994xh} is geometrically identified with the
curve \eqref{SWcurve} underlying the Calabi-Yau\@.
A $T$-duality along the compact circle in the $uv$-fiber, followed by a
lift to M-theory, translates \cite{Dijkgraaf:2007sw} this geometry into
a system of an M5-brane which wraps the Riemann surface $\Sigma_{\text{SW}}$
and fills $\bbR^{3,1}$.  In the IIA limit, this system is related to a
Hanany-Witten type brane configuration in type IIA, where one has two
NS5-branes with $N$ D4-branes stretching between them
\cite{Witten:1997sc, Giveon:1998sr}.
%
%
From this last IIA/M-theory point of view, it is easy to see the
relation of the system to $\CN=2$ $SU(N)$ super Yang-Mills as the
worldvolume theory on the D4-branes.  In particular, $\alpha_i$'s
correspond to the eigenvalues of the adjoint scalar $\Phi$ on the
Coulomb branch.
In passing we also note that the geometries \eqref{SWCY} are related to
toric geometries in IIA by mirror symmetry \cite{Katz:1996fh,
Katz:1997eq}.

\begin{figure}[htb]
  \begin{quote}
 \begin{center}
  \includegraphics[width=12.5cm]{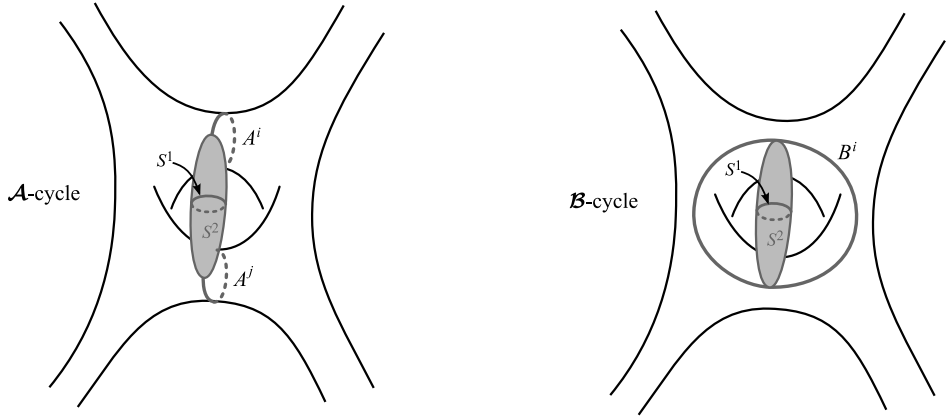} \\ \caption{The
 relation between 3-cycles in the Calabi-Yau $\CM_{\text{SW}}$ and 1-cycles on
 the Riemann surface $\Sigma_{\text{SW}}$ for the Seiberg-Witten geometry. For
 the $\CA$-cycle, by fibering $S^1$ over the line segment whose
 endpoints are at a point on $A^i$ and a point on $-A^j$, one obtains
 $S^2$.  By moving the endpoints over $A^i$ and $-A^j$, one obtains
 $S^2\times S^1$.  For the $\CB$-cycle, similarly moving the $S^2$
 ending on $B_i$, one obtains $S^3$.  \label{SW3cycles}}
 \end{center}
  \end{quote}
\end{figure}

Now let us look at the homological structure of the Seiberg-Witten
geometry \eqref{SWCY}, focusing on the relation between 1-cycles on the
hyperelliptic curve $\Sigma_{\text{SW}}$ \eqref{SWcurve} and 3-cycles in the
Calabi-Yau $\CM_{\text{SW}}$ \eqref{SWCY}.
%
The Riemann surface $\Sigma_{\text{SW}}$ may be compactified by adding two
points at infinity. If we represent the curve \eqref{SWcurve} as a
two-sheeted $x$-plane branched over $2N$ points, those infinities
correspond to $x=\infty$ on the two sheets.  It is thus a hyperelliptic
curve of genus $N-1$ with two punctures. Therefore, its first homology
$H_1(\Sigma_{\text{SW}})$ is formed by $N-1$ pairs of compact $A$ and
$B$-cycles, $(A^i,B_j)$, $i,j=1,\dots,N-1$, with in addition a closed
1-cycle $A^\infty$ around one of the punctures which is dual to an open
1-cycle $B_\infty$ connecting the two points.
How can these 1-cycles be lifted to 3-cycles in $\CM_{\text{SW}}$?  The fact that
$y \in \bbC^*$ means that A-cycles on $\Sigma_{\text{SW}}$ are not contractible
on the $x,y$-plane (recall that we are regarding $\Sigma_{\text{SW}}$ as
embedded in the $x,y$-plane). Instead, compact A-cycles in the
noncompact Calabi-Yau threefold will reduce to differences of A-cycles
on $\Sigma_{\text{SW}}$. Indeed, notice that a point on the 1-cycle $A^i$ and
one on another 1-cycle $-A^j$, with opposite orientation, are connected
by a $\bbP^1$ in the Calabi-Yau. The resulting 3-cycle therefore has the
topology of $S^2 \times S^1$. For the B-cycles this subtlety does not
arise, and compact B-cycles in the Calabi-Yau have $S^3$ topology and
reduce to compact 1-cycles connecting the two hyperelliptic planes.  See
Figure \ref{SW3cycles}.  This discussion is equivalent to page 10 of
\cite{Klemm:1996bj}, and in particular shows the equivalence between
3-cycles on the Calabi-Yau's and 1-cycles on the Seiberg-Witten curve.

As seen in section \ref{subsec:localCY}, the complex structure moduli
space is conveniently parametrized by the special coordinates
\eqref{spclcoordRS}, which in the Seiberg-Witten case is conventionally
denoted by $a^i$, $i=1,\dots, N-1$:\footnote{Because of the subtlety
mentioned above about how to take 1-cycles that lifts to compact
3-cycles in the Calabi-Yau, we should think of the $A^i$ appearing in
\eqref{spclcoordSW} {\it e.g.}\ as $\widetilde A^i\equiv A^i-A^N$, where
$i=1,\dots, N-1$. For simplicity of presentation, we write $\widetilde
A^i$ as $A^i$.}
\begin{align}
 a^i
 ={1\over 2\pi i}\int_{A^i}\lambda_{\text{SW}}
 ={1\over 2\pi i}\int_{A^i}x{dy\over y}.
 \label{spclcoordSW}
\end{align}
As in \eqref{spclmtrcRS}, the moduli space metric takes the special form
for these:
\begin{align}
 ds^2=(\Im\tau_{ij})\,da^i d\overline{a^j}.
 \label{SWmetrica}
\end{align}
Using $a^i$, the normalized basis of holomorphic 1-forms $\omega_i$ can
be obtained as follows. Differentiating \eqref{spclcoordSW} with respect
to $a^j$,
\begin{align}
 \delta^i_j&={1\over 2\pi i}{\partial \over \partial a^j}\int_{A^i}x{dy\over y}.
 \label{ddaofdefa}
\end{align}
Comparing with the first equation in \eqref{prdomegaRS}, this means that
\begin{align}
 \omega_i= {\partial \over \partial a^i}\left(x{dy\over y}+d\eta\right),
 \label{omegaIasader}
\end{align}
where the total derivative term $d\eta$ is fixed by requiring that
$\omega_i=\CO(x^{-2})dx$ as $x\to\infty$.  Specifically, this leads to
$d\eta= d(-x\log y)$ and $\omega_i$ is given by
\begin{align}
 \omega_i
 &
 ={\partial \over \partial a^i}(-\log y\,dx)
 =-{{\partial P_N(x)/ \partial a^i}\over\sqrt{P_N(x)^2-4\Lambda^{2N}}}dx.
 \label{omegaIexplct}
\end{align}
Although $\log y$ may appear problematic because it is not single-valued
on the Riemann surface, its $a^i$ derivative is single-valued and does
not cause any problem.

As we discussed in general in section \ref{subsec:localCY}, turning on
noncompact flux breaks $\CN=2$ supersymmetry to $\CN=1$ by inducing a
superpotential.  In the present case where the Riemann surface is
hyperelliptic, we can take $\{\xi_m\}$ and $\{\omega_i\}$ to be the
specific ones given in Appendix \ref{app:hyperelliptic}.  As in \eqref{1formRS}, the
3-form flux in $\CM_{\text{SW}}$ reduces to a harmonic 1-form on $\Sigma_{\text{SW}}$:
\begin{equation}
g = \sum_{m\ge 1} T^m  \xi_m 
 + \sum_{i=1}^{N-1} h^i \omega_i 
 + \sum_{i=1}^{N-1} \overline{l^i}\overline{\omega_i}.\label{1formgSW}
\end{equation}
Under the condition that the compact flux vanishes, \eqref{0cptflxRS},
this leads to the scalar potential \eqref{scalarpotRS2}.

We can write the superpotential we are adding to the system in a form
that will be useful later.  By manipulating the quantity $K_{jn}T^n$
appearing in \eqref{scalarpotRS2},
\begin{align}
 K_{jn}T^n
 &=T^n\oint_{B_j} \xi_n
 =-2T^n\oint_{\infty} x^n \omega_j
 =2T^n{\partial \over \partial a_j }
 \left(\oint_{\infty} x^n \log y\, dx\right)
 \notag\\
 &=-{2T^n\over N+1}{\partial \over \partial a_j }
 \left(\oint_{\infty} x^{n+1} {dy\over y}\right).
 \label{KJNTN}
\end{align}
Here we used \eqref{KIM_lclCY}, \eqref{KIMidentity}, and
\eqref{omegaIexplct}.  By examining \eqref{scalarpotRS2} and
\eqref{SWmetrica}, one sees that the superpotential is given by:
\begin{align}
 W_{\text{SW}}=\sum_{m} T^m u_{m+1},
 \label{supotRSgen}
\end{align}
where we defined
\begin{align}
 u_m\equiv {1\over 2\pi i m}\oint_{\infty} x^{m-1} \lambda_{\text{SW}}
 ={1\over 2\pi i m}\oint_{\infty} x^m {dy\over y}.
 \label{defuMgeom}
\end{align}

So far everything was about geometry.  Now let us turn to the gauge
theory interpretation of these.  As we mentioned above, the local Calabi-Yau
geometry \eqref{SWCY} without flux realizes $\CN=2$ Seiberg-Witten
theory, with the hyperelliptic curve \eqref{SWcurve} identified with the
$\CN=2$ curve of gauge theory.  The special coordinates $a^i$ defined in
\eqref{spclcoordSW} correspond to the $U(1)$ adjoint scalars in the IR
and parametrize the Coulomb moduli space.  The superpotential
\eqref{supotRSgen} also has a simple gauge theory interpretation.  To
see it, we need the relation between the vev of the adjoint scalar
$\Phi$ and the curve $\Sigma_{\text{SW}}$, given by \cite{Cachazo:2002ry,
Cachazo:2002zk}:
\begin{align}
 \Bracket{\tr{dx\over x-\Phi}}
 ={dy\over y}=
 {P'_N(x)\over\sqrt{P_N(x)^2-4\Lambda^{2N}}}dv.
 \label{vevrelSW}
\end{align}
In other words, $u_m$ defined geometrically in \eqref{defuMgeom} has an
interpretation in gauge theory as follows:
\begin{align}
 u_m^{}= {1\over m}\ev{\tr\Phi^m}.
 \label{uMgt}
\end{align}
From this, one immediately sees that  the superpotential
\eqref{supotRSgen} can be written as
\begin{align}
 W_{\text{SW}}
 =\sum_{m}  {T^m\over m+1} \tr \Phi^{m+1}
 =\tr[W(\Phi)],\label{spotpertSW}
\end{align}
where we defined
\begin{align}
  W(x)=\sum_m {T^m\over m+1} x^{m+1}.
 \label{defW(x)SW}
\end{align}
In \eqref{spotpertSW}, $\Phi$ is understood as the chiral superfield
whose lowest component is the adjoint scalar.

Therefore, the $\CN=2$ gauge theory perturbed by the single-trace
superpotential \eqref{spotpertSW} corresponds to the geometry
\eqref{SWCY} with the flux $g$ obeying the following asymptotic boundary
condition:
\begin{align}
 g\sim \sum_m m T^m x^{m-1}\,dx = W''(x)dx,
\end{align}
where we used \eqref{hypellxiMasym}.  Note that this equivalence holds
for any configurations, supersymmetric or nonsupersymmetric, because we
have shown the equality of the full off-shell scalar potential.  The
perturbed $\CN=2$ theory is precisely the system which was shown in
\cite{Ooguri:2007iu, Pastras:2007qr} to have nonsupersymmetric
metastable vacua if the superpotential is chosen
appropriately.\footnote{It was shown in \cite{Ooguri:2007iu} to be
possible to create metastable vacua by a single-trace superpotential of
the form \eqref{spotpertSW} at any point in the Coulomb moduli space for
$SU(2)$ and at least at the origin of the moduli space for $SU(N)$.}
Therefore, it tautologically follows that the IIB Seiberg-Witten
geometry with flux at infinity also has metastable vacua, if we tune the
parameters $T^m$ appropriately.

As we mentioned above, this IIB Seiberg-Witten geometry is dual to a IIA
brane configuration of NS5-branes and D4-branes which can be lifted to
an M5-brane configuration.  In \cite{Marsano:2008ts}, it was shown that
superpotential perturbation corresponds in the M-theory setup to
``curving'' the $\CN=2$ configuration of the M5-brane at infinity in a
way specified by the superpotential.  The metastable gauge theory
configuration of \cite{Ooguri:2007iu, Pastras:2007qr} was realized as a
metastable M5-brane configuration and its local stability was given a
geometrical interpretation.  The above proof of \eqref{spotpertSW} is
exactly in parallel to the one given in \cite{Marsano:2008ts} for the
M-theory system.
In passing, it is also worth mentioning that the M-theory analysis of
\cite{Marsano:2008ts} revealed that at strong coupling the
nonsupersymmetric configuration ``backreacts'' on the boundary condition
and it is no longer consistent to impose a holomorphic boundary
condition specified by a holomorphic superpotential, which is in accord
with \cite{Bena:2006rg}.  Therefore, also in the IIB flux setting, it is
expected that if we go beyond the approximation that the flux does not
backreact on the background metric, nonsupersymmetric flux
configurations will backreact and it will be impossible to impose a
holomorphic boundary condition of the type \eqref{1formgSW}.

Although we do not do it in the present paper, from the viewpoint of
flux compactification, it is a natural generalization to consider fluxes
through the compact 3-cycles.  Such flux will make additional
contribution to the superpotential of the form $e_i a^i+m^i F_i$ (see
\ref{subsec:OOPmech}).
On the gauge theory side, in the Seiberg-Witten theory, this can be
interpreted as perturbation one adds at the far IR, but its UV
interpretation is not clear \cite{Marsano:2007mt}.

\subsubsection{Dijkgraaf-Vafa (CIV-DV) Geometries}
\label{subsec:DVgeo}

Another example of geometries of the type \eqref{generalRSCY} is type
IIB on
\begin{equation}
  \CM_{\text{DV}}:\qquad uv - F_{\text{DV}}(x,y)=0, \qquad x,y \in \bbC,
   \label{DVCY}
\end{equation}
where the underlying Riemann surface $\Sigma_{\text{DV}}$ is a hyperelliptic curve
\begin{equation}
   \Sigma_{\text{DV}}: \qquad
    F_{\text{DV}}(x,y)\equiv w^2-\left[P_n(x)^2- f_{n-1}(x)\right]=0
\label{DVcurve}
\end{equation}
and $P_n(x)$ and $f_{n-1}(x)$ are polynomials of degree $n$ and $n-1$,
respectively. If we write
\begin{align}
 f_{n-1}(x)=\sum_{i=1}^{n-1} b_{i} x^i,\label{f(x)_and_bi}
\end{align}
then the coefficients of $P_n(x)$ as well as $b_{n-1}$ are
nonnormalizable and fixed\footnote{More precisely, $b_{n-1}$ is
log-normalizable and can be treated as a variable modulus if one wishes,
but in the present paper we treat it as a non-dynamical parameter.},
while $b_i$, $i=0,\dots,n-2$ are normalizable complex structure moduli.
The holomorphic 3-form is $\Omega_{\text{DV}}={du\over u}\wedge dx\wedge dy$
which reduces to
\begin{align}
 \lambda_{\text{DV}}=x\,dy  \label{DV1form}
\end{align} 
on the Riemann surface $\Sigma_{\text{DV}}$.  The geometry \eqref{DVCY} was
studied by Cachazo, Intriligator and Vafa (CIV) \cite{Cachazo:2001jy}
(see also \cite{Cachazo:2002pr}) in the context of large $N$ transition
\cite{Gopakumar:1998ki, Vafa:2000wi} and further generalized in
\cite{Cachazo:2001gh, Cachazo:2001sg}. The Dijkgraaf-Vafa (DV)
conjecture \cite{Dijkgraaf:2002fc, Dijkgraaf:2002vw, Dijkgraaf:2002dh}
was also based on the same geometry.  We will refer to this geometry as
the CIV-DV geometry \eqref{DVCY} or as the Dijkgraaf-Vafa geometry
henceforth.

The structure of the underlying hyperelliptic Riemann surface
$\Sigma_{\text{DV}}$ \eqref{DVcurve} is similar to the Seiberg-Witten case
\eqref{SWcurve}; $\Sigma_{\text{DV}}$ is a genus $n-1$ surface with two punctures
at infinity.  If we represent $\Sigma_{\text{DV}}$ as a two-sheeted $x$-plane
branched over $2n$ points, those infinities correspond to $x=\infty$ on
the two sheets.  The coefficients of $P_n(x)$, which are
nonnormalizable, determine the position of the $n$ cuts on the
$x$-plane, while the coefficients of $f_{n-1}(x)$, which are
normalizable, are related to the sizes of the cuts.  The first homology
$H_1(\Sigma_{\text{DV}})$ is spanned by $n-1$ pairs of compact $A$- and
$B$-cycles $(A^i,B_j)$, $i,j=1,\dots,n-1$ with in addition a closed
cycle $A^\infty$ around one of the infinities which is dual to the
noncompact $B$-cycle $B_\infty$ connecting two infinities.  Because
$x,y\in\bbC$, compact $A$- and $B$-cycles on $\Sigma_{\text{DV}}$ are all
contractible in the $x,y$-plane and hence all compact 1-cycles on
$\Sigma_{\text{DV}}$ lifts to 3-cycles in $\CM_{\text{DV}}$ with $S^3$ topology.

The special coordinates \eqref{spclcoordRS} in this case is conventionally
denoted by $S^i$, $\Pi_i$:
\begin{align}
 S^i
 ={1\over 2\pi i}\int_{A^i}\lambda_{\text{DV}}
 , \qquad
 \Pi_i
 ={1\over 2\pi i}\int_{B_i}\lambda_{\text{DV}}
 ,
 \label{spclcoordDV}
\end{align}
for which, as in \eqref{spclmtrcRS}, the moduli space metric takes the
special form:
\begin{align}
 ds^2=(\Im\tau_{ij})\,dS^i d\overline{S^j}.
 \label{DVmetricS}
\end{align}
Just as in \eqref{omegaIexplct}, we can write the basis of holomorphic
1-forms $\omega_i$ using $S^i$ as:
\begin{align}
 \omega_i
 &
 ={\partial \over \partial S^i}(-y\,dx)
 ={\partial f_{n-1}(x)/ \partial S^i \over 2\sqrt{P_n(x)^2-f_{n-1}(x)}}\,dx.
 \label{omegaIitfSderDV}
\end{align}

Adding flux at infinity and breaking $\CN=2$ supersymmetry to $\CN=1$ go
just as in the Seiberg-Witten case.  The Riemann surface $\Sigma_{\text{DV}}$
is hyperelliptic and we take $\{\xi_m\}$ and $\{\omega_i\}$ to be the
ones given in Appendix \ref{app:hyperelliptic}.  Just like
\eqref{1formRS} and \eqref{1formgSW}, the 3-form flux in $\CM_{\text{DV}}$ reduces
to a harmonic 1-form on $\Sigma_{\text{DV}}$:
\begin{equation}
g = \sum_{m\ge 1} T^m  \xi_m 
 + \sum_{i=1}^{N-1} h^i \omega_i 
 + \sum_{i=1}^{N-1} \overline{l^i}\overline{\omega_i}.\label{1formgDV}
\end{equation}
Under the condition that the compact flux vanishes (eq.\
\eqref{0cptflxRS}), the 1-form \eqref{1formgDV} leads to the scalar
potential \eqref{scalarpotRS2} which, just as we derived
\eqref{supotRSgen}, can be shown to correspond to the following
superpotential:
\begin{align}
 W_{\text{DV}}=\sum_m T^m \Sigma_{m+1},
 \label{supotDVgeo}
\end{align}
where we defined
\begin{align}
 \Sigma_m
 \equiv {1\over 2\pi i m}\oint_{\infty} x^{m-1} \lambda_{\text{DV}}
 = {1\over 2\pi i m}\oint_{\infty} x^{m}\, dy.
 \label{defSigmageo}
\end{align}
The 1-form $\lambda_{\text{DV}}$ depends on the complex structure moduli $S^i$
of the Riemann surface \eqref{DVcurve}. Therefore, by changing the
parameters $T^m$, we can generate a superpotential which is a quite
general function of $S^i$'s.
The OOP mechanism \cite{Ooguri:2007iu} states that, if one tunes
superpotential appropriately, one can create a metastable vacuum at any
point of the $\CN=2$ moduli space.  Therefore, also for this
Dijkgraaf-Vafa geometry, we expect to be able to create metastable vacua
by appropriately tuning $T^m$, {\it i.e.}, flux at infinity.  Indeed, in
the next subsection we will demonstrate the existence of metastable
vacua in a simple example.

We have been focusing on the case where there is flux at infinity but
there is \emph{no} flux through \emph{compact} cycles.  However, let us
digress a little while and think about the case where there \emph{is}
flux through compact cycles but there is \emph{no} flux at infinity.  In
this case, the IIB system has a standard interpretation
\cite{Cachazo:2001jy, Cachazo:2002pr, Dijkgraaf:2002fc,
Dijkgraaf:2002vw, Dijkgraaf:2002dh} as describing the IR dynamics of
$\CN=2$ $SU(N)$ theory\footnote{This is the case when we treat $b_{n-1}$
as non-dynamical.  If we regard this as dynamical, this system realizes
$U(N)$ theory.}  broken to $\CN=1$ by a superpotential
$W=\tr[W_n(\Phi)]$, $W'_n(x)=P_n(x)$, with the moduli $S^i$ identified
with glueball fields.  More precisely, if there are $N^i$ units of flux
through the cycle $A^i$, where $N=\sum_i N^i$, then the system
corresponds to the supersymmetric ground state of $SU(N)$ gauge theory
broken to $\bigl[\prod_i SU(N^i)\bigr]\times U(1)^{n-1}$.
It is important to note that this equivalence between the Dijkgraaf-Vafa
flux geometry and gauge theory is guaranteed to work only for
holomorphic dynamics, or for the $F$-term.  On the geometry side, one is
considering the underlying geometry \eqref{DVCY} determined by $P_n(x)$
and small flux perturbation on it.  On the gauge theory side, this
corresponds to the limit of large superpotential, where one has no
control of the $D$-term.  Therefore, there is no {\it a priori\/} reason
to expect that the $D$-term of the Dijkgraaf-Vafa geometry, which
governs {\it e.g.}\ existence of nonsupersymmetric vacua, and that of
gauge theory are the same, even qualitatively.  After all, two systems
are different theories and it is only the holomorphic dynamics that is
shared by the two.\footnote{Of course, it is a logical possibility that
even the $D$-terms of the two systems are identical, or closely related
to each other. }

Despite such subtlety, it is interesting to ask what is the gauge theory
interpretation of adding flux at infinity, in addition to flux through
compact cycles.
It is known that the curve \eqref{DVcurve} is related to the vev in
gauge theory as \cite{Dijkgraaf:2002fc, Dijkgraaf:2002vw,
Dijkgraaf:2002dh, Cachazo:2002ry, deBoer:2004he}:
\begin{align}
 -{1\over 32\pi^2}\Bracket{\tr{\CW^2\over x-\Phi}}dx
 ={y\,dx}
 =\sqrt{P_n(x)^2-f_{n-1}(x)}\,dx.
 \label{rel_vdw_W2Phi}
\end{align}
where $\CW^2=\CW_\alpha \CW^\alpha$ and $\CW_\alpha$ is the gaugino
field.  Comparing this with \eqref{defSigmageo}, one finds that the
quantity $\Sigma_m$ defined geometrically in \eqref{defSigmageo} has the
following interpretation:
\begin{align}
 \Sigma_m= {1\over 32\pi^2}\ev{\tr(\CW^2\Phi^{m-1})}.
 \label{V_M}
\end{align}
Therefore the superpotential \eqref{supotDVgeo} can be written as
\begin{align}
 W_{\text{DV}}
 ={1\over 32\pi^2} \sum_m T^m \tr[\CW^2\Phi^m]
 ={1\over 32\pi^2}\tr[\CW^2 M(\Phi)],
 \label{WDVgt}
\end{align}
where we defined
\begin{align}
 M(x)=\sum_m T^m x^{m}.
 \label{defM(x)DV}
\end{align}
Therefore,
flux at infinity of the  following
asymptotic form:
\begin{align}
 g\sim \sum_m m T^m x^{m-1}\,dx = M'(x)dx,
\end{align}
corresponds in gauge theory to adding a novel superpotential of the form
\eqref{WDVgt}.  Again, this correspondence must be taken with a grain of
salt, since it holds only for holomorphic physics.

Note also that flux through compact cycles will induce glueball
superpotential \cite{Cachazo:2001jy} of the form $\alpha_i S^i + N^i
\Pi_i(S)$ added to \eqref{supotDVgeo}.  Because this part does not contain
tunable parameters such as $T^m$ that can be made very small, it is
difficult, if not possible, to use the OOP mechanism to produce
metastable vacua in that case.

\bigskip
Now let us come back to the main focus of the present paper, the case
where there is no flux through compact cycles.  In this case, we do not
have an interpretation of the system as such an $SU(N)$ theory described
above, simply because $N=\sum_i N^i=0$.  Below, we take the Dijkgraaf-Vafa
geometry with flux at infinity and no flux through compact cycles as an
example, and see that we can generate metastable vacua by adjusting the
parameters $T^m$ using the OOP mechanism outlined in the previous
section.



\subsection{Metastable Flux Vacua in CIV-DV Geometries -- An Example}
\label{sec:OOPex}

To demonstrate that one can truly realize supersymmetry-breaking via the
OOP mechanism in type IIB Dijkgraaf-Vafa flux geometries, we turn our
attention now to a simple example, namely the geometry relevant for
$SU(2)$
\begin{equation}uv-F_{\text{DV}}(x,y)=0,\qquad \mbox{with} \quad F_{\text{DV}}(x,y)=y^2-\left[P_2(x)^2-b_1x-b_0\right],\end{equation}
where we choose
\begin{equation}P_2(x)=x^2-\frac{\Delta^2}{4}.\end{equation}
For simplicity, we will impose a $\mathbb{Z}_2$ symmetry on the
Calabi-Yau under which $x\leftrightarrow -x$, the effect of which is to
set the log-normalizable modulus $b_1$ to zero
\begin{equation}b_1=0.\end{equation}

As usual, we can focus our attention on the associated Riemann surface,
$\Sigma_{\text{DV}}$, which in this example has genus 1 and is determined by
the equation
\begin{equation}
F_{\text{DV}}(x,y)=y^2-\left[P_2(x)^2-b_0\right]=0.\label{DVg1}
\end{equation}
This geometry admits a single dynamical modulus, $b_0$.  This, in turn,
can locally be traded for the special coordinate $S^1$ which, for
notational simplicity, we refer to as $S$ in the remainder of this
section
\begin{equation}S\equiv S^1=\frac{1}{2\pi i}\oint_{A^1}\,x\,dy.\end{equation}
Alternatively, we can parametrize the moduli space by the globally
well-defined coordinate $\Sigma_2$ \eqref{defSigmageo}, which we choose
to denote simply by $\Sigma$
\begin{equation}\Sigma\equiv \Sigma_2=\frac{1}{4\pi i}\oint_{x=\infty}\,x^2\,dy.\end{equation}
 
 To this geometry, we now consider turning on flux given by
\begin{equation}g=\sum_{m\ge 1}T^m\xi_m+h\omega + \overline{l}\overline{\omega},\end{equation}
where $\omega$ is the unique holomorphic 1-form on $\Sigma_{\text{DV}}$.  As we
have seen, this induces a nontrivial potential for $\Sigma$ of the form
\begin{equation}
V\sim \left(\frac{\partial W_{\text{DV}}(\Sigma)}{\partial\Sigma}\right)
\left(\frac{\partial\Sigma}{\partial S}\right)
\left({1\over\Im\tau}\right)
\overline{\left(\frac{\partial\Sigma}{\partial S}\right)}\,\overline{\left(\frac{\partial W_{\text{DV}}(\Sigma)}{\partial\Sigma}\right)},\end{equation}
where
\begin{equation}W_{\text{DV}}(\Sigma_2)=\sum_m T^m\Sigma_{m+1}(\Sigma_2).\end{equation}
To engineer a metastable vacuum at a fixed point, $\Sigma^{(0)}$, we
need only choose the $T^m$ so that $W_{\text{DV}}(\Sigma)$ is a cubic
polynomial in $\Sigma$ obtained by truncating the expansion of a
K\"ahler normal coordinate associated to $\Sigma^{(0)}$ at cubic order.
To determine the requisite $T^m$, we proceed in two steps.  First, we
must determine the relation between $\Sigma$ and the higher $\Sigma_m$
\eqref{defSigmageo}.  This is rather trivial.  Second, however, we must
obtain an expression for the K\"ahler normal coordinate associated to a
generic point $\Sigma^{(0)}$.  This will be slightly messier.

\subsubsection{Relating $\Sigma_m$ and $\Sigma_2$}

Evaluating generic $\Sigma_m$ for $\Sigma_{\text{DV}}$ is relatively easy to do
given the defining equation \eqref{DVg1} and leads to the result
\begin{equation}\begin{split}\Sigma_{2q-1}&=0,\\
\Sigma_{4q}&=\sum_{n=0}^q\frac{\Gamma\left(2q-n+\frac{1}{2}\right)}{2\sqrt{\pi}\left(2q-2n+1\right)!\,n!}\left(\frac{\Delta^2}{2}\right)^{2(q-n)+1}\left(b_0-\frac{\Delta^4}{16}\right)^n,\\
\Sigma_{4q-2}&=\frac{1}{2q-1}\left(b_0-\frac{\Delta^2}{16}\right)^q\left[\frac{\Gamma\left(q+\frac{1}{2}\right)}{q!\sqrt{\pi}}\right]\\
&\qquad\qquad+
\sum_{n=0}^{q-1}\frac{\Gamma\left(2q-n-\frac{1}{2}\right)}{2\sqrt{\pi}\left(2q-2n\right)!n!}\left(\frac{\Delta^2}{2}\right)^{2(q-n)}\left(b_0-\frac{\Delta^4}{16}\right)^n.
\end{split}\end{equation}
From this, we first see that $\Sigma$ is proportional to $b_0$
\begin{equation}\Sigma=\frac{b_0}{2}.\end{equation}
More importantly, however, we are also able to immediately read off the
degree of each nonzero $\Sigma_m$ when viewed as a polynomial in
$\Sigma$
\begin{equation}\Sigma_{4q}\sim \Sigma^q+{\cal{O}}(\Sigma^{q-1}),\qquad \Sigma_{4q-2}\sim \Sigma^q+{\cal{O}}(\Sigma^{q-1}).\end{equation}
Consequently, the lowest $m$ for which $\Sigma_m$ contains a term
proportional to $\Sigma^q$ is $m=4q-2$.  This means that to introduce
terms of order $\Sigma^3$ into $W_{\text{DV}}(\Sigma)$, it will be
necessary to include $\Sigma_m$ up to $m=10$, leading to much more
singular flux than one might have otherwise thought.  This is the first
example of a general lesson we will have more to say about later, namely
that when engineering OOP vacua, the requisite noncompactly supported
flux can have a large degree of divergence which, to the best of our
knowledge, is not easy to determine by any simple arguments.

Because one can introduce quadratic (cubic) terms using any $\Sigma_m$
with $6\le m\le 9$ ($10\le m\le 13$) there is some choice as to which
$T^m$ we can turn on to achieve a particular desired
$W_{\text{DV}}(\Sigma)$.  For the purposes of our example, we will only
turn on $T^1$, $T^5$, and $T^{9}$, thereby adding terms proportional to
\begin{equation}\begin{split}\Sigma_2&\equiv\Sigma,\\
\Sigma_6&=\frac{1}{16}\left(\Sigma\Delta^4+8\Sigma^2\right),\\
\Sigma_{10}&=\frac{1}{256}\left(\Delta^8\Sigma+48\Delta^4\Sigma^2+128\Sigma^3\right).
\end{split}\label{sigmas2610}\end{equation}

\subsubsection{K\"ahler Normal Coordinate for $\Sigma_0$}

We now proceed to the second step, namely computing the first few terms
of the K\"ahler normal coordinate expansion about a generic point
$\Sigma^{(0)}$
\begin{equation}z=\left(\Sigma-\Sigma^{(0)}\right)+a_2\left(\Sigma-\Sigma^{(0)}\right)^2+a_3\left(\Sigma-\Sigma^{(0)}\right)^3+\ldots,\label{KNCg1}\end{equation}
where we have implicitly defined the coefficients
\begin{equation}\begin{split}a_2&=\frac{1}{2}\Gamma^{\Sigma}_{\Sigma\Sigma},\\
a_3&=\frac{1}{6}g^{\Sigma\overline{\Sigma}}\partial_{\Sigma}\left(g_{\Sigma\overline{\Sigma}}
  \Gamma^{\Sigma}_{\Sigma\Sigma}\right),\end{split}\end{equation}
with $g_{\Sigma\overline{\Sigma}}$ the metric associated to the $\Sigma$ coordinate
\begin{equation}g_{\Sigma\overline{\Sigma}}=\left|\frac{\partial S}{\partial\Sigma}\right|^2\Im\tau,\end{equation}
and $\Gamma^{\Sigma}_{\Sigma\Sigma}$ the associated nonvanishing
Christoffel symbol.  Computations of quantities such as $\partial
S/\partial\Sigma$ in Dijkgraaf-Vafa geometries are often performed using
a perturbative expansion about the singular point $S=0$.  For
engineering OOP type vacua, though, we need to consider instead the
neighborhood of a generic, nonsingular point $\Sigma_0$ away from $S=0$.
Fortunately, in the simple case of a genus 1 curve, we can actually
obtain exact results without too much work by taking advantage of the
parametric description reviewed in Appendix \ref{app:param}.  As
described there, one finds that both $S$ and $\Sigma$ can be expressed
directly as functions of $\tau$
\begin{equation}S=\frac{\Delta^3}{2\pi i \left[12\wp(\tau/2)\right]^{3/2}}\left(\frac{2g_2}{3}-4\wp(\tau/2)\left[\wp(\tau/2)-2\eta_1\right]\right),\label{Sresult}\end{equation}
\begin{equation}\Sigma=\frac{\Delta^4\left(12\wp(\tau/2)^2-g_2\right)}{288\wp(\tau/2)^2},\label{Sigresult}\end{equation}
where $\wp(z)$ is the Weierstrass $\wp$-function, $g_2$ is the
Weierstrass elliptic invariant appearing in the relation
\begin{equation}\left(\frac{\partial\wp(z)}{\partial z}\right)^2=4\wp(z)^3-g_2\wp(z)-g_3,\end{equation}
and $\eta_1$ is one of the half-periods of the Weierstrass $\zeta$-function
\begin{equation}\eta_1=\zeta\left(\frac{1}{2}\right).\end{equation}
From \eqref{Sresult} and \eqref{Sigresult}, we can apply the
differentiation formulae of Appendix \ref{app:param} to write both
$\partial\Sigma/\partial S$ and $g_{\Sigma\overline{\Sigma}}$ as
functions of $\tau$
\begin{equation}\frac{\partial\Sigma}{\partial S}=-\frac{i\pi \Delta}{\sqrt{3\wp(\tau/2)}}\qquad\implies\qquad g_{\Sigma\overline{\Sigma}}=\frac{3}{\pi^2}\left|\Delta^2\wp(\tau/2)\right|\Im\tau.\end{equation}
It is now straightforward to determine the coefficients of the K\"ahler
normal coordinate expansion \eqref{KNCg1} in terms of the value of
$\tau$ at $\Sigma^{(0)}$
\begin{equation}\begin{split}a_2&=\frac{36\wp(\tau/2)^2\left(g_2+12\eta_1\wp(\tau/2)-6\wp(\tau/2)^2\right)
      -\frac{216\pi\wp(\tau/2)^3}{\Im\tau}}{\Delta^4\left(g_2-3\wp(\tau/2)^2\right)\left(g_2-12\wp(\tau/2)^2\right)},\\
a_3&=\frac{864\wp(\tau/2)^4}{\Delta^8\left(g_2-3\wp(\tau/2)^2\right)^2\left(g_2-15\wp(\tau/2)^2\right)^2}\\
&\qquad\times\biggl[\frac{360\pi\wp(\tau/2)^3-48\pi g_2\wp(\tau/2)}{\Im\tau}\\
&\qquad\qquad\qquad +5g_2^2+96\eta_1g_2\wp(\tau/2)-63g_2\wp(\tau/2)^2-720\eta_1\wp(\tau/2)^3+252\wp(\tau/2)^4\biggr].\label{a2a3result}
\end{split}\end{equation}

\subsubsection{Noncompact Flux for Engineering OOP Vacuum}

We are finally ready to explicitly write the noncompact flux needed to
engineer an OOP vacuum at a generic point $\Sigma^{(0)}$.  In
particular, we seek to specify values for the coefficients $T^m$ which
render
\begin{equation}W_{\text{DV}}(\Sigma)=\sum_m\,T^m\Sigma_m(\Sigma)\end{equation}
equivalent, up to a constant shift, to a truncation of the K\"ahler
normal coordinate expansion \eqref{KNCg1} about $\Sigma^{(0)}$ at order
$\Sigma^3$.  Using \eqref{sigmas2610}, it is easy to see that the
following choice of nonzero $T^m$ does the job
\begin{equation}\begin{split}T^1&=2-\frac{a_2}{4}\left(\Delta^4+16\Sigma_0\right)+2a_3\left(\frac{5\Delta^8}{128}+\frac{3\Delta^4\Sigma_0}{8}+3\Sigma_0^2\right),\\
T^5&=\frac{12a_2}{5}-\frac{9a_3}{10}\left(\Delta^4+8\Sigma_0\right),\\
T^{9}&=\frac{20a_3}{9}, \label{Tex}\end{split}\end{equation} where $a_2$
and $a_3$ given by the expressions in \eqref{a2a3result} evaluated at
the value of $\tau$ corresponding to $\Sigma^{(0)}$.

\begin{figure}
\begin{center}
\epsfig{file=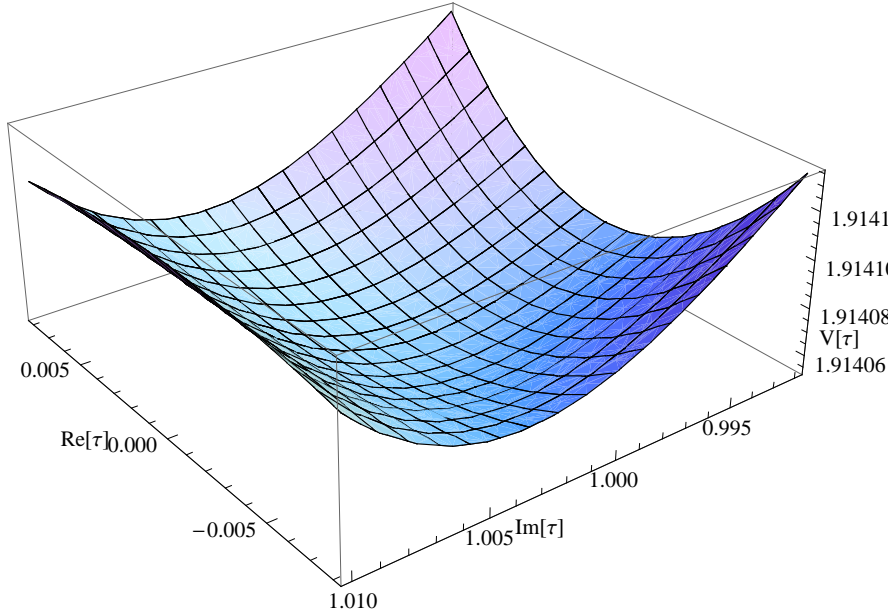,width=0.7\textwidth}
\caption{Plot of $V(\tau)$ in the neighborhood of our engineered OOP minimum at $\tau=i$}
\label{fig:potexample}
\end{center}
\end{figure}

These expressions, while nice and exact, are a little cumbersome so let
us also consider a special case where things simplify.  To that end, we
try to engineer an OOP vacuum at the special point $\tau=i$
corresponding to a square torus.  In this case, several elliptic
quantities simplify
\begin{equation}\left.\eta_1\right|_{\tau=i}=\frac{\pi}{2}\qquad \left.g_3\right|_{\tau=i}=0\implies \left.g_2\right|_{\tau=i}=\left.4\wp(\tau/2)^2\right|_{\tau=i}.\label{simpls}\end{equation}
The value of $\Sigma$ at $\tau=i$ can be obtained by applying \eqref{simpls} to \eqref{Sigresult}
\begin{equation}\Sigma_0=\frac{\Delta^4}{36}.\end{equation}
This means that the curve \eqref{DVg1} is given at this point by
\begin{align}
 y^2&=x^4-{\Delta^2\over 2}x^2+{\Delta^4\over 144}
 =
 \left[x^2-\left({1\over 4}+{1\over 3\sqrt{2}}\right)\Delta^2\right]
 \left[x^2-\left({1\over 4}-{1\over 3\sqrt{2}}\right)\Delta^2\right].
\end{align}
The coefficients $a_2$ and $a_3$ \eqref{a2a3result} appearing in the
K\"ahler Normal Coordinate expansion \eqref{KNCg1} simplify to
\begin{equation}\left.a_2\right|_{\tau=i}=\frac{9}{\Delta^4}\qquad\text{and}\qquad \left.a_3\right|_{\tau=i}=\frac{1080}{\Delta^8}\end{equation}
Inserting these into \eqref{KNCg1}, we find that our desired effective superpotential $W_{\text{DV}}(\Sigma)$ is given by
\begin{equation}W_{\text{DV}}(\Sigma)=\text{constant} + 3\Sigma-\frac{81\Sigma^2}{\Delta^4}+\frac{1080\Sigma^3}{\Delta^8},\end{equation}
while plugging into \eqref{Tex} yields the $T^m$ that do the job
\begin{equation}T^2=\frac{885}{8},\qquad T^6=-\frac{5832}{5\Delta^4},\qquad T^{10}=\frac{2400}{\Delta^8}.\end{equation}
While metastability of the vacuum at $\tau=i$ is guaranteed by the OOP procedure, it is also gratifying to see it graphically by explicitly plotting the potential near $\tau=i$ as in figure \ref{fig:potexample}.

\subsection{Degree of Superpotential Required for Metastable Vacua}
\label{subsec:tunability}

As we have seen in the above example, there is an issue about the degree
of superpotential we have to consider in order to create OOP metastable
vacua.  In this subsection, we analyze this issue.

As one can see from \eqref{KNCdef}, \eqref{WeffKNC}, in order to create
an OOP vacuum at a specific point $X^{(0)}$ in the moduli space, one must be
able to adjust the coefficients in the superpotential up to cubic terms
in $\Delta X=X-X^{(0)}$.  If the dimension of the moduli space is $d$, this means
that we generically need to tune
\begin{align}
 d+{d(d+1)\over 2}+{d(d+1)(d+2)\over 6}-d=
 {d(d+1)(d+5)\over 6}\equiv C_d
 \label{nopOOP}
\end{align}
parameters in the superpotential.\footnote{For having a metastable
vacuum, the superpotential does not have to be \emph{exactly} the same
as the ones given in \eqref{WeffKNC}; if the
coefficients are very close to the ones given in \eqref{KNCdef},
\eqref{WeffKNC}, one still expect to have metastable vacua.  However,
this does not generically affect the number of parameters we need to
tune.}  The last term is subtracting the degrees of freedom to choose
the vector $k_i$.

In the local Calabi-Yau geometries we have been considering, the
superpotential is parametrized by the coefficients $T^m$. For example,
in the Dijkgraaf-Vafa geometry, the superpotential was given by
\eqref{supotDVgeo}:
\begin{align}
 W_{\text{DV}}(S)=\sum_{m\ge 1} T^m \Sigma_{m+1}(S),\qquad
 \Sigma_m(S)= {1\over 2\pi i m}\oint_{\infty} x^{m} dy(S),
 \label{WDVitoSigmaS}
\end{align}
where we wrote the dependence of $\Sigma_{m}$'s on the moduli
$S=\{S^i\}$ explicitly.  Therefore, if $\Sigma_{m}(S)$'s are generic
functions of $S$ then, by tuning $C_{n-1}$ parameters\footnote{Note that
the number of moduli is $n-1$ because we are treating $b_{n-1}$
dynamical.} $T^2,T^3,\dots T^{C_{n-1}+1}$, one can create a metastable
vacuum at a generic point $S=S^{(0)}$.
More precisely, the OOP mechanism requires that, when we expand
$\Sigma_m(S)$'s around $S^{(0)}$ in $\Delta S=S-S^{(0)}$, the
coefficients of $\Delta S,(\Delta S)^2,(\Delta S)^3$ terms are all
independent and by taking linear combinations of $\Sigma_m(S)$'s we can
obtain the superpotential \eqref{WeffKNC}.

However, as we saw in the example above, the situation is not generic
for small $n$ and we need a more detailed analysis about how high
degrees one should go, which is done in Appendix
\ref{app:indepSigmap}\@.  The result (eq.\ \eqref{MminSU(N)}) is that,
if we would like to make a critical point at a generic point in the
moduli space, we have to tune on $T^m$ at least up to $m=m_{\text{min}}$,
where\footnote{This result is for the case where $b_{n-1}$ is treated
nondynamical.  For the result in the case where $b_{n-1}$ is regarded as
a modulus, see Appendix \ref{app:indepSigmap}.}
\begin{equation}
 \begin{split}
 n=2    &:~~ m_{\text{min}}=10,\\
 n=3    &:~~ m_{\text{min}}=15,\\
 n\ge 4 &:~~ m_{\text{min}}={n^3\over 6}+{n^2\over 2}+{n\over 3}+3.
\end{split}
\label{reqdegSigma}
\end{equation}
There is certain genericily assumption on the dependence of $\Sigma_m$
on the moduli (see Appendix \ref{app:indepSigmap}), and hence the actual
degree $m$ one must consider can be larger than the one given above.

Therefore, in order to stabilize metastable vacua made of $n$ cuts by
the OOP superpotential, we have to consider $\Sigma_m$'s up to rather
high degree $m_{\text{min}}$ given by \eqref{reqdegSigma} at least.  Because
the degree $m$ corresponds to the order of divergence of the flux at
infinity ($\xi_m$), the noncompactly supported flux must diverge at
infinity at the corresponding speed.


\section{Factorization}
\label{sec:factorization}
\subsection{The Basic Idea}
\label{subsec:factbasic}

In the previous sections we described how we can generate a
supersymmetry breaking potential for the complex structure moduli of a
local Calabi-Yau singularity by the introduction of 3-form flux which
has support at infinity. Allowing flux with noncompact support may
lead to various conceptual difficulties, such as the divergence of the
total energy density. To clarify these difficulties we would like to
sketch how such a 
system can be interpreted as an approximation of a larger Calabi-Yau
threefold with flux of compact support in a certain factorization limit.

More precisely, we start with a Calabi-Yau with a subset of cycles
pierced by usual 3-form flux of compact support. In another region of the
manifold we have a second subset of cycles. The flux from the
first cycles will generate a potential for the complex structure
moduli of the second set. In the limit where the cycles are separated
by a large distance, and where we zoom in towards the second set, the
flux from the first subset will look as if it is coming from
``infinity''\footnote{As we will see in more detail later, we also have
to scale the flux in an appropriate way.}. In this sense, the
noncompact setup considered in the previous section can be considered
as a small part of a larger Calabi-Yau with compactly supported flux.

In this section we would like to understand this embedding into a bigger
Calabi-Yau in more detail. Our goal is to see how the potential
\eqref{finalpot} arises starting from the standard Gukov-Vafa-Witten
superpotential for 3-form flux in the larger Calabi-Yau. 

For simplicity we will work with a noncompact Calabi-Yau $\CM$,
\begin{equation}
\CM:\quad  uv -F(x,y) =0,
\label{cyfact}
\end{equation}
which is based on a Riemann surface $\Sigma$ given by $F(x,y)=0$.  As we
explained before the complex parameters entering the defining equation
of the Riemann surface correspond to complex structure moduli of the
Calabi-Yau. Some of them are non-normalizable and can be considered as
external parameters. We want to tune these parameters to approach the
limit where the surface $\Sigma$ factorizes into two surfaces $\Sigma_L$
and $\Sigma_R$ connected by long tubes. This factorization lifts to the
entire Calabi-Yau $\CM$ and divides it into two regions $\CM_L$ and $\CM_R$
that are widely separated. We introduce 3-form flux $G_3$ of compact
support on the 3-cycles of $\CM_R$. The superpotential and scalar
potential are given by
\begin{equation}  
W = \int G_3 \wedge \Omega
 \qquad 
 \mbox{and} \qquad
 V = G^{I\overline{J}}\partial_I W
\overline{\partial_{J} W},
\label{pottwo}
\end{equation}
where the indices $I,J$ run over all complex structure moduli 
of the total threefold $\CM$. Using the properties of the K\"ahler
metric $G_{I\overline{J}}$ in the factorization limit we show
that the part of the potential \eqref{pottwo} which depends on the
complex structure moduli of $\CM_L$ is of the form
\eqref{finalpot}. Furthermore, we find an understanding of the
effective value of the parameters $T^m$.

\subsection{Geometry of Factorization}

\begin{figure}[h!]
\begin{quote}
 \begin{center}
 \epsfig{file=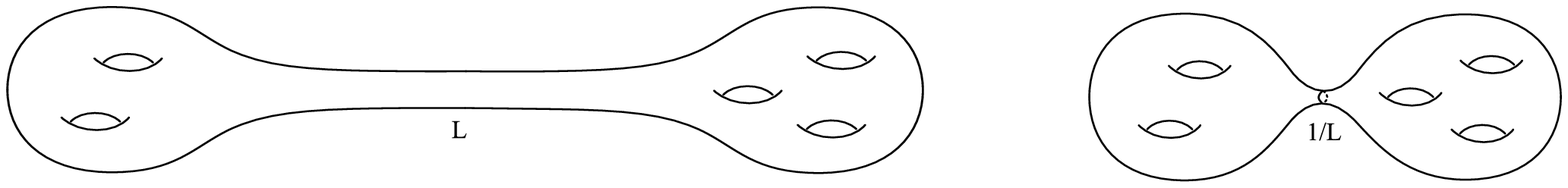,width=.85\textwidth} \caption{Two
 conformally equivalent ways of viewing the factorization of a Riemann
 surface into two parts. Physically though, we should distinguish both
 points of view, since particle masses depend on the size of the
 cycles. Because in our situation no new massless appears in the
 factorization limit, the left diagram represents our point of view
 best. } \label{fig:section6plumb}
 \end{center}
\end{quote}
\end{figure}

In this subsection we study the degeneration of a
Riemann surface $\Sigma$ into two components $\Sigma_L$ and
$\Sigma_R$, depicted in figure~\ref{fig:section6plumb}.\footnote{In
  general these components  could be connected  in a non-trivial
  way. We restrict our computations in this section to the case in 
  which they are linked by just one long tube. These should be
  easily extendible to more general cases.}  In this factorization
data of the full Riemann surface is expressed in terms of the complex 
structure of the individual surfaces. It is well known that in the
limit where the length of the tubes $L= 1/\epsilon$ goes to
infinity the period matrix of the full surface becomes block diagonal
\begin{equation}
  \tau = \left(\begin{array}{cc} \tau^{LL} & 0\\ 0& \tau^{RR}
    \\ \end{array}\right) + {\cal O}\left(\epsilon\right).
\label{periodfactorized}
\end{equation}
While the off-diagonal components $\tau^{LR}$ go to zero in the
factorization limit, their subleading behavior is quite important
in our analysis since it expresses the weak interaction between the
two sectors. The period matrix $\tau^{LR}$ can be computed
systematically in an expansion in $\epsilon$ from data on each of
the two surfaces as we explain below.

Technically, we describe the factorization of the Riemann surface with
the \emph{plumbing fixture} method \cite{Vafa:1987ea}. So consider two
Riemann surfaces $\Sigma_L$ and $\Sigma_R$ of genus $g_L$ and $g_R$
respectively.  On the left surface $\Sigma_L$ we have $g_L$ holomorphic
differentials $\omega_i$, while on the right surface $\Sigma_R$
similarly $g_R$ holomorphic differentials $\omega_{i'}$. The complex
structure of the left surface is determined by the periods of the
holomorphic differentials
\begin{equation}
  {1\over 2\pi i}\int_{A^i}\omega_j = \delta^i_{j},\qquad {1\over 2\pi i}\int_{B_i}\omega_j = \tau^{LL}_{ij},
\label{periodholdif}
\end{equation}
where $\tau^{LL}_{ij}$ is the period matrix of $\Sigma_L$, and we
choose our definitions similarly for the right surface.

The plumbing fixture method works after choosing a puncture $P$ on
$\Sigma_L$ and $P'$ on $\Sigma_R$. It
connects the two surfaces by a long tube of length $L$ 
which is glued onto neighborhoods of the punctures $P$ and $P'$. More
precisely, we pick a local holomorphic coordinate $z$ 
around the puncture $P$ such that $z(P)=0$ and a holomorphic coordinate
$z'$ near $P'$ with $z'(P')=0$. Then we identify points in these
neighborhoods as
\begin{equation}
  z z' = \epsilon.
\label{plumbing}
\end{equation}

Now we want to compute the period matrix of the full Riemann surface in
terms of complex structure data of the two surfaces. For this we need
to understand how the differentials $\omega_i$ and $\omega_{i'}$ extend to
well-defined holomorphic differentials on the full surface $\Sigma=
\Sigma_L \cup \Sigma_R/\sim$, where $\sim$ is the above
identification. Let us first consider how to lift
the differential $\omega_i$. Around the puncture $P$ it may be
expanded as
\begin{equation}
  \omega_i = \sum_{m=1}^\infty K^{P}_{im} z^{m-1} dz,
\label{bla1}
\end{equation}
where the functions $K_{im}^P$ are given by \eqref{intBxiP=KIM}.
Once we write  this in terms of $z'$ we observe that, as seen from the
right surface, the differential has a Laurent expansion. So $\omega_i$ will 
be written as a linear combination of the meromorphic differentials 
$\xi_{m}^{P'}$ of the right surface. A meromorphic differential
has the following expansion around the puncture
\begin{equation}
  \xi_m^P = \left({m\over z^{m+1}} + \sum_{n=1}^\infty h^P_{mn}
  z^{n-1}\right)dz.
\label{bla2}
\end{equation}
Here we have introduced the functions $h_{mn}^P$, which depend on the
complex structure moduli of the surface and the position of $P$.
So in general the differential $\omega_i$ will lift to a differential 
$\widetilde{\omega_i}$ on the full surface which can be written as
\begin{equation} 
\widetilde{\omega_i} = \left\{ \begin{array}{ll} \omega_i +
    \displaystyle\sum_{m=1}^\infty x_{i m} \xi^P_m &
\text{on}\quad \Sigma_L,\\[1ex]
  \displaystyle\sum_{m=1}^\infty y_{i m} \xi^{P'}_{m} &
\text{on}\quad \Sigma_R. \end{array} \right.
\label{bla3}\end{equation}
for some coefficients $x_{i m}$ and $y_{i m}$. Matching the differential on the two
sides we find the following conditions
\begin{equation} \begin{split}
 x_{im} &= - {\epsilon^m\over m} \sum_{n=1}^\infty  y_{in} h^{P'}_{nm},
		  \qquad
 y_{im}  = - {\epsilon^m\over m} \left( K^P_{im} + \sum_{n=1}^\infty x_{in}
  h^{P}_{nm} \right).
\end{split}\label{bla4}\end{equation}
This allows us to compute the cross-period matrix as
\begin{equation}
\begin{split}
  \tau^{LR}_{ij'} = \int_{b_{j'}} \omega_i 
 &= \sum_{m=1}^\infty  K^{P'}_{j'm} y_{im}  
 = -\sum_{m,n=1}^\infty {\epsilon^n\over n} K_{im}^P G^{-1}_{mn} K^{P'}_{j'n},\\
 G_{mn}&\equiv\delta_{mn}-\sum_{l=1}^\infty {\epsilon^{n+l}\over n l}h'_{ml}h_{ln}.
\end{split}
 \label{crossperiod}
\end{equation}
From this equation we can
read off all order $\epsilon$-corrections to the 
off-diagonal piece of the period matrix when a surface $\Sigma$ degenerates. 

Also, this procedure gives a clear understanding of the term ``flux at
infinity''. We see that the flux at infinity is generated by regular
forms on the degenerated surface, and therefore will at most have finite
order poles at the punctures.

Notice that for a Calabi-Yau threefold that is based on a Riemann surface,
the factorization region is described by the
deformed conifold geometry
\begin{equation}
 uv+ x^2 + y^2 = \epsilon, \quad \mbox{or equivalently}  \quad uv + zz' = \epsilon. 
\end{equation}
Usually, this is described as a 3-sphere shrinking to zero-size when
$\epsilon \to 0$. However, as for the complex 1-dimensional plumbing
fixture case we want the two sectors to be far apart from each other. Therefore
we consider the conformally  
equivalent setup where the 3-sphere is scaled to be of
finite size, while the transverse directions are made very
large. The finite size three-sphere reduces to the cross-section of
the tube on the left in figure~\ref{fig:section6plumb}, whereas the
transverse directions reduce to the tube-length.  

To describe the left and right neighborhoods of the degeneration, we
can fix $x= \sqrt{\epsilon - y^2 -uv}$ on the left and $x=-
\sqrt{\epsilon - y^2 -uv}$ on the right. In the limit that $\epsilon
\to 0$ these neighborhoods will not just intersect in a point, but in
the divisor $uv + y^2 = 0$. This is the region where regular forms on
the total threefold will develop poles when the degeneration
starts.



\subsection{Dynamics}
\label{subsec:dynamics}

\begin{figure}[h!]
\begin{center}
\epsfig{file=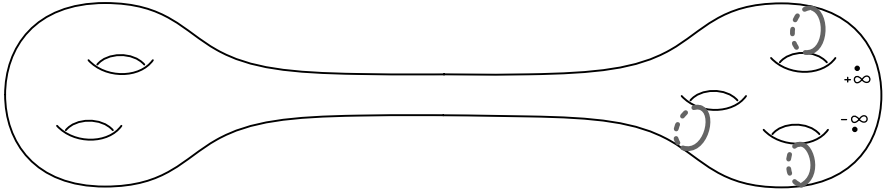,width=.70\textwidth}
\caption{Turning on flux on the right part of the factorized Calabi-Yau.}
\label{fig:section5degflux}
\end{center}
\end{figure}

Now we consider turning on flux on the threefold. For simplicity we
again take a Calabi-Yau \eqref{cyfact} that is
based on a factorized Riemann surface. We turn on 3-form flux $G_3 =
F_3 - \tau H_3$ which is only piercing the set of A-cycles
corresponding to $\Sigma_R$, as can be seen in
figure~\ref{fig:section5degflux}, and write down the corresponding
(super) potential. For regularization issues later, we take
two more punctures on the right surface labeled by 
$\pm\infty$ and turn on some flux $\alpha$ through the noncompact ${\cal
  B}_{\infty}$ cycle running from $+\infty$ to $-\infty$. 

A basis of $\cal{A}$ and $\cal{B}$ cycles is given by the compact
3-cycles on the left and the right, together with the lift
$\cal{A}^{\infty}$ of the $A$-cycle enclosing $+\infty$ and
$\cal{B}_{\infty}$. So the flux is determined by
\begin{equation} \begin{split}
&\int_{\CA^{i}} G_3 = 0, \qquad  \int_{\CA^{i'}} G_3 = N^{i'}, \qquad
\int_{\CA^{\infty}} G_3 = 0, \\
& \int_{\CB_i} G_3= 0,\qquad \int_{\CB_i'} G_3 = 0, \qquad \quad
\int_{\CB_{\infty}} G_3 
  = \alpha. \end{split}
\end{equation}

Let us denote the complex structure moduli and their duals by $X^I$ and
$F_I$, which are the $\CA^I$ resp.\ $\CB_I$ periods of
the holomorphic 3-form $\Omega$.  Here we use the capital indices
$I=\{i,i', \infty\}$ to run over both the left and the right sides.  Then
the GVW superpotential for the complex structure moduli is given by
\begin{equation}
  W = \int G_3 \wedge \Omega = \alpha  X^{\infty}  + \sum_{i'} N^{i'} F^R_{i'},
\label{gvwfact}
\end{equation}
and the corresponding scalar potential by
\begin{equation}
  V = \sum_{I,J} G^{I\overline{J}} \partial_I W \overline{\partial_J W}.
\label{scalarfact}
\end{equation}
Since $X^{\infty}$ corresponds to a log-normalizable period and the
derivatives in the above potential just correspond to normalizable
modes, the $\alpha$-factor decouples. This shows that
\begin{align}
  V = &
  \sum_{i,j,k',l'}  
  \left(N^{k'}\tau_{k'i}^{LR}\right)
  \left({1\over \Im\tau}\right)_{\!LL}^{ij} 
  \overline{\left(N^{l'}\tau_{l'j}^{LR}\right)}
  +\sum_{i,j',k',l'}
  \Re\left[ \left( N^{k'}\tau_{k'i}^{LR} \right)
  \left({1\over \Im\tau}\right)_{\!LR}^{ij'} 
  \overline{\left(N^{l'}\tau_{l'j'}^{RR}\right) }\right]\notag\\
  &\qquad
  + \sum_{i',j',k',l'}
  \left( N^{k'}\tau_{k'i'}^{RR}\right)
  \left({1\over \Im\tau}\right)_{\!RR}^{i'j'} 
  \overline{ \left(N^{l'}\tau_{l'j'}^{RR}\right)}.
\label{longpot}
\end{align}
Thus the total potential is the sum of three terms, which we
denote in the obvious way by $V=V_1+V_2+V_3$. 

Next we consider what happens in the limit where the distance $L$
between the two sets of 3-cycles gets very large. As explained
before the period matrices $\tau^{LL}$ and $\tau^{RR}$
remain of order one in this limit and become almost independent of the
moduli $X_R$ 
and $X_L$, respectively. 

On the other hand, $\tau^{LR}$ goes to zero which would make the first term
$V_1$ in the potential vanish in the limit that
$\epsilon \to 0$, at least if we don't scale the fluxes $N^{i'}$  appropriately.
Since $V_1$ describes the interaction between the
two sides of the Calabi-Yau, we really want to scale the fluxes
$N^{i'}$ to go to infinity in such a way that the term $V_1$
remains finite. 

Then it becomes clear that the term $V_3$ of the
potential dominates over the other two contribution to $V$. 
This implies that in the limit $\epsilon \to 0$ the term $V_3$ should
be minimized first, {\it i.e.},
\begin{equation}
  \sum_{k'} N^{k'}\tau_{k'i'}^{RR} =0, \quad \forall i',
\label{susyright}
\end{equation}
which is a set of $n_R$ equations for the $n_R$ moduli $x^{j'}$. The solutions
of this system correspond to supersymmetric vacua for the 3-cycles on
the right side. 
Once we have fixed all $X^{j'}$ to their supersymmetric values $\widehat{X}^{j'}$, 
we can consider the effect of the backreaction of the right side to the
left. This is purely expressed through the potential $V_1$, since
the term $V_2$ vanishes as well at the supersymmetric point. 

So effectively the potential for the complex structure
moduli $X_L^i$ of the left surface is 
\begin{equation}
  V_1 = \sum_{i,j,k',l'}  
   \left(N^{k'}\tau_{k'i}^{LR}\right) 
   \left({1\over \Im\tau}\right)_{\!LL}^{ij} 
   \overline{\left(N^{l'}\tau_{l'j}^{LR}\right)}.
   \label{v1ff}
\end{equation}
This may be written as
$  V_1 =\sum_{i,j} \partial_i W_{\text{eff}} (1/\Im\tau)_{LL}^{ij} \overline{\partial_j W_{\text{eff}}},$
where we define the effective ``superpotential'' for the left complex
structure moduli as 
\begin{equation}
 \partial_i W_{\text{eff}} \equiv \sum_{k'} N^{k'} \tau_{k'i}^{LR}.
\label{effpotential}
\end{equation}

Comparing with expression \eqref{crossperiod} it is clear that the
fluxes on the right should be scaled in such a way that the coefficients
\begin{equation}
T^m =  \epsilon^m   \sum_{k'} N^{k'} K'_{k'm}
\label{tms}
\end{equation}
remain constant. In that situation the effective superpotential is
\begin{equation}
  \partial_i W_{\text{eff}}  = \sum_m  T^m K_{im}
\label{finalsup}
\end{equation}
to leading order in $\epsilon$, which is precisely of the form \eqref{finalpot}.

\subsection{Genericity of Potential and Metastable Vacua}
\label{subsec:fivepointthree}

Let us summarize what we have demonstrated so far. We started with a
large Calabi-Yau that consists of two parts $\CM_L$ and $\CM_R$ separated
by a large distance, and turned on a large 3-form flux on one of the
sides, say $\CM_R$. This flux generates a large potential for the
complex structure moduli of $\CM_R$, which are therefore set to their
supersymmetric minima. The flux on $\CM_R$ is also weakly backreacting
to the other side $\CM_L$, inducing a small superpotential for the
complex structure moduli of $\CM_L$. We computed this superpotential in
equations \eqref{effpotential} and \eqref{finalsup} and found that it
is of the form \eqref{finalpot}. The main point is that the side $\CM_L$
only knows about $\CM_R$ via the parameters $T^m$ given by \eqref{tms}.

In this section we discuss two questions. The first to which 
degree we can tune the parameters $T^m$ independently. And the
second is whether these $T^m$'s can be chosen to
realize an OOP supersymmetry breaking superpotential.

As we can see from \eqref{tms}, the values of the parameters $T^m$
depend on the fluxes $N^{l'}$ on the cycles of $\CM_R$ and also on the
value of the (generalized) period matrix $K'_{l'm}$. The last one
depends on the choice of the supersymmetric vacuum $\widehat{X}^{j'}$ on the
right side. For given large fluxes $N^{l'}$ there is a huge number of
supersymmetric vacua, or solutions of \eqref{susyright}, with different
values of $\widehat{X}^{j'}$ and consequently of $K'_{l'm}$. The density of
such supersymmetric vacua over the complex structure moduli space of
$\CM_R$ has been studied before \cite{Douglas:2003um, Ashok:2003gk,
Denef:2004cf, Denef:2004ze, Torroba:2006kt}, and it is believed that the
vacua become dense in the moduli space in the limit where the fluxes are
very large.

The coefficients $K'_{l'm}$ are holomorphic functions over the complex
structure moduli space of $\CM_R$. So naively one would conclude that
when the dimension of this moduli space is large enough, meaning that
the number of 3-cycles in $\CM_R$ is large, we can always find
supersymmetric points where the $K'_{l'm}$'s have the desired
values. However the functions $K'_{l'm}$ are not ``generic'' and there
may be relations between them which affect the naive counting. We have
not analyzed this problem in detail but we think the following statement
is true. Any number of the $T^m$'s in the superpotential
\eqref{finalsup} can be tuned by considering a Calabi-Yau whose right
side $\CM_R$ has a sufficiently large number of 3-cycles, and there will
be some supersymmetric vacua with right values of $K'_{l'm}$ to
reproduce the desired $T^m$'s to good accuracy.

This claim is made more intuitive by the following physical
interpretation of equation \eqref{tms}. Start by turning on fluxes
$N^{l'}$ on the cycles of $\CM_R$, which is based on the Riemann surface
$\Sigma_R$. When reduced on the Riemann surface the flux looks like the
electric field produced by a charge in two dimensions. The set of fluxes
$N^{l'}$ resembles a charge distribution on the cycles of the Riemann
surface.  To compute the field produced by these charges in the distant
region of the other set of cycles $\Sigma_L$, one has to consider a
multipole expansion. Since the matrix $K'_{l'm}$ computes the $m$th
multipole expansion of a charge distributed along the $l'$th cycle, the
coefficients $T^m$ are exactly the multipole moments of the charge
distribution. In this formulation our first question reads whether we
can arrange a charged distribution to have the desired multipole moments
given by the coefficients $T^m$. We expect that the answer is positive.

The second question is more subtle. To realize a metastable
nonsupersymmetric vacuum via the OOP mechanism, one has to tune the
superpotential in a way which is determined by properties of the
K\"ahler metric at that point. As we saw in section \ref{sec:OOPex} one
has to tune the coefficients of the effective superpotential only up to
cubic order in an expansion around the candidate metastable point. Since
we have a very large number of parameters $T^m$ at our disposal it seems
that generically we should be able to tune them to generate metastable
vacua at most points on the moduli space.  However we do not have a
proof of this statement and it is possible that various relations
between the period matrices and the K\"ahler metric invalidate the naive
counting\footnote{This question is similar to whether one can realize
the OOP mechanism with a single trace superpotential for the adjoint
scalar in an $SU(N)$ gauge theory. In \cite{Ooguri:2007iu} it was
demonstrated that for $SU(2)$ a metastable vacuum can be generated
anywhere on the moduli space by a single trace superpotential, and for
$SU(N)$ at the center of the moduli space. It was not fully analyzed
whether this is possible in generality.}.

\section{Factorization II: An Example}
\label{sec:ex2}

In the previous section, we argued, based on the factorization of the
Riemann surface and Calabi-Yau, that it is possible to embed the
nonsupersymmetric metastable vacua we found in
\ref{sec:localCYnoncptflux} in a ``larger'' Calabi-Yau, the idea being
that the flux threading compact cycles on one side of the Calabi-Yau
looks like flux coming from infinity from the viewpoint of the other
side of the Calabi-Yau\@.
In this section, we will discuss the Dijkgraaf-Vafa geometries of
subsection \ref{subsec:DVgeo}:
\begin{align} 
\Sigma_{\text{DV}}: \quad y^2 &= P_n(x)^2 - f_{n-1}(x),\qquad
P_n(x) = \prod_{I=1}^n (x- \alpha_I),\label{DVgeo_factn}
\end{align}
as an example where our proposal can in principle be implemented, and
make some steps towards actually confirming our proposal.




\subsection{Factorization Limit in Practice}
\label{subsec:keepSfinite}

As explained in \ref{subsec:DVgeo}, $\alpha_I$ are non-normalizable
parameters which represent the positions of the cuts on the $x$-plane,
while the coefficients in $f_{n-1}(x)$, or equivalently variables $S^I$
defined in \eqref{spclcoordDV}, are normalizable (or at least
log-normalizable) and hence are dynamical variables describing the size
of those cuts.  Therefore, in this Dijkgraaf-Vafa case
\eqref{DVgeo_factn}, $\alpha_I$ are the parameters we want to adjust in
order to approach the factorization limit where $\Sigma_{\text{DV}}$
degenerates into two subsectors.

So, what we should do is clear: we divide the $n$ cuts into two parts as
$n= n_L + n_R$, the ones on the left indexed by $i$ and on the right by
$i'$, and send these two groups apart from each other by a large factor
$L= 1/\epsilon$ so that
\begin{align}
\alpha_i - \alpha_{i'} = \mathcal{O}(L) \qquad \mbox{(when}~
L \to \infty).  
\end{align}
In the $L\to\infty$ limit, the left and right sides will be very far
apart and the factorization we discussed in the previous section must be
achieved.  For example, the period matrix of the total Riemann surface
must diagonalize as in \eqref{periodfactorized} up to $1/L$ correction.

There is one thing we should be careful about when taking the
$L\to\infty$ limit.  If we try to separate the two sets of cuts by
naively taking the typical difference between $\alpha_i$ and
$\alpha_{i'}$ to be of order $L$ while keeping the size of the cuts
fixed, then a simple estimate of the scaling of $S_i^L,S_{i'}^R$ using
\eqref{spclcoordDV} shows that the physical size of the 3-cycles in the
Calabi-Yau blows up.  What we want instead is to end up with two sets of
3-cycles of finite size, separated by a large distance, so that we are
left with nontrivial dynamics of $S_i^L,S_{i'}^R$.  To achieve this we
must also scale the size of the cuts, as we send $L\to\infty$. Let $x_L$
and $x_R$ be local coordinates in the left and right sectors,
respectively, and set
\begin{equation}
  \widetilde x_L = L^{r} x_L,\qquad  \widetilde x_R = L^{r'} x_R, 
   \label{scalecuts}
\end{equation}
where
\begin{equation}
  r = {n_R \over n_L+1} ,\qquad r' = {n_L \over n_R+1}.
   \label{scalecuts2}
\end{equation}
Then, from \eqref{spclcoordDV}, it is not difficult to see that we can
keep $S_i^L,S_{i'}^R$ finite if we keep $\widetilde x_L$, $\widetilde
x_R$ finite while taking the $L\to\infty$ limit.
A similar rescaling of local coordinates must be also necessary when
taking a factorization limit in any other examples than
\eqref{DVgeo_factn}.

\subsection{Computation of Period Matrix}
\label{subsec:comptau}

In the Dijkgraaf-Vafa geometry \eqref{DVgeo_factn}, the period matrix is
given by
\begin{equation}
 \tau_{IJ} = 
  {\partial^2 \mathcal{F}_0 \over \partial S^I \partial S^J},\label{tauIJF0}
\end{equation}
Here, $\CF_0$ is the B-model prepotential, which by the Dijkgraaf-Vafa
relation \cite{Dijkgraaf:2002fc, Dijkgraaf:2002pp} is related to matrix
models.
The precise way to scale various quantities to take the factorization
limit being understood from subsection \ref{subsec:keepSfinite}, it is
in principle possible to confirm our proposal for the Dijkgraaf-Vafa
geometry using \eqref{tauIJF0}.
For doing that, it is important to be able to compute the prepotential
$\CF_0$ for a large number of cuts $n$.  The results from section
\ref{sec:OOPlocalCY} show that generating a metastable vacuum requires
quite a lot of coefficients $T^m$.  Since we roughly need the same
number of cuts on the right as the number of tuned $\Sigma_m$'s on the
left, the total Riemann surface must have quite a large number of cuts.
So, in this subsection we will explain the way to compute $\CF_0$ and
thus $\tau_{IJ}$ for an arbitrary $n$.

For Dijkgraaf-Vafa geometries \eqref{DVgeo_factn} the prepotential
$\CF_0$ may in fact be computed for any number of cuts $n$ in a number
of ways. The most direct way is evaluating the period integrals on the
hyperelliptic curve. This has been done up to cubic order in $S^I$ in
\cite{Itoyama:2002rk}.  Duality with a $U(N)$ matrix model
\cite{Dijkgraaf:2002fc, Dijkgraaf:2002pp}
\begin{equation}
 Z =  \exp \left[\sum_{g=0}^\infty
  g_s^{2g-2} \mathcal{F}_g(S) \right] =\int d^{N^2}\! \Phi \,\exp \left[
  \frac{1}{g_s} \tr W(\Phi) \right], 
\end{equation}
where the matrix model action is given by 
\begin{equation}
  W'(x) = P_n(x) = \prod_{I=1}^n (x-\alpha_I)
\end{equation}
makes this computation quite a bit simpler. Let us quickly show this
argument \cite{Dijkgraaf:2002pp}.

The field $\Phi$ is an $N \times N$ matrix. Say $N^I$ eigenvalues of
$\Phi$ are placed at the critical point $x=\alpha_I$ and divide the
matrix $\Phi$ into $N^I \times N^J$ blocks $\Phi_{IJ}$, where
$\sum_{I=1}^n N^I=N$. One can go to the gauge $\Phi_{IJ}=0$ for $I \neq
J$ by introducing fermionic ghosts in the matrix model action. This
produces the following extra term in the action, where $\Phi_I\equiv
\Phi_{II}$:
\begin{align}
 W_{\text{ghost}}=\sum_{I \neq J}\tr(B_{JI}\Phi_{I}C_{IJ}+C_{JI}\Phi_{I}B_{IJ}).
\end{align}

To write down Feynman diagrams, we expand $\Phi_I$ around
$x=\alpha_I$ as $\Phi_{I}=\alpha_{I}+\phi_{I}.$ A Taylor series of
$W(\Phi_{I}) = W(\alpha_{I} + \phi_{I})$ around $\alpha_{I}$ yields
the propagator and $p$-vertices for $\phi_{I}$. In particular, this
shows that the propagator for $\phi_{I}$ is given by
\begin{align}
 \ev{\phi_{I}\phi_{I}}={1\over W''(\alpha_I)}={1\over \Delta_I},
\end{align}
where $\Delta_I =W''(\alpha_I)=\prod_{J\neq I}^n
\alpha_{IJ}$. Moreover, expanding the ghost action determines the ghost
propagator to be 
\begin{align}
 \ev{B_{JI}C_{IJ}}={1\over \alpha_{IJ}},
\end{align}
and gives the Yukawa interactions between $\phi_{I}$,
$B_{JI}$ and $C_{IJ}$.\\

\begin{figure}[htb]
\begin{quote}
  \begin{center}
  \includegraphics[width=12.5cm]{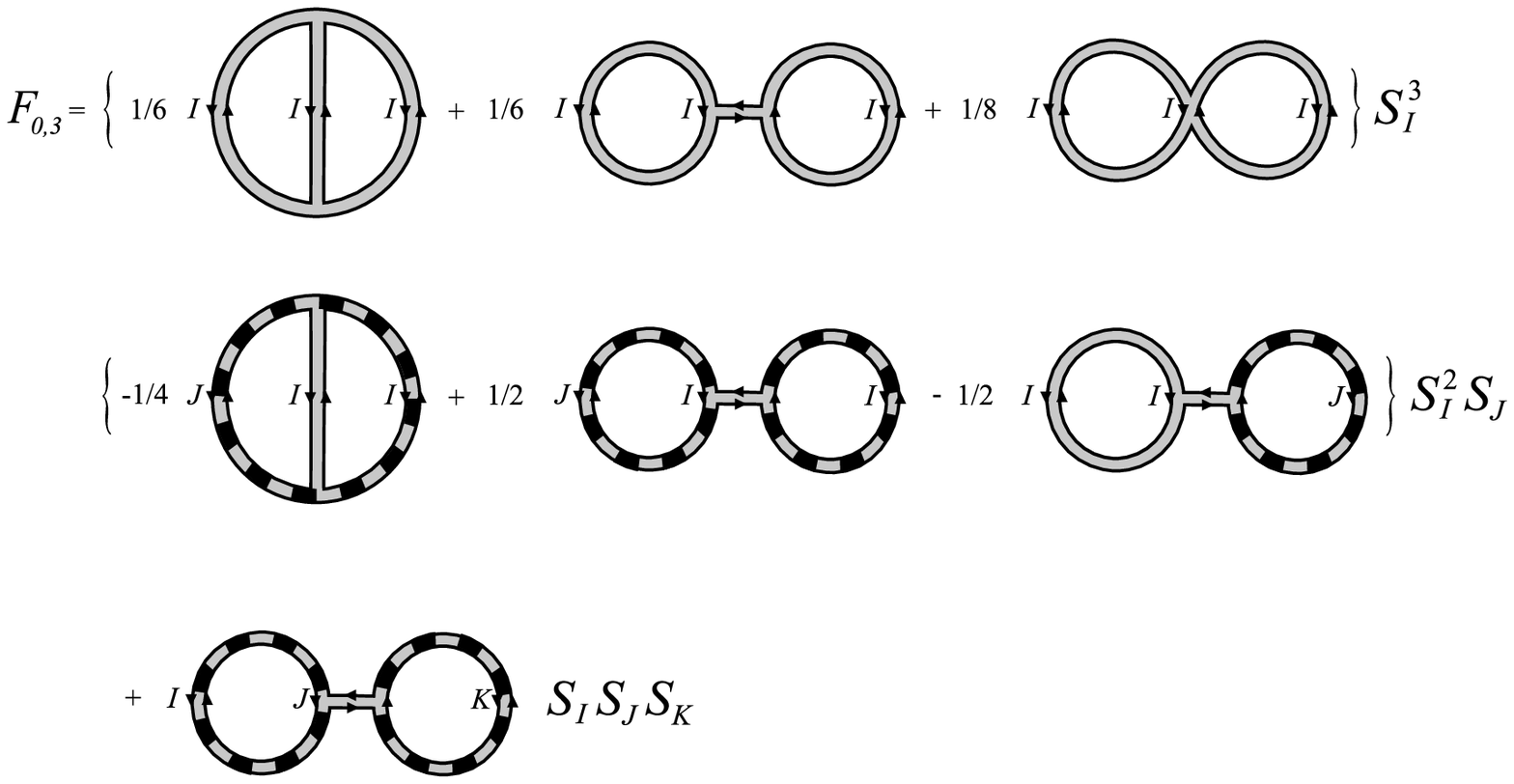} \\
 \caption{The contribution to $\mathcal{F}_{0,3}$ given in terms of
 matrix diagrams.  Gray double lines represent $\phi_I$ fields, while
 black-and-gray double lines represent $BC$ ghosts.  \label{F03}}
 \end{center}
\end{quote}
\end{figure}

The contribution to the prepotential $\mathcal{F}_0$ of order three in
the $S^I$'s is given by planar diagrams with three holes.
 Writing down the expressions $g_{I,3}$ and $g_{I,4}$ in terms of
$\alpha$'s and $\Delta$'s shows that
\begin{equation}
 \CF_{0,3}=\sum_{I=1}^n u_I S_I^3+\sum_{I\neq J}^n u_{I;J}S_I^2S_J
 +\sum_{I< J< K}^n u_{IJK} S_I S_J S_K,
\end{equation}
where
\begin{align*}
 u_I&={2\over 3}\biggl(
 -\sum_{J\neq I}{1\over \alpha_{IJ}^2\Delta_J}
 +{1\over 4\Delta_I}\sum_{J<K\atop J,K\neq i}{1\over \alpha_{IJ}\alpha_{IK}}
 \biggr), \\
 u_{I;J}&=
 -{3\over \alpha_{IJ}^2\Delta_I}+{2\over \alpha_{IJ}^2\Delta_J}
 -{2\over \alpha_{IJ}\Delta_I}\sum_{K\neq I,J}{1\over \alpha_{IK}}
 \quad  \mbox{and}\\
 u_{IJK}&=4\left({1\over \alpha_{IJ}\alpha_{IK}\Delta_I}
 +{1\over \alpha_{JI}\alpha_{JK}\Delta_J}
 +{1\over \alpha_{KI}\alpha_{KJ}\Delta_K}
 \right).
\end{align*}

In appendix \ref{app:prepot}, we discuss the generalization of this
result to higher order in $S^I$.  In particular, we compute $\CF_0$ up
to $S^5$ terms.

\subsection{Scaling of Period Matrix} 

The method explained in subsection \ref{subsec:comptau} allows one in
principle to compute the period matrix to any order in $S^I$ for general
Dijkgraaf-Vafa curves \eqref{DVgeo_factn}.  Then the factorization limit
can be achieved simply by taking the $L \to \infty$ limit of the result
and one can start looking for metastable vacua.
In this subsection, as a step towards it, let us pursue a more modest
goal of seeing the factorized behavior of the period matrix,
\eqref{periodfactorized}.

The form of the scaling can be elegantly derived for any possible
contributing matrix model diagram to $\mathcal{F}_0$. First 
note that $\Delta_{i}$ scales as $L^{2r}$ as $L \to \infty$, and
$\Delta_{i'}$ as $L^{2r'}$. All propagators with indices from
either side of the surface have an expansion in terms of
$\alpha_{IJ}$'s and $\Delta_I$'s, and thus a scaling in $L$ which is
easy to determine. The total scaling of a planar diagram with an
arbitrary number of these elements turns out to depend just on the
number of ghost vertices that connect the left side to the right
side. It is given by
\begin{equation}
\frac{1}{L^{(1+r)N_{ii'}+(1+r')N_{i'i}}},
\end{equation}
where $N_{ii'}$ is the number of ghost vertices with external ghost
lines indexed by $(i,i')$ and the external $\phi$-line by $(i,i)$.  Note
that in deriving this we assumed the scaling \eqref{scalecuts} and thus
$S^i_L,S^{i'}_R$ are of order one.

This shows that a diagram with only indices on the left (or on the
right) will be of order 1 in $L$. Since such diagrams contribute to the
period matrix $\tau_{ij}$ (or $\tau_{i'j'}$), so this shows that the
period matrix is of order 1 in $L$, with corrections in $1/L$ from
diagrams that contain at least two loops indexed by $i$ and $j$. On the
other hand, the off-diagonal pieces of the period matrix $\tau_{ii'}$
and $\tau_{i'i}$ contain at least one ghost cross-vertex with indices
$i$ and $i'$. These parts will therefore scale at least as $1/L$. In
particular, for large $L$ the properties of the full Riemann surface
$\Sigma$ are determined by those of the two factors $\Sigma_L,\Sigma_R$,
and the period matrix $\tau_{IJ}$ indeed diagonalizes as in
\eqref{periodfactorized}.
%

\bigskip

Having checked the diagonalization \eqref{periodfactorized}, the problem
of actually finding an example of a metastable vacuum then just amounts
to solving equation (\ref{tms}) together with (\ref{susyright}) using
the data from matrix model, for $T^m$ giving a metastable vacuum.
Solving these equations is nontrivial, since the relation between the
flux parameters $N^{i'}$ on the right and the coefficients in the
superpotential $T^m$ we want on the left are non-linear, although we
expect that the solutions do exist by the multipole argument we gave in
section \ref{sec:factorization}. We leave matrix model computations up
to requisite orders as well as finding the actual metastable vacua by
solving those equations for the future work.

\section{Conclusion and Generalizations}
\label{sec:conclusions}


Summarizing, we found that turning on flux with support at infinity in
local Calabi-Yau in type IIB induces superpotential for the moduli in the local
Calabi-Yau, thus breaking $\CN=2$ of the Calabi-Yau compactification down to $\CN=2$.
Then we demonstrated that one can create metastable vacua by tuning the
flux at infinity using the OOP mechanism, using a Dijkgraaf-Vafa
(CIV-DV) geometry as a primary example.  The metastable vacua known to
exist \cite{Ooguri:2007iu, Pastras:2007qr} in perturbed Seiberg-Witten
theory can also be understood in terms of metastable flux configuration.


Flux diverging at infinity may appear problematic, but in reality a
local Calabi-Yau must be regarded as a local approximation of a larger compact Calabi-Yau
and the flux at infinity has a natural interpretation there; there is
flux floating around in the rest of the Calabi-Yau, which ``leaks'' into our
local Calabi-Yau and just appear to be coming in from infinity.  This,
furthermore, motivates a more natural setting to realize metastable flux
vacua: in a part, say on the right side, of the full Calabi-Yau $\CM$, there are
some 3-cycles threaded by flux (and possibly O-planes to cancel net
charge if $\CM$ is compact) and on the left side there are some 3-cycles
without flux through them.  If the distance between the left and right
sectors is large, the full Calabi-Yau $\CM$ factorizes into an almost decoupled
system of $\CM_L$ and $\CM_R$, and the flux in $\CM_R$ appears to be flux at
infinity from the viewpoint of $\CM_L$ and induces superpotential in
$\CM_L$.  By adjusting the number of fluxes in $\CM_R$, we can tune the
superpotential and generate metastable vacua in $\CM_L$. This is a very
well controlled setting to analyze flux vacua, which may shed light on
the structure of the nonsupersymmetric landscape of string vacua.
We also made some steps toward actually embedding metastable vacua in a
larger Calabi-Yau as sketched above in the case of Dijkgraaf-Vafa geometry by
computing certain matrix model amplitudes.  Actually finding explicit
vacua along that line is an interesting open problem.

%

Note that we needed just two main ingredients to achieve this
result. The OOP mechanism requires that the complex structure moduli
space is special K\"ahler, and it is important that a superpotential for
flux is very much controllable by tuning the flux, such as the
Gukov-Vafa-Witten superpotential. This means that we can generalize the
above story to any setting which fulfills these two requirements. Other
possibilities therefore include M-theory and F-theory on Calabi-Yau
fourfolds \cite{Becker:1996gj, Gukov:1999ya}. Let us finish by saying a
few words on these two setups.  \\

%
%

Compactifying M-theory on a Calabi-Yau fourfold $\CM_4$ with fluxes yields
a three-dimensional low energy theory with 4 supercharges. The complex
structure moduli of the Calabi-Yau are part of the chiral
supermultiplets and are described by variations of the holomorphic
$(4,0)$-form $\Omega$. In the local limit where the fourfold becomes
noncompact, the K\"ahler potential on the moduli space is given by
\begin{equation} \label{eqn:MthKahlerpot}
  K = \int_{\CM_4} \Omega \wedge \overline{\Omega},
\end{equation}
so that the metric on the moduli space is indeed special
K\"ahler. Moreover, it is well-known that the complex moduli may be
stabilized by turning on 4-form flux $F_4$, which introduces the
superpotential 
\begin{equation}
  W = \int_{\CM_4} F_4 \wedge \Omega.
\end{equation}
The condition for unbroken supersymmetry is $W=dW=0$, so that $F_4$ has
to be a $(2,2)$-form. Stabilizing the K\"ahler moduli as 
well requires that the flux is primitive under the Lefschetz
decomposition (and in particular self-dual). Turning on primitive
$(2,2)$ flux on some compact 4-cycles, we can now follow an equivalent
procedure as in IIB.    
 
M-theory compactified on $\CM_4$ is equivalent to compactifying
F-theory on $\CM_4 \times S^1$, at least if $\CM_4$ is an elliptically
fibered Calabi-Yau. This leads to a four-dimensional space-time with 4
supercharges. So again, the K\"ahler potential is given by
(\ref{eqn:MthKahlerpot}), and the flux $F_4$ is a primitive
$(2,2)$-form. The relation with IIB consistently reduces $F_4$ to a
harmonic $(2,1)$-flux $G_3$. The extra seven-branes that must be inserted
in IIB when reducing over a singular $T^2$ do not contribute to the
superpotential and thus don't play an important role here.   

In particular, consider as an example the local Calabi-Yau fourfold
\begin{equation}
 u^2 + v^2 +  w^2 + F(x,y) =0,
\end{equation}
where all variables are $\mathbb{C}$ (or $\mathbb{C}^*$) valued, and
$F(x,y)$ defines a smooth curve in the $x,y$-plane. Its holomorphic
four-form is given by  
\begin{equation}
\Omega = \frac{du \wedge dv}{w} \wedge dx \wedge dy.
\end{equation}
The $u,v,w$--fiber defines a two-sphere over each point in the
$x,y$-plane, which shrinks to zero-size over the curve
$F(x,y)=0$\footnote{Like in the Calabi-Yau threefold case, the real
  part of $F(x,y)$ changes sign when crossing the Riemann
  surface. This flop changes the parametrization of the compact $S^2$
  in the $T^*S^2$-fiber from a  ``real'' $S^2$ into an ``imaginary'' $S^2$.}.  
Four-cycles can be constructed as an $S^2$ fibration over some disk $D$
ending on the curve and have the topology of a four-sphere (when $x$
and $y \in \mathbb{C}$). Notice that the intersection lattice is
symmetric now and not simply symplectic anymore, so that the bilinear
identity takes a more complicated form. However, like in the threefold
case all relevant quantities reduce to the Riemann surface, and the
analysis is similar as before.

\section*{Acknowledgments}

We would like to thank J.~de~Boer, R.~Dijkgraaf, Y.~Ookouchi, and K.~Saraikin for
valuable discussions.
J.M. would also like to thank the SITP at Stanford University for their
kind hospitality during the completion of this work.
The work of L.H. and M.S. was supported by an NWO Spinoza grant.  The
work of J.M. was supported in part by Department of Energy grant
DE-FG03-92ER40701 and by a John A McCone postdoctoral fellowship.  The
work of K.P. was supported by Foundation of Fundamental Research on
Matter (FOM).

\appendix

\section{Some Basic Results on Riemann Surfaces}\label{app:RiemannS}

In this appendix we summarize some basic properties of Riemann surfaces
\cite{GriffithsHarris}.

A compact Riemann surface $\Sigma_g$ is a one-dimensional compact
complex manifold and its topology is completely characterized by its
genus $g$.  The middle cohomology group has $\dim H^1(\Sigma_g) =
2g$. The intersection form on $H_1(\Sigma_g,\mathbb{Z})$ is
antisymmetric and by Poincar\'e duality unimodular, meaning that we can
pick a basis of 1-cycles $A^i,B_j$ with intersection:
\begin{equation}
  A^i\cap A^j =0,\quad B_i\cap B_j = 0,\quad A^i \cap B_j = \delta^i_{j},\qquad
   i,j=1,\dots,g.
   \label{ABintersect}
\end{equation}
Such a basis is unique up to a symplectic transformation in $Sp(2g,\mathbb{Z})$.

$\Sigma_g$ has a complex structure moduli space ${\cal M}_g$ with
$\dim{\cal M}_g = 3g-3, \, g\geq 2$.

A 1-form $\omega$ on a Riemann surface is called a 
\emph{holomorphic differential} if in a local coordinate patch it has the form:
\begin{equation}
  \omega = f(z) dz,\qquad f(z): \text{~holomorphic}.
   \label{holoomega}
\end{equation}
We will also consider \emph{meromorphic differentials},
for which we allow the function $f(z)$ to have poles at certain 
points on the surface. Now we present a standard basis for holomorphic and
meromorphic differentials on a general Riemann surface:

\paragraph{Holomorphic differentials\protect\footnote{These are also called
meromorphic differentials of the first kind.} $\omega_i$:} Once we pick
a symplectic basis of one-cycles, there is a canonical basis of
holomorphic differentials $\omega_i$, $i=1,..,g$, with the following
periods:
\begin{equation}
  {1\over 2\pi i}\oint_{A^i} \omega_j = \delta^i_j,\qquad 
   {1\over 2\pi i}\oint_{B_i} \omega_j = \tau_{ij}.
   \label{holoperiods_app}
\end{equation}
The (symmetric) matrix $\tau_{ij}$ is the period matrix of the surface,
which depends on the complex structure of $\Sigma_g$.

\paragraph{Meromorphic differentials of the second kind,\protect\footnote{A more common notation in the literature for meromorphic differentials of the second and third kinds is $d\Omega_m^P$ and $d\Omega_0^{P,P'}$.} $\xi_{m\ge 1}^P$:}
These are characterized by a point $P$ on the surface where the
differential has a pole of order $m+1$ with $m\geq 1$. They are
normalized so that in local complex coordinates $z$ where $z(P) = 0$
they have the Laurent expansion:
\begin{equation}
  \xi^P_m \sim  m{dz \over z^{m+1}}  + \text{regular}.
\label{meroexpand}
\end{equation}

\paragraph{Meromorphic differentials of the third kind, $\xi_0^{P,P'}$:}
characterized by two points $P,P'$, where the differential has first 
order poles with opposite residues. Around $P$ we have:
\begin{equation}
  \xi_0^{P,P'} \sim {dz\over z}  + \text{regular}
\label{meroexpandb}
\end{equation}
and similarly around $P'$ with the opposite sign.

Notice that we can always shift a meromorphic differential by
a holomorphic differential without changing the singular part of the 
Laurent expansions \eqref{meroexpand}, \eqref{meroexpandb}. We can eliminate 
this ambiguity by demanding that the $A$ periods of the meromorphic
differentials vanish:
\begin{equation}
  \oint_{A^i} \xi_m^P = 0.
   \label{intAxiP=0}
\end{equation}
In general, it is not possible to simultaneously set the $B$ periods 
to zero. Instead we have:
\begin{equation}
  \oint_{B_i} \xi_m^P = K^P_{im},
   \label{intBxiP=KIM}
\end{equation}
where the matrix $K^P_{im}$ depends on the complex structure moduli of the
Riemann surface and the position of the puncture $P$.


\subsection{Hyperelliptic Case}
\label{app:hyperelliptic}

Let us consider the case where $\Sigma_g$ is hyperelliptic.  For
example, the curve appearing in the Dijkgraaf-Vafa case,
\eqref{DVcurve}, can be written as:
\begin{align}
 y^2=P_n(x)^2-f_{n-1}(x), \qquad P_n(x)=\prod_{i=1}^n(x-\alpha_i).
\end{align}
This curve can be regarded as a two-sheeted $x$-plane with $n$ cuts and
two punctures, the latter corresponding to infinities on the two
$x$-planes.  Let us denote these points by $\infty$ and
$\widetilde\infty$.

A basis of holomorphic differentials $\omega_i$, $i=1,\dots,n-1$ can be
constructed by
\begin{align}
 \omega_i&={Q_i(x)\over y}dx={Q_i(x)\over \sqrt{P_n(x)^2-f_{n-1}(x)}}dx,
\label{hypellomegaI}
\end{align}
where $Q_i(x)$ is a polynomial of degree up to $n-2$ chosen so that
\eqref{holoperiods_app} holds.  Note that this $\omega_i$ goes as $\sim
\CO(x^{-2})dx$ as $x\to\infty,\widetilde\infty$, which means that this
is regular at $x=\infty,\widetilde\infty$.

In the hyperelliptic case, it is convenient to take the meromorphic
differentials of the second kind, $\xi_m$, as
\begin{align}
 \xi_m&={R_m(x)\over y}dx={R_m(x)\over \sqrt{P_n(x)^2-f_{n-1}(x)}}dx,\qquad m\ge 1.
\label{hypellxiM}
\end{align}
Here, $R_m(x)=mx^{m+n-1}+\dots$ is a polynomial and the coefficients of
$x^{m+n-2},\dots,x^{n-1}$ are chosen so that
\begin{align}
 \xi_m&=\pm \left[m x^{m-1}+\CO(x^{-2})\right]dx,\qquad
 x\sim\infty,\widetilde\infty
\label{hypellxiMasym}
\end{align}
is satisfied.  This condition is similar to \eqref{meroexpand}, but this
$\xi_m$ has poles at two points, $x=\infty,\widetilde\infty$, instead of
one.  The coefficients of $x^{n-2},\dots,x^0$ are chosen so that
\eqref{intAxiP=0} is satisfied.

The meromorphic differentials of the third kind, $\xi_0$, can be defined
likewise using a polynomial $R_0(x)=x^{n-1}+\dots$, where the
coefficients are chosen so that
\begin{align}
 \xi_0&={R_0(x)\over y}dx=\pm \left[{1\over x}+\CO(x^{-2})\right]dx,\qquad
 x\sim\infty,\widetilde\infty
\label{hypellxi0}
\end{align}
holds and \eqref{intAxiP=0} is satisfied.

Let us derive a formula that will be useful in the main text.  By
expanding the right hand side of the trivial identity $0=\int_{\Sigma_g}
\omega_i\wedge \xi_m$ by the Riemann bilinear identity, one finds
\begin{align}
 0&=\sum_j \left(\int _{A^j}\omega_i \int_{B_j}\xi_m
 -\int _{A^j}\xi_m \int_{B_j}\omega_i
 \right)+\sum_{p=\infty,\widetilde\infty}\oint_{p}\omega_i\,d^{-1}\xi_m
 \notag\\
 &=K_{im}+ \sum_{p=\infty,\widetilde\infty}\oint_{p}\omega_i\,d^{-1}\xi_m.
\end{align}
Because the behaviors of $\omega_i,\xi_m$ at $x=\infty$ is the same as
those at $x=\widetilde\infty$ up to a sign,
\begin{align}
 K_{im}
 &=-\sum_{p=\infty,\widetilde\infty}\oint_{p}\omega_i\,d^{-1}\xi_m
 =-2\oint_{\infty}\omega_i\,d^{-1}\xi_m
 =-2\oint_{\infty}x^m\omega_i.
\label{KIMidentity}
\end{align}

\section{Parametric Representation of Genus 1 Curves and Sample Computations}
\label{app:param}

\begin{figure}
\begin{center}
\subfigure[Depiction of $\Sigma_{\text{DV}}$ as double cover of the $x$-plane with compact $A$ and $B$ cycles indicated]
{\epsfig{file=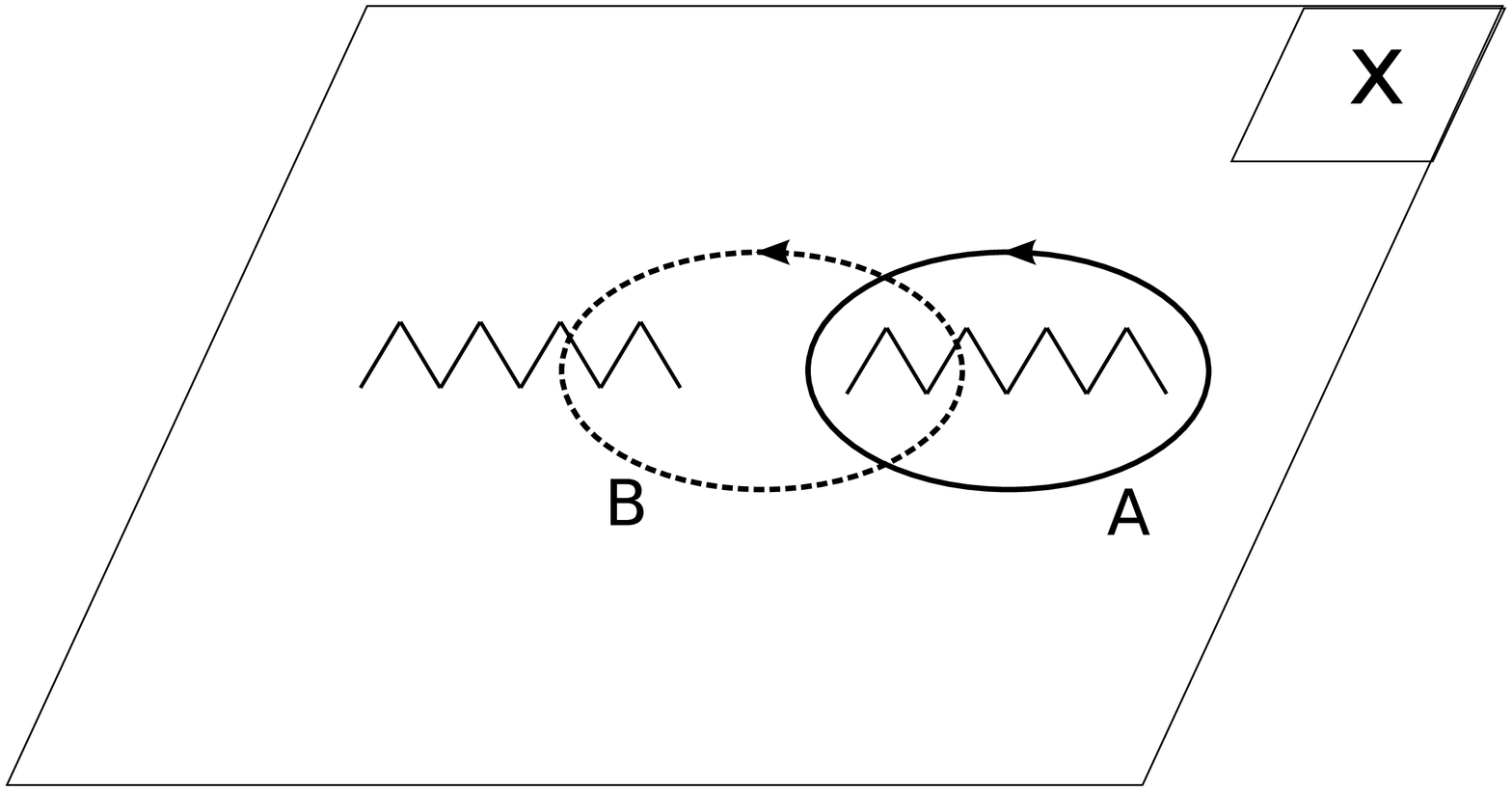,width=0.4\textwidth}\label{DVxplane}}
\hspace{0.1\textwidth}
\subfigure[Depiction of $\Sigma_{\text{DV}}$ as fundamental domain in $z$-plane with the corresponding $A$ and $B$ cycles indicated]
{\epsfig{file=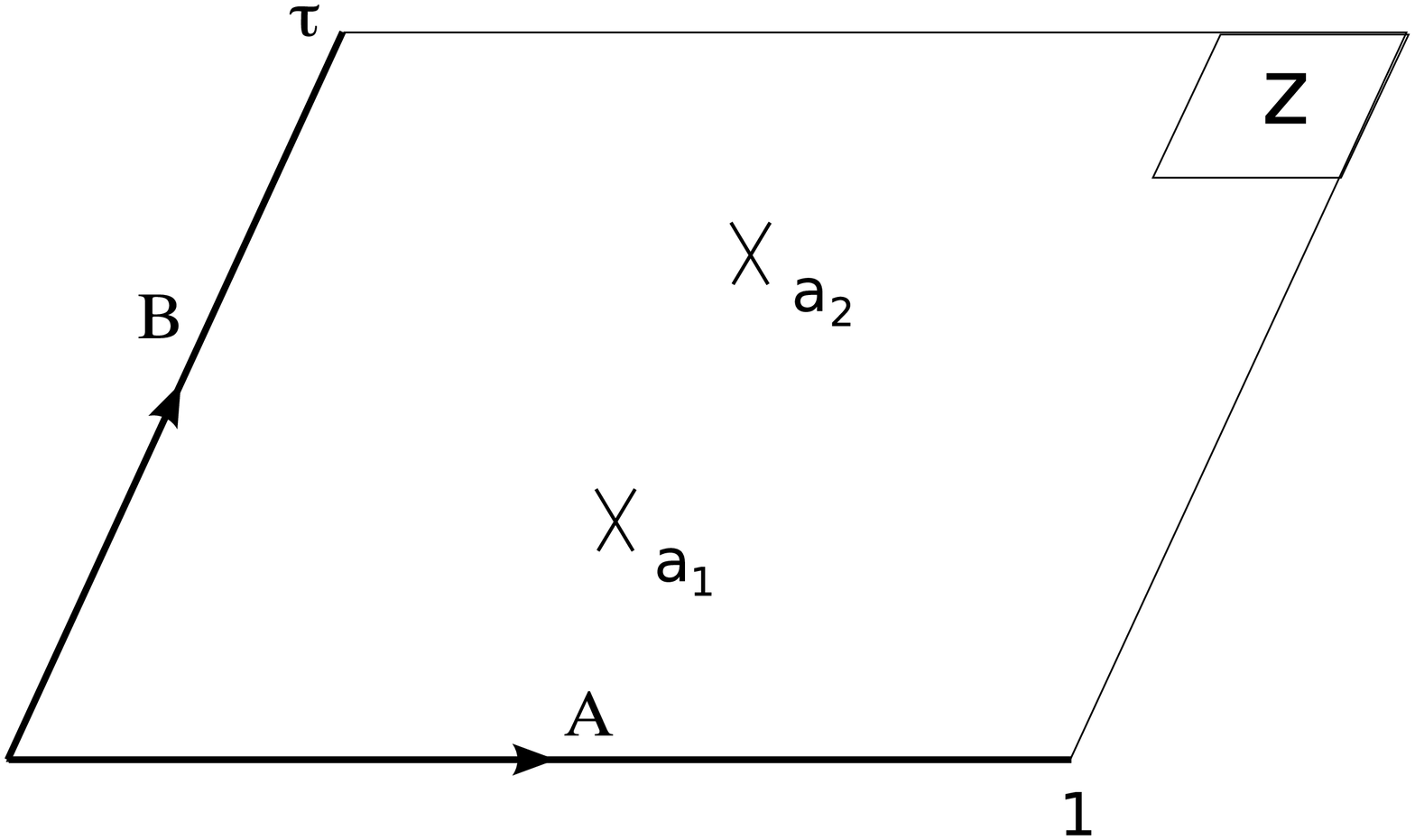,width=0.37\textwidth}\label{DVzplane}}
\end{center}
\label{DVfigs}
\end{figure}

In this appendix, we review the parametric representation of the genus 1 Riemann surface $\Sigma_{\text{DV}}$ of section \ref{sec:OOPex} defined by
\begin{equation}0=F_{\text{DV}}(x,y)=y^2-\left[\left(x^2-\frac{\Delta^2}{4}\right)^2-b_0\right]\label{FDVapp}\end{equation}
and its application to obtaining some of the results used therein.  In particular, we think of $\Sigma_{\text{DV}}$ as a copy of the standard fundamental domain with two marked points, $a_1$ and $a_2$, corresponding to the points at infinity on the two sheets.  In figures \ref{DVxplane} and \ref{DVzplane}, we depict both the standard visualization of $\Sigma_{\text{DV}}$ as a double-sheeted cover of the $x$-plane as well as the parametric one, identifying the standard $A$ and $B$ cycles in the former and their realization in the latter.

\subsection{Building Blocks}

The embedding of $\Sigma_{\text{DV}}$ into $xy$ space is obtained by specifying functions $x(z)$ and $y(z)$ which satisfy \eqref{FDVapp}.  The basic building blocks that we use to construct $x(z)$ and $y(z)$ are Janik's functions $F_i(z)$ \cite{Janik:2003hk}
\begin{equation}F_i(z)\equiv \ln\theta(z-a_i-\tilde{\tau})\qquad\tilde{\tau}=\frac{\tau+1}{2}\end{equation}
and their derivatives
\begin{equation}F_i^{(n)}(z)\equiv \left(\frac{\partial}{\partial z}\right)^nF_i(z)\end{equation}
A detailed description of these functions, their properties, and several sample computations can be found in Appendix C of \cite{Marsano:2007fe}.  For now, we simply note a few elementary facts.  First, we point out that $F_i(z)$ introduces a branch point at $a_i$ while $F_i^{(n)}$ introduces a pole of order $n$.  For $n\ge 2$ these functions are elliptic while for $n=0,1$ they have the following monodromies
\begin{equation}\begin{split}F_i(z+1)&=F_i(z)\\
F_i(z+\tau)&=F_i(z)+i\pi-2\pi i(z-a_i)\\
F_i^{(1)}(z+1)&=F_i^{(1)}(z)\\
F_i^{(1)}(z+\tau)&=F_i^{(1)}(z)-2\pi i\end{split}\label{monods}\end{equation}
It is also useful to record the relation between these functions and the Weierstrass $\sigma$, $\zeta$, and $\wp$ functions
\begin{equation}\begin{split}
F(z)&=\ln\sigma(z)-\eta_1z^2+i\pi z + \ln\theta'(\tilde{\tau})\\
F^{(1)}(z)&=\zeta(z)+i\pi -2\eta_1z\\
F^{(2)}(z)&=-\wp(z)-2\eta_1\\
F^{(n)}(z)&=-\left(\frac{\partial}{\partial z}\right)^{n-2}\wp(z)\qquad\text{n>2}
\end{split}\end{equation}
where
\begin{equation}\eta_1=\zeta\left(\frac{1}{2}\right)\end{equation}
Finally, we also recall the differential equation satisfied by $\wp(z)$
\begin{equation}\left(\frac{\partial\wp(z)}{\partial z}\right)^2=4\wp(z)^3-g_2\wp(z)-g_3\end{equation}
which can also be taken as an implicit definition of the Weierstrass elliptic invariants $g_2$ and $g_3$.

\subsection{The Embedding Functions $x(z)$ and $y(z)$}

Using the building block functions $F_i^{(n)}(z)$, it is fairly easy to write down embedding functions $x(z)$ and $y(z)$ satisfying \eqref{FDVapp}.  Because $x(z)$ should be locally one-to-one near the marked points, we must construct it from functions with single poles, namely $F_1^{(1)}$ and $F_2^{(1)}$.  On the other hand, $y(z)\sim x(z)^2$ near the marked points so it must contain functions with double poles, $F_1^{(2)}$ and $F_2^{(2)}$.  This leads us to write{\footnote{The constant term that we add to $x(z)$ is added for later convenience.}}
\begin{equation}\begin{split}
x(z)& = X\left(F_1^{(1)}-F_2^{(1)}-\left[F^{(1)}(a)-i\pi\right]\right)\\
y(z)&= X^2\left(F_1^{(2)}-F_2^{(2)}\right)
\end{split}\end{equation}
where
\begin{equation}a\equiv a_2-a_2\end{equation}
Because elliptic functions such as $x(z)$ and $y(z)$ are completely determined by their pole structure, it is in fact quite easy to verify that
\begin{equation}y(z)^2=\left[\left(x(z)^2-\frac{\Delta^2}{4}\right)^2-b_1x(z)-b_0\right]\end{equation}
where
\begin{equation}\begin{split}\Delta^2&=12X^2\wp(a)\\
b_1&=-4X^3\wp'(a)\\
b_0&=X^4\left[12\wp(a)^2-g_2\right]
\label{Db1b0}\end{split}\end{equation}
In order to obtain $b_1=0$ we set
\begin{equation}a=\frac{\tau}{2}\end{equation}
We can solve for $X$ in \eqref{Db1b0}
\begin{equation}X^2=\frac{12\wp(\tau/2)b_0}{\Delta^3\left[12\wp(\tau/2)^2-g_2\right]}\end{equation}
and use this to eliminate $X$, thereby obtaining a direct relationship between $\Delta$, $b_0$, and the complex modulus $\tau$
\begin{equation}b_0=\frac{\Delta^4\left[12\wp(\tau/2)^2-g_2\right]}{144\wp(\tau/2)^2}\label{b0rel}\end{equation}

\subsection{Two Sample Computations}

We now describe two sample computations which illustrate the power of this approach.  First, we will reproduce a result that is more easily obtained using the explicit representation \eqref{FDVapp}.  Next, we will consider a computation for which the parametric approach is simpler.

As our first example, let us consider the quantity
\begin{equation}\Sigma=\frac{1}{4\pi i}\oint_{x=\infty}\,x^2\,dy\end{equation}
As we saw in section \ref{sec:OOPex}, this can be done quite easily using the explicit representation \eqref{FDVapp} with the result
\begin{equation}\Sigma=\frac{b_0}{2}\label{sigresapp}\end{equation}
We can also write this directly in terms of $\tau$ using \eqref{b0rel}
\begin{equation}\Sigma=\frac{\Delta^4\left[12\wp(\tau/2)^-g_2\right]}{288\wp(\tau/2)^2}\end{equation}
Let us now see how the result \eqref{sigresapp} can be obtained using the parametric representation.  For this, we write
\begin{equation}\Sigma=\frac{1}{4\pi i}\oint_{a_1}\,x(z)^2\frac{\partial y(z)}{\partial z}\,dz=\frac{1}{4\pi i}\oint_{a_2}\,X^4\left(F_1^{(1)}-F_2^{(1)}-\left[F^{(1)}(a)-i\pi\right]\right)^2\left(F_1^{(3)}-F_2^{(3)}\right)\,dz\end{equation}
and expand the integrand near $a_1$.  This is straightforward and leads to
\begin{equation}x(z)^2\frac{\partial y(z)}{\partial z}\sim \frac{b_0}{12\wp(\tau/2)^2-g_2}\left(\frac{2}{(z-a_1)^5}+\frac{4\wp(\tau/2)}{(z-a_1)^3}+\frac{12\wp(\tau/2)^2-g_2}{z-a_1}+{\cal{O}}([z-a_1]^0)\right)\end{equation}
where we have used \eqref{Db1b0}.  The residue appearing in $\Sigma$ is now easily read off with the desired result
\begin{equation}\Sigma=\frac{b_0}{2}\end{equation}

Next, let us turn our attention to the computation of
\begin{equation}S\equiv \frac{1}{2\pi i}\oint_{A^1}\,y\,dx\end{equation}
In the parametric formalism, we write this as
\begin{equation}S\equiv \frac{1}{2\pi i}\int_{A^1}\,y(z)\,\frac{\partial x(z)}{\partial z}\,dz=\frac{1}{2\pi i}\int_{A^1}\,X^3\left(F_1^{(2)}-F_2^{(2)}\right)^2\end{equation}
To evaluate this, we will write the integrand as a sum of quasi-elliptic functions and use their known monodromies \eqref{monods}.  Given that the integrand has poles of degree at most 4 with even (odd) poles at $a_1$ and $a_2$ entering with identical (opposite) signs, the general form of this expansion is relatively simple
\begin{equation}\left(F_1^{(2)}-F_2^{(2)}\right)^2=a\left(F_1^{(4)}+F_2^{(4)}\right)+b\left(F_1^{(3)}-F_2^{(3)}\right)+c\left(F_1^{(2)}+F_2^{(2)}\right)+d\left(F_1^{(1)}-F_2^{(1)}+i\pi\right)+e\label{genexpns}\end{equation}
In terms of these expansion coefficients, the monodromies \eqref{monods} lead to the simple result
\begin{equation}S=\frac{X^3}{2\pi i}\left(i\pi d + e\right)\end{equation}
In practice, the coefficients $a,\ldots,e$ can be by comparing pole structures on the two sides of \eqref{genexpns} with the following result when $a=\tau/2$
\begin{equation}a=-\frac{1}{6},\quad b=0,\quad c=2\wp(\tau/2),\quad d=0,\quad
e= \frac{2g_2}{3}+8\eta_1\wp(\tau/2)-4\wp(\tau/2)^2\end{equation}
This means that $S$ is actually given by the relatively simple expression
\begin{equation}S = \frac{\Delta^3}{2\pi i\left[12\wp(\tau/2)\right]^{3/2}}\left(\frac{2g_2}{3}+8\eta_1\wp(\tau/2)-4\wp(\tau/2)^2\right)\end{equation}

\subsection{Some Useful Identities}

Finally, we close this Appendix by listing a few derivative identities that were useful in section \ref{sec:OOPex}.  First, some derivative identities
\begin{equation}\begin{split}\frac{\partial\zeta(z)}{\partial\tau}&=-\frac{1}{2\pi i}\left[\frac{1}{2}\wp'(z)+\zeta(z)\wp(z)-\frac{g_2z}{12}+2\eta_1\left(\zeta(z)-z\wp(z)\right)\right]\\
\frac{\partial\wp(z)}{\partial\tau}&=\frac{1}{2\pi i}\left[2\wp(z)^2+\zeta(z)\wp'(z)-\frac{g_2}{3}-2\eta_1\left(z\wp'(z)+2\wp(z)\right)\right]\\
\frac{\partial\eta_1}{\partial\tau}&=-\frac{1}{2\pi i}\left(2\eta_1^2-\frac{g_2}{24}\right)\\
\frac{\partial g_2}{\partial\tau}&=\frac{1}{2\pi i}\left(6g_3-8g_2\eta_1\right)
\end{split}\end{equation}
Several of these can be combined in order to derive the additional useful result
\begin{equation}\frac{\partial}{\partial\tau}\left[\wp(\tau/2)\right]=\frac{1}{2\pi i}\left[2\wp(\tau/2)^2-\frac{g_2}{3}-4\eta_1\wp(\tau/2)\right]\end{equation}

We also remind the reader that the partial differential equation
\begin{equation}\left(\frac{\partial \wp(z)}{\partial z}\right)^2=4\wp(z)^3-g_2\wp(z)-g_3\end{equation}
combined with the fact that
\begin{equation}\wp'(\tau/2)=0\end{equation}
implies that the elliptic invariant $g_3$ can be written in terms of $g_2$ and $\wp(\tau/2)$ as
\begin{equation}g_3=4\wp(\tau/2)^3-g_2\wp(\tau/2)\end{equation}

\section{Independence of $\Sigma_m$'s}\label{app:indepSigmap}

In this appendix, we consider the Dijkgraaf-Vafa geometry and examine
the dependence of $\Sigma_m$'s on the moduli $S=\{S^i\}$, or
equivalently, on the coefficients $b=\{b_i\}$ of the polynomial
$f_{n-1}(x)$ as defined in \eqref{f(x)_and_bi}.  To apply the OOP
mechanism and generate a metastable vacuum at a point $b_i=b_{i}^{(0)}$,
it is needed that, when we expand $\Sigma_m(b)$'s around a point in
$\Delta b_i\equiv b_i-b_i^{(0)}$, the coefficients of $\Delta b_i$,
$\Delta b_i\,\Delta b_j$, $\Delta b_i\,\Delta b_j \Delta b_k$ terms are
all independent and by taking linear combinations of $\Sigma_m(b)$'s we
can obtain the OOP superpotential \eqref{WeffKNC}.

For simplicity, let us first discuss the case where we treat $b_{n-1}$,
which is log-normalizable, as a dynamical modulus.  In this case the
number of moduli is $n$ and the number of coefficients we would like to
tune is, from \eqref{nopOOP},
\begin{align}
 C_n&={n(n+1)(n+5)\over 6}.
 \label{Cn_app}
\end{align}
Explicitly, $\Sigma_m(b)$ is given by
\begin{align}
 \Sigma_m(b)
 &= {1\over 2\pi i m}\oint_\infty x^m dy
 = {1\over m}\Res_{x=\infty}
 \left[
   x^m{2P_n(x)P_n'(x)-f_{n-1}'(x)\over 2\sqrt{P_n(x)^2-f_{n-1}(x)}}
 \right]\notag\\
 &= {1\over M}\Res_{x=\infty}
 \biggl[
   x^m
  \biggl(P_n'(x)-{f_{n-1}'(x)\over 2P_n(x)}\biggr)\notag\\
 &\qquad\qquad\qquad
 \times
  \biggl(
   1+{1\over 2}{f_{n-1}(x)\over P_n(x)^2}
   +{3\over 8}\biggl({f_{n-1}(x)\over P_n(x)^2}\biggr)^2
   +{5\over 16}\biggl({f_{n-1}(x)\over P_n(x)^2}\biggr)^3
   +\cdots
  \biggr)
 \biggr].
\label{Sigma(b)expn}
\end{align}
So, $\Sigma_m(b)$ are polynomials in $b_i$'s.  If they are generic
polynomials in $b$ with high enough degree, then the expansion of
$\Sigma_m(b)$ around a generic point $b^{(0)}$ in $\Delta b$ will have
different coefficients of $\Delta b$, $(\Delta b)^2$, $(\Delta b)^3$
terms, for different values of $m$.  If this were the case, then the
minimum number of $\Sigma_m$'s we need to consider would be $C_n$ in
\eqref{Cn_app}.

However, for small $m$, $\Sigma_m(p)$ is not a generic polynomial in
$b_i$ and we need to be careful.  From \eqref{Sigma(b)expn}, one can
read off the following pattern of dependence of $\Sigma_m$'s on $b_i$'s:
\begin{itemize}
 \item $\Sigma_{-n},\dots,\Sigma_{0}$ do not depend on $b_k$'s, because
       the only contributions come from $P_n'$.
 \item A term with just one $b_i$ ($i=0,\dots n-1$) appears in
       $P_n'(f_{n-1}'/P_n^2)$ and $f_{n-1}'/P_n$.  Such a term has
       degree $i-n-1$ in both cases and hence $b_i$ first shows up in
       $\Sigma_{n-i}$.  $b_{n-1}$ appears in $\Sigma_1$ and $b_0$
       appears in $\Sigma_n$.
 \item The combination $b_i b_j$ ($i,j=0,\dots n-1$) appears in
       $P_n'(f_{n-1}'/P_n^2)^2$ and $(f_{n-1}'/P_n)(f_{n-1}'/P_n^2)$.
       These terms have degree $i+j-3n-1$ and hence $b_i b_j$ first
       shows up in $\Sigma_{3n-i-j}$.  $b_{n-1}^2$ appears in
       $\Sigma_{n+2}$ and $b_0^2$ appears in $\Sigma_{3n}$.
 \item The combination $b_i b_j b_k$ ($i,j,k=0,\dots n-1$) appears in
       $P_n'(f_{n-1}'/P_n^2)^3$ and $(f_{n-1}'/P_n)(f_{n-1}'/P_n^2)^2$.
       These terms have degree $i+j+k-5n-1$ and hence $b_i b_j b_k$
       first shows up in $\Sigma_{5n-i-j-k}$.  $b_0^3$ appears in
       $\Sigma_{2n+3}$ and $b_0^3$ appears in $\Sigma_{5n}$.
\end{itemize}
From these, we can see that we have to satisfy some requirements.  Let
us call $m$ of $\Sigma_m$ ``order.''
\begin{itemize}
 \item We need all combinations of $\Delta b_i\,\Delta b_j\,\Delta b_k$, but $b_0^3$, which
       contains $(\Delta b_0)^3$, does not appear until order $A_n=5n$.
 \item The number of possible cubic terms, $\Delta b_i\,\Delta b_j\,\Delta b_k$, is
       $n(n+1)(n+2)/6$.  Cubic terms start to appear at order $2n+3$ and
       therefore, for all possible cubic terms to have chance of all
       showing up in a linear independent way, we need to wait until
       order $(2n+3)+n(n+1)(n+2)/6-1\equiv B_n$.
 \item The number of possible quadratic terms and cubic terms is
       $n(n+1)/2+n(n+1)(n+2)/6$.  Quadratic terms start to appear at order $n+2$ and
       cubic terms appear at higher order.  Therefore, for all possible
       quadratic and cubic terms to have chance of all showing up in a
       linear independent way, we need to wait until order
       $(n+2)+n(n+1)/2+n(n+1)(n+2)/6-1\equiv \widetilde{B}_n$.
 \item From \eqref{Cn_app}, we need $C_n$ independent
       coefficients.  So, we need to wait until at least order $C_n$.
\end{itemize}
By looking at which of $A_n,B_n,\widetilde{B}_n,C_n$ is largest for
given $n$, we find that we need $\Sigma_m$'s at least up to $m_{\text{min}}$,
where
\begin{equation}
 \begin{split}
 n=1    &~~\rightarrow~~ m_{\text{min}}=5,\\
 n\ge 2 &~~\rightarrow~~ m_{\text{min}}=B_n={(n+1)(n+2)(n+3)\over 6}.
\end{split}
\label{MminU(N)}
\end{equation}

By a similar analysis, if $b_{n-1}$ is regarded as a nondynamical
parameter, we find the following:
\begin{equation}
 \begin{split}
 n=2    &~~\rightarrow~~ m_{\text{min}}=10,\\
 n=3    &~~\rightarrow~~ m_{\text{min}}=15,\\
 n\ge 4 &~~\rightarrow~~ m_{\text{min}}={n^3\over 6}+{n^2\over 2}+{n\over 3}+3.
\end{split}
\label{MminSU(N)}
\end{equation}

\section{Prepotential for Dijkgraaf-Vafa (CIV-DV) Geometries}\label{app:prepot}

In this appendix, we first review different approaches to computing the
prepotential $\CF_0$ for the Dijkgraaf-Vafa (CIV-DV) geometries
\cite{Cachazo:2001jy, Cachazo:2002pr, Dijkgraaf:2002fc,
Dijkgraaf:2002vw, Dijkgraaf:2002dh} given in eqs.\ \eqref{DVCY},
\eqref{DVcurve}:
\begin{gather}
  uv - F_{\text{DV}}(x,y)=0,   \label{DVCYapp}\\
  F_{\text{DV}}(x,y)\equiv w^2-\left[P_n(x)^2- f_{n-1}(x)\right],     \label{DVcurveapp}\\
 P_n(x)=W'(x)=g_{n+1}\prod_{i=1}^n(x-\alpha_i),
\end{gather}
for arbitrary number of cuts $n$.  Moreover, we will present $\CF_0$ for
general $n$ up to $S^5$ terms.  In the present paper, the prepotential
is used in section \ref{sec:ex2} to evaluate the period matrix of the
underlying hyperelliptic Riemann surface.  However, the content of this
appendix is almost independent of the main text and can be read
separately.

The prepotential is physically important, because by putting fluxes in
the Dijkgraaf-Vafa geometry one can realize supersymmetric $\CN=1$
$U(N)$ gauge theory, and its glueball superpotential which governs low
energy dynamics can be computed from the prepotential
\cite{Cachazo:2001jy, Cachazo:2002pr}.
Furthermore, by the Dijkgraaf-Vafa relation \cite{Dijkgraaf:2002fc,
Dijkgraaf:2002vw, Dijkgraaf:2002dh}, the prepotential is related to
unitary matrix models.  The relation to matrix models was studied also
using supergraphs \cite{Dijkgraaf:2002xd} and Konishi anomaly
\cite{Cachazo:2002ry}.  The same prepotential also underlies the physics
of metastable brane-antibrane systems studied recently
\cite{Aganagic:2006ex, Heckman:2007wk, Marsano:2007fe}.

The first computation of the prepotential was performed in
\cite{Cachazo:2001jy} for $n=2$ (two cuts) up to $S^5$ terms by directly
evaluating period integrals, where $S$ is the glueball.  For small values
of $n$, the computation of $\CF_0$ up to several orders in $S$ is
relatively easy, but computations for general number of cuts $n$ require
more systematic approaches.
One such approach is to evaluate period integrals systematically; ref.\
\cite{Itoyama:2002rk} established a methodology, computing $\CF_0$ for
general $n$ up to $S^3$ terms.
Another approach is to use the relation to matrix models.  One can
evaluate the matrix integrals directly \cite{Naculich:2002hi} or by a
more sophisticated diagrammatic technique \cite{Dijkgraaf:2002pp}.  This
matrix model approach turns out to be rather efficient in actual
computations and indeed, in section \ref{app:F0_mm}, we will compute
$\CF_0$ up to $S^5$ terms.
Yet another approach is to use the relation to the Whitham hierarchy
\cite{Itoyama:2003qt}.  For other work on computations of $\CF_0$ for
general $n$, see \cite{Naculich:2002hr, GomezReino:2004dr,
Aoyama:2005jp}.

We believe that the result of this appendix has various practical
applications, including search for nonsupersymmetric vacua in $\CN=1$
gauge theories.


\subsection{Matrix Model}\label{app:F0_mm}

By the Dijkgraaf-Vafa relation \cite{Dijkgraaf:2002fc, Dijkgraaf:2002vw,
Dijkgraaf:2002dh}, the prepotential $\CF_0(S)$, $S=(S^1,\dots,S^n)$, of
the geometry \eqref{DVCYapp} is related to the free energy of the
associated $U(N)$ matrix model,
\begin{align}
 Z
 =e^{-F_{mm}(g_s,N)}
 =\int d^{N^2}\!\Phi\,\exp\left[ -{1\over g_s}\tr W(\Phi) \right],
 \label{MMint_app}
\end{align}
where $\Phi$ is an $N\times N$ matrix.\footnote{Here, the argument
``$N$'' in $F_{mm}(g_s,N)$ denotes $(N^1,\dots,N^n)$ collectively, not
to be confused with the rank $N=\sum_i N^i$ of the matrix $\Phi$.}  This matrix
integral is performed around the vacuum where $N^i$ eigenvalues of
$\Phi$ sit at $\alpha_i$.  If we replace $N^i$ in $F_{mm}(g_s,N)$ by
$S^i$ by the relation
\begin{align}
 g_s N^i&=S^i,
\end{align}
then the free energy organize itself into a genus ('t Hooft)
expansion. Namely,
\begin{align}
 F_{mm}\!\left(g_s,{S\over g_s}\right)=\sum_{g=0}^\infty g_s^{2g-2} \CF_g(S).
\end{align}

As reviewed in section \ref{sec:ex2}, one can evaluate the matrix
integral \eqref{MMint_app} by perturbation theory using diagrams
\cite{Dijkgraaf:2002pp}, as far as the perturbative part of $F_{mm}$ is
concerned.  However, this quickly gets out of hand, particularly because
for general $n$ one can have $p$-point interaction vertices with
arbitrarily large $p$, which makes the number of diagrams explode.
Namely, if we expand $\Phi$ around the critical point $x=\alpha_i$, each
coefficient $g_{i,p}$ in the expansion
\begin{align}
 W(\alpha_i+x)
 =
 W(\alpha_i)
 +{m_i^2\over 2}x^2
 +\sum_{p=3}^{n+1}{g_{i,p}\over p}x^p
 \label{gIp}
\end{align}
gives a $p$-vertex interaction,  and $p$ can be arbitrarily large
for general $n$.  Here,
\begin{align}
 m_i^2=W''(\alpha_i),\qquad
 g_{i,p}={1\over (p-1)!}W^{(p)}(\alpha_i)\label{defmIgIp}
\end{align}
and $W^{(p)}$ is the $p$th derivative.

A more efficient method amenable to computer was proposed in
\cite{Kraus:2003jf, Kraus:2003jv, Intriligator:2003xs}, and here we
generalize it to the case with an arbitrary number of cuts $n$.  First
note that the perturbative part of the matrix model free energy $F_{mm}$
can be written as an expansion in the coupling constant $g_s$ as:
\begin{align}
 F_{mm,pert}(g_s,N)&=\sum_{k=1}^\infty g_s^k f_k(N).
\end{align}
Here, the order $k$ amplitude $f_k(N)$ is a polynomial of degree $k+2$
in $N^i$'s, which in turn has a genus expansion as follows:
\begin{align}
 f_k(N)
 &=\sum_{g=0}^{\left[{k+1\over 2}\right]}
    A_{i_1  \dots i_{k-2g+2}}^{(k,g)}
    N^{i_1} \cdots N^{i_{k-2g+2}},
\end{align}
where the coefficients $A_{i_1 i_2 \dots}^{(k,g)}$ are totally symmetric
in $i_1,i_2,\dots$, and $[x]$ is the integer part of $x$.  For a given
finite $k$, the number of coefficients $A_{i_1 i_2 \dots}^{(k,g)}$ in
$f_k(N)$ is of course finite. Therefore, if we compute $f_k(N)$ for some
small values of $\{N^i\}$ by computer, we can determine the coefficients
$A_{i_1 i_2 \dots}^{(k,g)}$.  Furthermore, there is symmetry under
exchange of eigenvalues; for example, if we know $A_{1123}^{(k,g)}$, we
can obtain $A_{2214}^{(k,g)}$ by the manipulation:
\begin{align*}
 (\alpha_1,m_1,g_{1,p})\leftrightarrow (\alpha_2,m_2,g_{2,p}),\qquad
 (\alpha_3,m_3,g_{3,p})\to (\alpha_4,m_4,g_{4,p}).
\end{align*}
This symmetry significantly reduces the number of ``data points''
$\{N^i\}$, for which we should evaluate the matrix integral in order to
determine $f_k(N)$. In particular, this means that, if one knows
$f_k(N)$ for $n=k+2$ cuts, then one can determine $f_k(N)$ for arbitrary
number of cuts $n$ by symmetry.

For actually evaluating matrix integrals, it is convenient to go to the
eigenvalue basis \cite{Brezin:1977sv}:
\begin{align}
 e^{-F_{mm}(g,N)}
=\int d^N \lambda\left[\prod_{a<b}^N(\lambda_a-\lambda_b)^2\right]
 \exp\left[-{1\over g_s}\sum_{a=1}^N W(\lambda_a)\right],
\label{MMint_eign_app}
\end{align}
where the Van der Monde determinant is from the change of variables
\cite{Brezin:1977sv}.  We would like to compute this perturbatively
around the vacuum where $N^i$ of the eigenvalues $\lambda_a$'s are equal
to $\alpha_i$, where $i=1,\dots,n$.  So, let us divide $\lambda_a$'s
into $n$ groups and expand around $\alpha_i$ as:
\begin{align}
 \lambda_{ia}=\alpha_i+\mu_{ia}.
 \qquad i=1,\dots,n, \qquad a=1,\dots,N^i.
\end{align}
Then the matrix integral \eqref{MMint_eign_app} is, up to a
multiplicative constant,
\begin{equation}
\begin{split}
  &\int d^N \mu
 \left[\prod_{i=1}^n \prod_{a< b}^{N^i}(\mu_{ia}-\mu_{ib})^2\right]
 \left[\prod_{i<j}^n \prod_{a=1}^{N^i} \prod_{b=1}^{N^j} (\mu_{ia}-\mu_{jb}+\alpha_{ij})^2\right]\\
 &\qquad\qquad\qquad\qquad\qquad\qquad
 \times
 \exp\left[-{1\over g_s}
 \sum_{i=1}^n\sum_{a=1}^{N^i} \left({m_i^2\over 2}\mu_{ia}^2+\sum_{p=3}^n {g_{i,p}\over p}\mu_{ia}^p\right)\right],
\end{split}
\end{equation}
where we used the expansion \eqref{gIp} and
$\alpha_{ij}\equiv\alpha_i-\alpha_j$.  Given $\{N^i\}$, we can evaluate
this using computer by power expansion in $g_s$ which, following the
procedure sketched above, allows us to determine $f_k(N)$ order by
order.

\subsection{Result}
\label{app:mmresult}

By setting $\Delta_i\equiv m_i^2$, the first order result ($\CO(N^3)$) is
\begin{align}
 f_1(N)
 =&\sum_i \left({g_{i,4}\over 2\Delta_i^2} -{2g_{i,3}^2\over 3\Delta_i^3}\right)(N^i)^3
 + \sum_{i\neq j}
 \left({2g_{i,3}\over \Delta_i^2\alpha_{ij}}+{1\over \Delta_i\alpha_{ij}^2}-{2\over \Delta_j\alpha_{ij}^2}\right)(N^i)^2 N^j
 \notag\\
 &\qquad\qquad\qquad
 +4\sum_{i<j<k} \left( 
    {1\over \Delta_i^{} \alpha_{ij}^{}\alpha_{ki}^{}}
   +{1\over \Delta_j^{} \alpha_{jk}^{}\alpha_{ij}^{}} 
   +{1\over \Delta_k^{} \alpha_{ki}^{}\alpha_{jk}^{}} 
 \right)N^i N^j N^k
 \notag\\
 &+\sum_i \left( {g_{i,4}\over 4\Delta_i^2}- {g_{i,3}^2\over 6\Delta_i^3}\right)N^i,
 \label{f1_gen}
\end{align}
The coupling constants $g_{i,p}$ can be expressed in terms of $\alpha^i$
using \eqref{defmIgIp}.  The terms cubic in $N^i$ are planar amplitude,
while the ones linear in $N^i$ are genus 1 (torus) amplitude.  This
agrees with the known result \cite{Itoyama:2002rk}, if we use identities
\begin{equation}
 \begin{split}
 {g_{i,4}\over \Delta_i^2}
  &= g_{n+1} {1\over \Delta_i}\sum_{j<k\atop j,k\neq i}{1\over \alpha_{ij}\alpha_{ik}},\qquad
 {g_{i,3}\over \alpha_{ij}\Delta_i^2}
 =
  g_{n+1}
  \biggl(
  {1\over \alpha_{ij}^2\Delta_i}+{1\over \alpha_{ij}\Delta_i}\sum_{k\neq i,j}{1\over \alpha_{ik}}
  \biggr),
\\
 {g_{i,3}^2\over \Delta_i^3}
&=g_{n+1}^2  
  \biggl(
  -\sum_{j\neq i}{1\over \alpha_{ij}^2 \Delta_j}
  +{1\over \Delta_i}\sum_{j<k\atop j,k\neq i} {1\over \alpha_{ij}\alpha_{ik}}
  \biggr),
\end{split}\label{g3g4rel}
\end{equation}
upon using which \eqref{f1_gen} becomes, 
after setting $g_{n+1}=1$,
\begin{align}
 f_1(N)
 =&{2\over 3}
 \sum_i \biggl(
 \sum_{j\neq i}{1\over \alpha_{ij}^2\Delta_j}
 -{1\over 4\Delta_i}\sum_{j<k\atop j,k\neq i}{1\over \alpha_{ij}\alpha_{ik}}
 \biggr)(N^i)^3
 \notag\\
 &\qquad\qquad
 + \sum_{i\neq j}
 \biggl({3\over \Delta_i\alpha_{ij}^2}-{2\over \Delta_j\alpha_{ij}^2}
 +{2\over \alpha_{ij}\Delta_{i}}\sum_{k\neq i,j}{1\over \alpha_{ik}}\biggr)(N^i)^2 N^j
 \notag\\
 &\qquad\qquad
 +4\sum_{i<j<k} \biggl( 
    {1\over \Delta_i \alpha_{ij}\alpha_{ki}}
   +{1\over \Delta_j \alpha_{jk}\alpha_{ij}} 
   +{1\over \Delta_k \alpha_{ki}\alpha_{jk}} 
 \biggr)N^i N^j N^k
 \notag\\
 &+\sum_i \biggl(
 {1\over 6}\sum_{j\neq i}{1\over \alpha_{ij}\Delta_j^2}
 +{1\over 12\Delta_i}\sum_{j<k \atop j,k\neq i}{1\over \alpha_{ij}\alpha_{ik}}
 \biggr)N^i.
\end{align}

The second order result ($\CO(N^4)$) is much more lengthy.  Let us write
it as:
\begin{equation}
 \begin{split}
  f_2(N) =& 
 \sum_i a_{iiii} N_i^4 + \sum_{i\neq j} a_{iiij} N_i^3 N_j + \sum_{i<j} a_{iijj} N_i^2 N_j^2
 \\
 &\qquad\qquad\qquad
 +\sum_{i,j,k\atop j<k} a_{iijk} N_i^2 N_i N_j +\sum_{i<j<k<l} a_{ijkl} N_i N_j N_k N_l
 \\
 &\qquad
 +\sum_{i} b_{ii} N_i^2 + \sum_{i<j} b_{ij} N_i N_j.
\end{split}
\end{equation}
Then the coefficients $a_{iiii}$, etc.\ are:
\begin{align}
 a_{iiii}=&
  +\frac{5 g_{i,6}          }{6 \Delta_i^3}
  -\frac{3 g_{i,5} g_{i,3}  }{  \Delta_i^4}
  -\frac{9 g_{i,4}^2        }{8 \Delta_i^4}
  +\frac{6 g_{i,4} g_{i,3}^2}{  \Delta_i^5}
  -\frac{8 g_{i,3}^4        }{3 \Delta_i^6},
  \notag\\
 a_{iiij}=&
  +\frac{4 g_{i,5}          }{\alpha_{ij}   \Delta_i^3}
  -\frac{12g_{i,4}   g_{i,3}}{\alpha_{ij}   \Delta_i^4}
  -\frac{2 g_{i,4}          }{\alpha_{ij}^2 \Delta_i^3}
  +\frac{8 g_{i,3}^3        }{\alpha_{ij}   \Delta_i^5}
  +\frac{4 g_{i,3}^2        }{\alpha_{ij}^2 \Delta_i^4}
  +\frac{8 g_{i,3}          }{3 \alpha_{ij}^3 \Delta_i^3}
  -\frac{8 g_{j,3}          }{3 \alpha_{ij}^3 \Delta_j^3}
  -\frac{4 g_{i,3}          }{\alpha _{ij}^3 \Delta_i^2 \Delta _j}
 \notag\\
  &
  +\frac{1}{\alpha_{ij}^4 \Delta _i^2}
  +\frac{4}{\alpha_{ij}^4 \Delta _j^2}
  -\frac{4}{\alpha_{ij}^4 \Delta _i \Delta _j},
 \notag\\
 a_{iijj} =&
  +\frac{6 g_{i,4}        }{\alpha_{ij}^2 \Delta _i^3}
  +\frac{6 g_{j,4}        }{\alpha_{ij}^2 \Delta _j^3}
  -\frac{8 g_{i,3}^2      }{\alpha_{ij}^2 \Delta_i^4}
  -\frac{8 g_{j,3}^2      }{\alpha_{ij}^2 \Delta_j^4}
  -\frac{2 g_{i,3} g_{j,3}}{\alpha_{ij}^2 \Delta_i^2 \Delta_j^2}
 \notag\\
 & -\frac{8 g_{i,3}        }{\alpha_{ij}^3 \Delta_i^3}
  +\frac{8 g_{j,3}        }{\alpha_{ij}^3 \Delta_j^3}
  +\frac{2 g_{i,3}        }{\alpha_{ij}^3 \Delta_i^2 \Delta_j}
  -\frac{2 g_{j,3}        }{\alpha_{ij}^3 \Delta_i \Delta_j^2}
  -\frac{ 5}{\alpha_{ij}^4 \Delta _i^2}
  -\frac{ 5}{\alpha_{ij}^4 \Delta _j^2}
  +\frac{11}{\alpha_{ij}^4 \Delta _i \Delta _j},
\notag\\
  a_{iijk}=&
  +\frac{12g_{i,4}  }{\alpha_{ij}   \alpha_{ik}   \Delta_i^3}
  -\frac{16g_{i,3}^2}{\alpha_{ij}   \alpha_{ik}   \Delta_i^4}
  \notag\\
  &
  -\frac{8 g_{i,3}  }{\alpha_{ij}^2 \alpha_{ik}   \Delta_i^3}
  -\frac{8 g_{i,3}  }{\alpha_{ik}^2 \alpha_{ij}   \Delta_i^3}
  +\frac{8 g_{j,3}  }{\alpha_{ij}^2 \alpha_{jk}   \Delta_j^3}
  -\frac{8 g_{k,3}  }{\alpha_{ik}^2 \alpha_{jk}   \Delta_k^3}
  +\frac{4 g_{i,3}  }{\alpha_{ij}^2 \alpha_{jk}   \Delta_i^2 \Delta _j}
  -\frac{4 g_{i,3}  }{\alpha_{ik}^2 \alpha_{jk}   \Delta_i^2 \Delta _k}
  \notag\\
  &
  -\frac{8}{\alpha_{ij}   \alpha_{ik}   \alpha_{jk}^2 \Delta_j \Delta_k}
  -\frac{8}{\alpha_{ij}^3 \alpha_{jk}   \Delta_j^2}
  +\frac{8}{\alpha_{ij}^3 \alpha_{ik}   \Delta_i \Delta _j}
  +\frac{8}{\alpha_{ik}^3 \alpha_{jk}   \Delta_k^2}
  +\frac{8}{\alpha_{ik}^3 \alpha_{ij}   \Delta_i \Delta _k}
  \notag\\
  &
  -\frac{4}{\alpha_{ij}^3 \alpha_{ik}   \Delta_i^2}
  +\frac{4}{\alpha_{ij}^3 \alpha_{jk}   \Delta_i \Delta _j}
  -\frac{4}{\alpha_{ik}^3 \alpha_{ij}   \Delta_i^2}
  -\frac{4}{\alpha_{ik}^3 \alpha_{jk}   \Delta_i \Delta _k}
  \notag\\
  &
  +\frac{4}{\alpha_{ij}^2 \alpha_{jk}^2 \Delta_j^2}
  -\frac{2}{\alpha_{ij}^2 \alpha_{ik}^2 \Delta_i^2}
  +\frac{4}{\alpha_{ik}^2 \alpha_{jk}^2 \Delta_k^2} ,
\notag\\
  a_{ijkl}=&
  +\frac{16 g_{i,3}}{\alpha _{ij} \alpha_{ik} \alpha_{il} \Delta _i^3}
  -\frac{16 g_{j,3}}{\alpha _{ij} \alpha_{jk} \alpha_{jl} \Delta _j^3}
  +\frac{16 g_{k,3}}{\alpha _{ik} \alpha_{jk} \alpha_{kl} \Delta _k^3}
  -\frac{16 g_{l,3}}{\alpha _{il} \alpha_{jl} \alpha_{kl} \Delta _l^3}
 \notag\\
  &
  -\frac{8}{\alpha_{ij}^2 \alpha_{il}   \alpha_{jk}   \Delta _i \Delta _j}
  -\frac{8}{\alpha_{ij}^2 \alpha_{ik}   \alpha_{jl}   \Delta _i \Delta _j}
  +\frac{8}{\alpha_{ik}^2 \alpha_{il}   \alpha_{jk}   \Delta _i \Delta _k}
  -\frac{8}{\alpha_{ik}^2 \alpha_{ij}   \alpha_{kl}   \Delta _i \Delta _k}
 \notag\\
  &
  +\frac{8}{\alpha_{il}^2 \alpha_{ik}   \alpha_{jl}   \Delta _i \Delta _l}
  +\frac{8}{\alpha_{il}^2 \alpha_{ij}   \alpha_{kl}   \Delta _i \Delta _l}
  +\frac{8}{\alpha_{jk}^2 \alpha_{ik}   \alpha_{jl}   \Delta _j \Delta _k}
  +\frac{8}{\alpha_{jk}^2 \alpha_{ij}   \alpha_{kl}   \Delta _j \Delta _k}
 \notag\\
  &
  -\frac{8}{\alpha_{jl}^2 \alpha_{ij}   \alpha_{kl}   \Delta _j \Delta _l}
  +\frac{8}{\alpha_{jl}^2 \alpha_{il}   \alpha_{jk}   \Delta _j \Delta _l}
  -\frac{8}{\alpha_{kl}^2 \alpha_{il}   \alpha_{jk}   \Delta _k \Delta _l}
  -\frac{8}{\alpha_{kl}^2 \alpha_{ik}   \alpha_{jl}   \Delta _k \Delta _l}
 \notag\\
  &
  +\frac{8}{\alpha_{ij}^2 \alpha_{ik}   \alpha_{il}   \Delta _i^2}
  +\frac{8}{\alpha_{ij}^2 \alpha_{jk}   \alpha_{jl}   \Delta _j^2}
  +\frac{8}{\alpha_{ik}^2 \alpha_{ij}   \alpha_{il}   \Delta _i^2}
  -\frac{8}{\alpha_{ik}^2 \alpha_{jk}   \alpha_{kl}   \Delta _k^2}
 \notag\\
  &
  +\frac{8}{\alpha_{il}^2 \alpha_{ij}   \alpha_{ik}   \Delta _i^2}
  +\frac{8}{\alpha_{il}^2 \alpha_{jl}   \alpha_{kl}   \Delta _l^2}
  -\frac{8}{\alpha_{jk}^2 \alpha_{ij}   \alpha_{jl}   \Delta _j^2}
  -\frac{8}{\alpha_{jk}^2 \alpha_{ik}   \alpha_{kl}   \Delta _k^2}
 \notag\\
  &
  -\frac{8}{\alpha_{jl}^2 \alpha_{ij}   \alpha_{jk}   \Delta _j^2}
  +\frac{8}{\alpha_{jl}^2 \alpha_{il}   \alpha_{kl}   \Delta _l^2}
  +\frac{8}{\alpha_{kl}^2 \alpha_{ik}   \alpha_{jk}   \Delta _k^2}
  +\frac{8}{\alpha_{kl}^2 \alpha_{il}   \alpha_{jl}   \Delta _l^2}
\notag\\ 
b_{ii}=&
  +\frac{ 5 g_{i,6}          }{3 \Delta_i^3}
  -\frac{ 4 g_{i,5} g_{i,3}  }{  \Delta_i^4}
  +\frac{13 g_{i,4} g_{i,3}^2}{2 \Delta_i^5}
  -\frac{15 g_{i,4}^2        }{8 \Delta_i^4}
  -\frac{ 7 g_{i,3}^4        }{3 \Delta_i^6}
 \notag\\
b_{ij}=&
  +\frac{2 g_{i,5}          }{\alpha _{ij} \Delta _i^3}
  -\frac{2 g_{j,5}          }{\alpha _{ij}   \Delta _j^3}
  -\frac{4 g_{i,4}   g_{i,3}}{\alpha _{ij} \Delta _i^4}
  +\frac{4 g_{j,3}   g_{j,4}}{\alpha _{ij} \Delta _j^4}
  -\frac{  g_{i,4}          }{\alpha _{ij}^2 \Delta   _i^3}
  -\frac{  g_{j,4}          }{\alpha _{ij}^2 \Delta _j^3}
 \notag\\
  &
  +\frac{2 g_{i,3}^3        }{\alpha _{ij} \Delta _i^5}
  -\frac{2 g_{j,3}^3        }{\alpha   _{ij} \Delta _j^5}
  +\frac{  g_{i,3}^2        }{\alpha _{ij}^2 \Delta _i^4}
  +\frac{  g_{j,3}^2        }{\alpha _{ij}^2 \Delta _j^4}
  +\frac{2 g_{i,3}          }{3 \alpha _{ij}^3 \Delta _i^3}
  -\frac{2 g_{j,3}          }{3 \alpha _{ij}^3 \Delta _j^3}
 \notag\\
  &
  -\frac{2}{\alpha _{ij}^4 \Delta _i \Delta _j}
  +\frac{1}{2 \alpha _{ij}^4 \Delta _i^2}
  +\frac{1}{2 \alpha   _{ij}^4 \Delta _j^2}
 \label{f2coeffs}
\end{align}

The third order result ($\CO(N^5)$) is too lengthy to be included here.
The interested reader can find the result in the Mathematica file
included in the source file for the current paper at arXiv.org.

It is easy to identify the matrix model diagrams corresponding to each
term in the above result.  For example, the first term in $a_{iiii}$ in
the second order result \eqref{f2coeffs}, ${5g_{i,6}\over 6\Delta_i^3}$
comes from the following two planar diagrams:
\begin{center}
 \begin{tabular}{c@{\hspace{1cm}}c}
  \includegraphics[width=2.5cm]{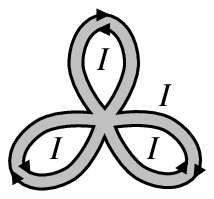} &  \includegraphics[width=2.5cm]{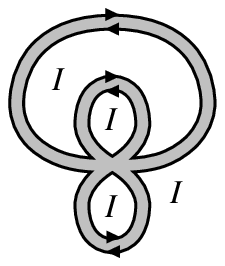} \\
  $-2\cdot (-{g_{i,6}\over 6})\cdot {1\over \Delta_i^3}$ & $-3\cdot (-{g_{i,6}\over 6})\cdot {1\over \Delta_i^3}$
 \end{tabular}
\end{center}
Note that the definitions \eqref{MMint_app} means that the coupling
constants $g_{i,p}$ enter the free energy $F_{mm}$ with a sign as
follows:
\begin{align}
 -\prod_{i,p} (-g_{i,p}).
\end{align}
This is in addition to the signs coming from the fermionic ghosts.

\end{document}